\documentclass[12pt]{article}
\pdfoutput=1

\usepackage{amssymb,amsmath,mathrsfs,epstopdf,slashed,color}
\usepackage{tikz}
\usetikzlibrary{matrix}
\usetikzlibrary{positioning}
\usepackage{fancyhdr} % Required for custom headers
\usepackage{lastpage} % Required to determine the last page for the footer
\usepackage{extramarks} % Required for headers and footers
\usepackage{graphicx} % Required to insert images
\usepackage{lipsum} % Used for inserting dummy 'Lorem ipsum' text into the template
\usepackage{amsmath}
\usepackage{amsfonts}
\usepackage{amssymb}
\usepackage{latexsym}
\usepackage{color}
\usepackage{xypic}
\usepackage{cancel,slashed}
\usepackage[linktocpage]{hyperref}
\usepackage{booktabs}
\usepackage{caption}
\usepackage{comment}
\usepackage{csquotes} % for quotes like conjectures etc
\usepackage{makecell}
\usepackage{graphicx}
\usepackage{placeins}
\usepackage{hyperref}
\usepackage{cite}
\usepackage{braket}
\usepackage{bm}
\usepackage{diagbox}
\usepackage{multirow}

\graphicspath{{./figure/}}
\usepackage{xcolor}
\usepackage{xspace}

\def\d{{\rm d}}

\allowdisplaybreaks[1]

\makeatletter

\@addtoreset{equation}{section}
\makeatother

\begin{document}
	
	\begin{titlepage}
		
		\setcounter{page}{0}
		
		\begin{flushright}
			\small
			\normalsize
		\end{flushright}

		%\begin{titlepage}
		
		\setcounter{page}{0}
		
		\begin{flushright}
			\small
			\normalsize
		\end{flushright}

		\vskip 1cm
		\begin{center}
			
			%    {\Large {\bf DRAFT}}
			
			\vskip 0.7cm
			
			%   {\Large An Axion and The Higgs}
			
			%\vskip 0.5cm
			
			{\Large \textbf{Axi-Higgs Cosmology}}
			
			%\vskip 0.5cm
			
			%{\Large Cosmological Tensions Resolved ?}
			
			%\vskip 0.4cm
			
			%{\Large An Upshift of the Electroweak Higgs Vacuum Expectation Value in Early Universe}
			
			\vskip 0.6cm
			
			{Leo WH Fung${}^{1,a}$,  Lingfeng Li${}^{1,b}$,  Tao Liu${}^{1,c}$, \\ Hoang Nhan Luu${}^{1,d}$, Yu-Cheng Qiu${}^{1,e}$, 
			S.-H. Henry Tye${}^{1,2,f}$}
			
			\vskip 0.6cm

			${}^1$ Department of Physics and Jockey Club Institute for Advanced Study, \\
			Hong Kong University of Science and Technology, Hong Kong S.A.R., China\\
			${}^2$ Department of Physics, Cornell University, Ithaca, NY 14853, USA \\
			
			\vskip 0.6cm
			
			Email: ${}^a$~\href{whfungad@connect.ust.hk}{whfungad@connect.ust.hk}, ${}^b$~\href{iaslfli@ust.hk}{iaslfli@ust.hk}, ${}^c$~\href{taoliu@ust.hk}{taoliu@ust.hk}, \\ ${}^d$~\href{hnluu@connect.ust.hk}{hnluu@connect.ust.hk}, ${}^e$~\href{yqiuai@connect.ust.hk}{yqiuai@connect.ust.hk}, ${}^f~$\href{iastye@ust.hk}{iastye@ust.hk}

			\vskip 0.7cm
			
			\abstract{\normalsize
	If the electroweak Higgs vacuum expectation value $v$ in early universe is $\sim 1 \%$ higher than its present value $v_0=246$ GeV, the $^7$Li puzzle in BBN and the CMB/$\Lambda$CDM tension  with late-universe measurements on Hubble parameter are mitigated. We propose a model of an axion coupled to the Higgs field, named ``axi-Higgs'', with its mass $m_a \sim 10^{-30} - 10^{-29}\,{\rm eV}$ and decay constant $f_a \sim 10^{17} - 10^{18}\,{\rm GeV}$, to achieve this goal. The axion initial value  $a_{\rm ini}$ yields an initial $\Delta v_{\rm ini}/v_0 \sim 0.01$ throughout the BBN-recombination epoch and a percent level contribution to the total matter density today. Because of its very large de Broglie wavelength, this axion matter density $\omega_a$ suppresses the matter power spectrum, alleviating the CMB/$\Lambda$CDM $S_8/\sigma_8$ tension with the weak-lensing data. It also explains the recently reported isotropic cosmic birefringence by its coupling with photons. Adding the axion ($m \sim 10^{-22}\,$eV) in the fuzzy dark matter model to the axi-Higgs model allows bigger $\Delta v_{\rm rec}$ and $\omega_a$ to address the Hubble and $S_8/\sigma_8$ tensions simultaneously. The model predicts that $\Delta v$ may be detected by the spectral measurements of quasars, while its oscillation may be observed in the atomic clock measurements.
	\\

			}
			\vspace{5mm}
			
			\vspace{0.6cm}
			\begin{flushleft}
%				\today
			\end{flushleft}
			
		\end{center}
	\end{titlepage}

	\setcounter{page}{1}
	\setcounter{footnote}{0}
	
	\tableofcontents
	
	\parskip=5pt

	\section{Introduction}
	
Cosmology has made tremendous progress since the mid-20th century, moving from a speculative to a precision science. The inflationary universe scenario, big bang nucleosynthesis (BBN), cosmic microwave background (CMB) and structure formation have merged theory and observational data into a generally accepted picture of our universe.

Two prominent successes in precision cosmology are the measurement of BBN and the determination of Hubble parameter $H_0$. However, as more and better data becomes available while theoretical understanding is progressing, tensions (or frictions/conflicts) emerge. They include in particular the four cases listed below. 

\begin{enumerate}
	
	\item While theoretical estimates for the primordial abundances of helium $^4\rm He$ and deuterium $\rm D$ in BBN are consistent	with the observational data, the theoretical prediction for the primordial Lithium abundance, $^7\rm Li/H  = (5.62\pm 0.25) \times 10^{-10}$,  is too big compared to its observed value $^7\rm Li/H^{\rm obs} = (1.6 \pm 0.3) \times 10^{-10}$. This $\sim 9 \sigma$ discrepancy is known as the $^7$Li puzzle~\cite{Kneller:2003xf}.
	
	\item The determination of the Hubble parameter value from the CMB measurement in Planck 2018 (P18) within the $\Lambda$ cold dark matter ($\Lambda$CDM) model (early universe), namely $H_{0, {\rm P18}}=67.36 \pm 0.54$ km/s/Mpc~\cite{Aghanim:2018eyx},  is smaller than $H_{0,\text{late}}=73.3 \pm 0.8$  km/s/Mpc, the Hubble parameter value obtained from late-time (with redshift $z < 2$) measurements~\cite{Verde:2019ivm}. This $\sim 4-6 \, \sigma$ discrepancy is referred to as the Hubble tension.
	
	\item Recently, a measurement of isotropic cosmic birefringence (ICB) was reported, based on the cross-power (parity-violating) $C^{EB}_l$ data in CMB~\cite{Minami:2020odp}. It excludes the null hypothesis at 99.2\% confidence level (C.L.). This needs to be explained too.
	
		\item The weak lensing measurement of $S_8$ together with the clustering parameter $\sigma_8$~\cite{Troxel:2017xyo} yields a value smaller than that given by the CMB/$\Lambda$CDM value.
	This $\sim 2-3 \,\sigma$~\cite{Hildebrandt:2016iqg, Handley:2019wlz} discrepancy poses another problem to our understanding of the universe.

\end{enumerate}

In this paper, we present a simple model, with an axion coupled to the Higgs field and hence named ``axi-Higgs'', to solve or alleviate these four tensions. Let us consider the possibility that the Higgs vacuum expectation value (VEV) in the standard model (SM) of particle physics, $v_0=246$ GeV today, is $\sim 1 \%$ higher in the early universe, $i.e.$, $\delta v_{\rm ini} = (v_{\rm ini} -v_0)/v_0 \sim 1 \%$~\footnote{In this paper, we will take a set of shorthand notations, including $\Delta X = X - X_{\rm ref}$, $\delta X = d\ln X = \frac{X - X_{\rm ref}}{X_{\rm ref}}$ and $Y_{|X} =\frac{\partial \ln Y}{ \partial \ln X}$, $Y_{||X} =\frac{d \ln Y}{d \ln X}$, unless otherwise specified. If $X= \omega_{b}$, the notations of $Y_{|X}$ and $Y_{||X}$ will be further simplified as $Y_{|b}$ and $Y_{||b}$ etc.}. If the massive gauge bosons, quarks and charged leptons in the SM all have masses of about $\delta v_{\rm ini}$ higher than their today's values, the discrepancies in the first two cases will be substantially reduced. We propose that a $\delta v >0$  is the leading effect in modifying the $\Lambda$CDM model in the early universe. 

That a $\delta v_{\rm BBN} \gtrsim 1\%$ at BBN time solves the $^7$Li problem is known~\cite{Kneller:2003xf, Li:2005km,Coc:2006sx,Dent:2007zu, Browder:2008em, Bedaque:2010hr,Cheoun:2011yn,Berengut:2013nh,Hall:2014dfa,Heffernan:2017hwa,Mori:2019cfo}. That an electron mass
$m_e \propto v$ about $1\%$ higher at recombination time ($i.e.$, $\delta m_e \simeq \delta v_{\rm rec}$) has been suggested to alleviate the Hubble tension~\cite{Ade:2014zfo,Hart:2019dxi}. To implement both, the Higgs VEV with $\delta v \sim 1 \%$ needs to stay throughout the BBN-recombination epoch (from seconds/minutes to 380,000 years after the Big-Bang) and then drops to its today's value where its drift rate is $\lesssim 10^{-16}{\rm yr}^{-1}$, to satisfy the observational bounds~\cite{Huntemann:2014dya,Godun:2014naa,Lange:2020cul}. 

Such a setup can be naturally achieved in string theory. Consider the scenario of brane world in Type IIB string theory, where anti-D3-branes span our three spacial dimensional universe. The SM particles are open-string modes inside the branes. It is known that the electroweak-scale interactions will shift the cosmological constant $\Lambda$ by many orders of magnitude above its exponentially-small observed value, so fine-tuning is needed to have the right value. In the supergravity (SUGRA) model proposed recently~\cite{Li:2020rzo}, a superpotential $W = X(m_s^2F(A) - \kappa H_u H_d) + \cdots$ is introduced. Here $A$ stands for complex-structure (shape) moduli and dilaton that describe the compactification of extra dimensions and $X$ is a nilpotent superfield which projects the two electroweak Higgs doublets $H_u$, $H_d$ to the single Higgs doublet $\phi$. This leads to the axi-Higgs model,
\begin{align}\label{model}
	V = m_a^2f_a^2\left( 1-\cos \frac{a}{f_a} \right)+ \left|m_s^2F(a) - \kappa \phi^{\dagger} \phi\right|^2 \ , \  \  {\rm with} \  \ 
	F(a) = 1 + C \frac{a^2}{M_{\rm Pl}^2} \ . 
\end{align}
In this model, the axion-like field $a$ is a pseudo-scalar component in $A$. This axion starts with an initial value $a_{\rm ini}$ in the early universe. We normalize $F(a)$ to be $F(a =0)=1$, such that the Higgs VEV $v_0=\sqrt{2}m_s/\sqrt{\kappa} =246\, {\rm GeV}$ and the Higgs boson mass $m_\phi=2 m_s \sqrt{\kappa}=125\, {\rm GeV}$. So this model is characterized by four parameters, namely $m_a, f_a, C$ and $a_{\rm ini}$. The perfect square form of the Higgs potential, where the Higgs contribution to $\Lambda$ is completely screened by the supersymmetry (SUSY) breaking anti-D3-brane tension $m_s^4$, allows a naturally small $\Lambda$~\cite{Sumitomo:2013vla,Qiu:2020los}. Notably, this perfect square form of the Higgs potential, together with the damping effect of the Higgs decay width ($\Gamma_\phi \simeq 4$ MeV), is crucial in yielding the desirable feature of the model: the effect of the Higgs field evolution is totally negligible in the axion evolution, but the axion evolution significantly affects the evolution of the Higgs VEV~\footnote{The name "axi-Higgs" has also been used to refer to a boson in a model~\cite{Coriano:2005own} different from the one described by Eq.~(\ref{model}).}. Note that all parameters in the standard electroweak model are unchanged. In particular, the electron Yukawa coupling is unchanged, so $\delta v = \delta m_e$.

Starting with an initial $\delta v_{\rm ini} = C a_{\rm ini}^2/2M_{\rm Pl}^2$ for $a_{\rm ini}\ne 0$, via the mis-alignment mechanism~\cite{Preskill:1982cy,Abbott:1982af,Dine:1982ah}, $\delta v$ evolves after the recombination epoch ($z \sim 10^3$) when $3H(t)$ drops below $m_a$. We find the favored axion mass
\begin{equation}\label{eq:a1mass}
	m_a \sim 10^{-30} - 10^{-29} \, {\rm eV}  \ .
\end{equation}
Here the upper limit of $m_a$ is determined by whether $\delta v$ will drop too much by the time of recombination, which happens for $m_a >  3.3 \times 10^{-29}\,{\rm eV}$. The lower limit of $m_a$, instead, is set by the late-time measurements of $\delta v(t)$ or its drift rate. The current atomic clock (AC) measurements on $d(\delta v)/d t|_{t_0}$~\cite{Lange:2020cul} excludes $m_a \lesssim 1.6\times 10^{-30}\, {\rm eV}$ at 95\% C.L. Such a mass scale is compatible with string theory and typical axion masses~\cite{Hui:2016ltb,Tye:2016jzi}. Note that it is very difficult to satisfy the AC bound today if we introduce a scalar field $\varphi$ instead, as $F(\varphi)$, a counterpart of $F(a)$ in Eq.~\eqref{model}, will contain a linear term with a coefficient too big in the absence of fine-tuning.

Physically, an upward variance of the Higgs VEV will reduce $Y_{\rm p}$ but raise D/H. The current experimental bounds on $Y_{\rm p}$ and D/H are still compatible with a change of percent level in $v$ if $\eta$ is also $1-2\%$ larger than its reference value $6.127\times 10^{-10}$~\cite{Pitrou:2018cgg,Aghanim:2018eyx}. Beyond that, it is  suggested in~\cite{Li:2005km,Dent:2007zu,Cheoun:2011yn,Mori:2019cfo} that the $^7$Li problem can be greatly alleviated if the light quarks are $\sim 1\%$ heavier during the BBN epoch. Following this, we find that addressing the $^7$Li problem yields 
\begin{equation} \label{BBN_linear1}
	\delta v_{\rm BBN} = (1.1 \pm 0.1)\%,  \ \ \ \ \delta \eta = (1.7\pm 1.3)\%  \ .
\end{equation}
Here the baryon density $\omega_b$ is about $1.7\%$ higher than the value obtained from the P18 data. 

We then introduce a semi-analytical formalism to study the impacts of $\delta v_{\rm rec} = \delta v_{\rm BBN} \sim 1.1\% $ for the combined P18+BAO (Baryon Acoustic Oscillation) predictions~\cite{Pogosian:2020ded} using the original fitting results of $\Lambda$CDM as a reference~\cite{Aghanim:2018eyx}. Since $\delta v_{\rm rec}$ is small, we treat its effects perturbatively. We demand the angular sound horizon $\theta_*$ at the recombination to be preserved while the $r_d h$ value to be shifted from its CMB value to its BAO value when $\delta v_{\rm rec}$ is turned on. Here $r_d$ is the sound horizon at the end of baryon drag epoch and $h$ is the dimensionless Hubble parameter. 
Then with the inputs from the linear BBN analysis, namely $\delta v_{\rm rec} = \delta v_{\rm BBN}$ and $\delta \omega_b =\delta \eta$, we eventually find 
\begin{equation}\label{3percent}
	H_0=H_{0,\text{P18}} (1+ h_{||v} \, \delta v_{\rm rec}) = 69.03 \pm 0.61 \, {\rm km/s/Mpc}   \ .
\end{equation}
Here $H_0$ is deviated from its reference value $H_{0,\text{P18}}$ (see Sec.~\ref{sec:cmb}) and its error mostly comes from the BAO uncertainties. This $H_0$ value alleviates the tension with its late-time measurements.  It is consistent with the numerical analysis taken by Planck 2015~\cite{Ade:2014zfo} and Hart \& Chluba~\cite{Hart:2019dxi}. However, as we shall see, this may not be the whole story on $H_0$ in the axi-Higgs model. 

The existence of this axion introduces an axion relic density $\omega_a>0$, hence contributes to the CDM today. Because of its huge de Broglie wavelength ($\sim 10^3$~Mpc), this axion tends to suppress the matter power spectrum. The $S_8/\sigma_8$ prediction by the CMB data is then shifted down from its previously determined value. The weak-lensing and the CMB measurements are thus reconciled to some extent, with  
\begin{equation}
	x\equiv \frac{\omega_a}{\omega_m} \sim 1\%\; 
\end{equation}
or equivalently
\begin{equation}
a_{\rm ini} \sim 10^{17} - 10^{18}\  {\rm GeV} \ . \label{eq:aini}
\end{equation}

Being an ultra-light axion, the coupling of $a$ to two photons naturally introduces an ICB effect in the CMB, at a level in agreement with the recently observed $C^{EB}_l$ spectrum~\cite{Minami:2020odp}. In data fitting, we determine (as in $aF_{\mu \nu}{\tilde{F}}^{\mu \nu}/32 \pi^2f_a$)~\footnote{Fitting the ICB data requires the $a_{\rm ini}$ values to be negative. Here we simply drop this minus sign for the convenience of presentation, since the cosmological problems to be addressed in this paper, except this one, are not sensitive to this sign (protected by the $Z_2$ symmetry of $a\to -a$ in the axi-Higgs model (\ref{model})).\label{ft:z2}}
\begin{equation}
	 \frac{a_{\rm ini}}{f_a} \simeq 1.0 \pm 0.3  \label{eq:afa}
\end{equation}
and hence 
\begin{equation}
	 f_a \simeq 10^{17} - 10^{18}\  {\rm GeV} \ ,  \label{eq:fa}
\end{equation}
with an input of Eq.~(\ref{eq:aini}). Since $a_{\rm ini}/{f_a} < \pi$, the axion rolls towards $a=0$ instead of $a=2\pi$. This explanation as one possibility is already known~\cite{Carroll:1989vb,Carroll:1991zs,Harari:1992ea}, though our determination of the parameters comes from an entirely different direction and is much more precise in terms of the axion properties. Here, $f_a \lesssim M_{\rm Pl} = 2.4 \times 10^{18}\,$\,GeV.

\begin{figure}
	\begin{center}
		\includegraphics[width=0.7\textwidth]{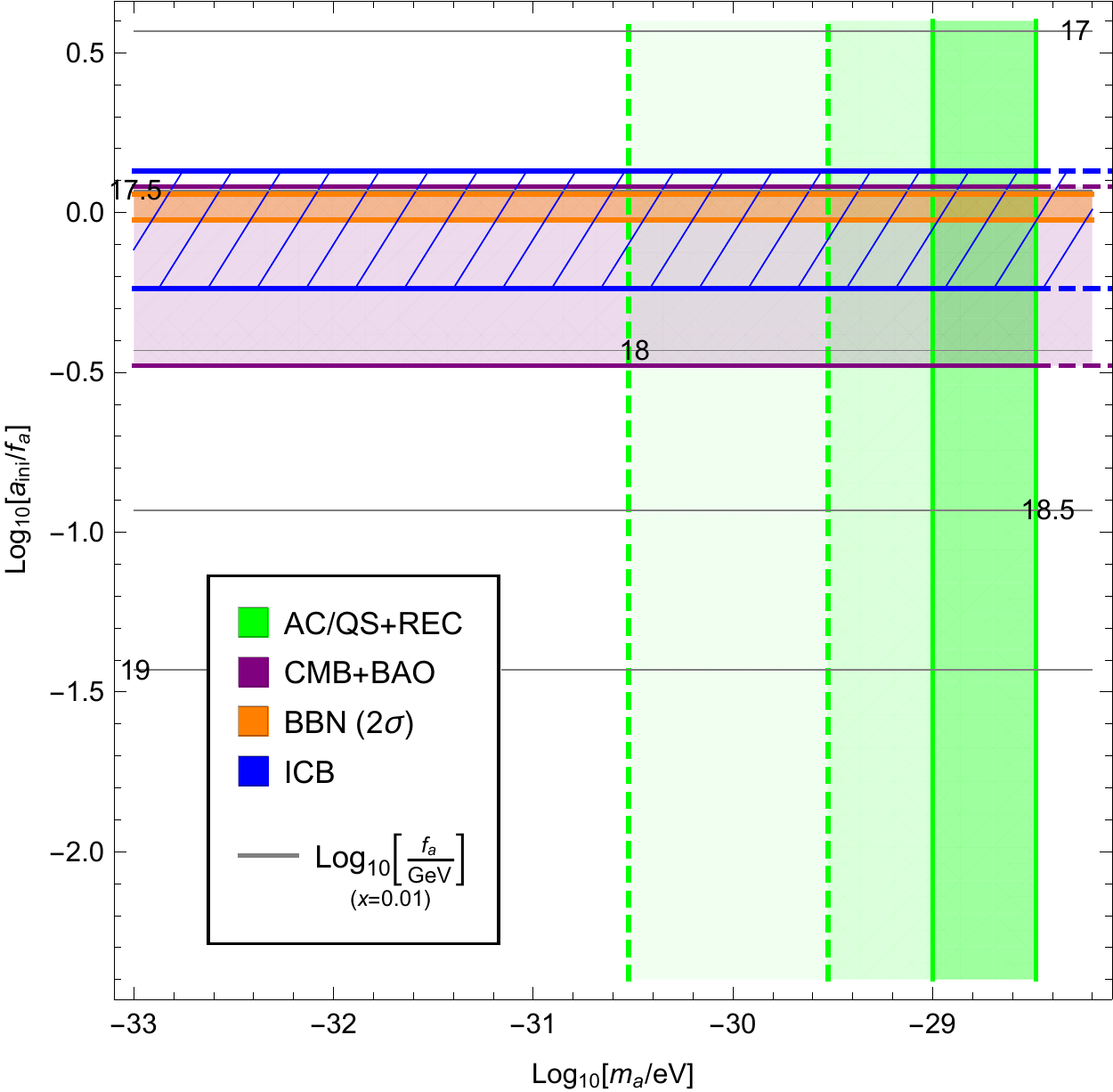}
		\caption{Overall picture on the axi-Higgs cosmology with a single axion. The axion mass is bounded from above by requiring the axion not to roll down until near or after the recombination, and limited from below by the AC measurements on $m_e/m_p$ drift rate~\cite{Lange:2020cul} (solid-green). The projected lower limits from astronomical  observations of molecular absorption spectra, in terms of the present and the two-order improved precisions for eighteen known quasars~\cite{Levshakov:2020ule}, are also presented (dashed-green). The CMB+BAO data, previously encoded in the $\Lambda$CDM+$m_e$ context to address the Hubble tension~\cite{Hart:2019dxi}, is recast in this axi-Higgs model (with $C'=0.01$). The BBN data fitting is shown at $2\sigma$ C.L. ($1\sigma$ C.L. is taken for the others) for better demonstration. In the favored parameter region, the $^7$Li puzzle is largely solved. The recently reported ICB anomaly~\cite{Minami:2020odp} also gets explained in this model. We draw the contours of $f_a$ with $x=0.01$, the value suggested to mitigate the $S_8/\sigma_8$ tension. In the intersection region of all, $f_a$ is favored to be $\sim 10^{17} - 10^{18}\,$~GeV.}
		\label{fig:overall}
	\end{center}
\end{figure}

In summary, the axion density $\omega_a$ determines the value of $a_{\rm ini}$, while the initial variation of the Higgs VEV $\delta v_{\rm ini}$ determines the value of $C a_{\rm ini}^2$.  The axion mass $m_a$ is determined by the requirements that $\delta v_{\rm ini}$ stays unchanged (or mildly changed) until near or after the recombination and oscillates with a highly-suppressed amplitude at low redshift and today, while $f_a$ is determined by the ICB data. The four parameters parametrizing this axi-Higgs model are determined up to an order of magnitude at $1\sigma$ C.L. Note that the impact of $\delta v_{\rm ini}$ in BBN is mostly in  quark (and nucleon) and $W$-boson masses, while its impact on the CMB is mostly via electron mass $m_e$. Fortunately, they are intimately linked in the Standard Model of particle physics, where the particle masses are proportional to $v$. Overall, the properties of the axi-Higgs model in addressing the four issues are presented in Fig.~\ref{fig:overall}. For the convenience of presentation, we redefine $F(a)$ in Eq.~(\ref{model}) as 
\begin{eqnarray}
F(a) = 1 + 2C' \frac{a^2}{f_a^2}, \quad {\rm with} \quad C' = C \frac{f_a^2}{2M_{\rm Pl}^2} \ .  \label{eq:Cp}
\end{eqnarray}

Note that the evolution of $\delta v$ is described by the physics of a damped oscillator. The oscillating feature of $\delta v$ may be detected by the AC measurements~\cite{Huntemann:2014dya,Godun:2014naa,Lange:2020cul}, while its non-zero value may be detected by the quasar (QS) spectral measurements~\cite{Levshakov:2020ule}. With further improvements in their precisions in the near future, the axi-Higgs model should be seriously tested.

Notably, though the Hubble tension and the $S_8/\sigma_8$ tension can be alleviated in the single-axion case, by turning on $\delta v_{\rm rec}$ and $x$ respectively, some trade-off effect exists between relaxing the Hubble and $S_8/\sigma_8$ tensions. Turning on $\delta v_{\rm rec}$ alone exacerbates the $S_8/\sigma_8$ tension while turning on $x$ alone exacerbates the Hubble tension. This friction can be alleviated by allowing a larger $\delta v_{\rm rec}$ if we introduce a second axion. Recall the fuzzy dark matter (FDM) scenario~\cite{Hu:2000ke,Schive:2014dra,Marsh:2015xka,Hui:2016ltb}, in which an axion $a_2$ with mass $m_2 \sim 10^{-22}$~eV comprises the CDM $\omega_c$; here, the problems such as cusp-core, too many satellites, etc confronting the weakly-interacting-massive-particle scenario are generically absent.  In the axi-Higgs model with two axions, $F(a)$ extends to
\begin{equation}\label{model2}
F(a_1, a_2) = (1+\delta v)^2 = 1+ C_1 \frac{a_1^2}{M_{\rm Pl}^2} +C_2 \frac{a_2^2}{M_{\rm Pl}^2} \ ,
\end{equation}
where $a_1$ should be recognized as the counterpart of the $a$ field (see Eq.~(\ref{eq:a1mass})). The FDM axion $a_2$ starts with $a_{2, {\rm ini}}$ at the BBN time and rolls down at a redshift $z_2$ with $z_{\rm rec} \ll z_2\simeq 2.0 \times 10^{6} \ll z_{\rm BBN}$. The present CDM density $\omega_c$ determines the value of $a_{2, {\rm ini}}$. So the $a_2$ contribution in $F(a_1,a_2)$ is important at the BBN epoch but becomes negligible at the recombination time. In this context, $\delta v_{\rm rec} > \delta v_{\rm BBN}$ is allowed with a negative $C_2$. With this additional parameter ($C_2$ or $\delta v_{\rm rec}$) and $\delta v_{\rm BBN} $ remaining at $1.1\%$, we roughly find that a choice of
   \begin{equation}
 \delta v_{\rm rec} \sim 4\,\% \,   \quad {\rm and} \quad x \sim 2\,\%\,  
   \end{equation}
helps to resolve both the Hubble and the $S_8/\sigma_8$ tensions. An analysis pinning down more precise values is forthcoming.

The rest of the paper goes as follows: Sec.~\ref{sec:bbn} covers the BBN epoch. Choosing $\delta v_{\rm ini}=1.1 \%$ reduces the theoretical prediction for $^7\rm Li/H$, mostly due to the caused modifications to the strong/nuclear interaction rates. Sec.~\ref{sec:cmb} discusses the $H_0$ value with the input of $\delta v_{\rm rec} = \delta v_{\rm BBN}$. We present a semi-analytical approach, referring to~\cite{Hart:2019dxi} for a numerical analysis. Substituting in the BBN values for $\delta v$ and $\omega_b$ as inputs, we determine the upshift of $H_0$ from its P18 value. Sec.~\ref{sec:model} presents a simple axi-Higgs model suggested by string theory on how the Higgs VEV evolves from $v_{\rm ini}=v_0(1 + \delta v_{\rm ini})$ to $v_0$ today, predicting the existence of an axion with mass $m_{a} \sim 10^{-30} - 10^{-29}\; {\rm eV}$. Sec.~\ref{sec:s8} discusses the impact of the axion density $\omega_a$ on the CMB measurements of $S_8/\sigma_8$. Sec.~\ref{sec:HS8} discusses the trade-off effect between relaxing the Hubble and  $S_8/\sigma_8$ tensions, where the two-axion model comes in handy. Sec.~\ref{sec:icb} discusses how this axion explains the recently reported ICB anomaly, with $f_{a}\simeq a_{\rm ini} \sim 10^{17}-10^{18} \;{\rm GeV}$. Section~\ref{sec:test} discusses the testing of the axi-Higgs model in the near future, via the AC and/or QS measurements. Sec.~{\ref{sec:con} contains the conclusion and some remarks. The appendix provides some auxiliary information and technical details on the approach adopted in Sec.~\ref{sec:cmb}.}

\vspace{4mm}

	\section{Big-Bang Nucleosynthesis}\label{sec:bbn}
	
	BBN occurs during the radiation-dominant epoch, with a typical temperature scale of $\mathcal{O}$(1-0.1)~MeV, when the radiation becomes too soft to significantly break the generated light chemical elements or bound states of nucleons. Locally, the primordial abundances of these elements can be extrapolated from optical observations, such as the absorption lines of ionized hydron region in compact blue galaxies~\cite{Aver:2015iza}, the QS light passing through distant clouds~\cite{Cooke:2017cwo}, and the spectra of metal-poor main-sequence stars~\cite{Sbordone:2010zi}.  Most of the measured values match with their theoretical prediction based on the standard $\Lambda$CDM model with very high precision, except a discrepancy about $9\sigma$ appearing for $^7$Li. This is often named the $^7$Li puzzle~\cite{Fields:2011zzb}. We present the primordial abundances of $^4$He,  D and $^7$Li, including their theoretical predictions and astrophysical measurements, in Tab.~\ref{tab:theory}. Notably, the consistency between the observed $^4$He and D primordial abundances and their theoretical predictions strongly constrains the model space to address this puzzle (for some recent efforts, see e.g.~\cite{Hayakawa:2020bjr,Ishikawa:2020fbm,Clara:2020efx,Iliadis:2020jtc,Li:2005km,Gupta:2020wgz,Mori:2019cfo}), in which the proposal that $\delta v \sim 1\%$ can solve the $^7$Li puzzle has been studied in some detail. Below we will discuss the impacts of a percent-level shift in Higgs VEV for BBN.

	\begin{table}[h!]
		\begin{footnotesize}
			\begin{center}
				\begin{tabular}{ccc}
					\hline
					& Prediction~\cite{Pitrou:2018cgg} & Observation~\cite{Zyla:2020zbs} \\ 
					\hline$Y_{\rm p}$ & 0.2471$\pm$ 0.0002  & 0.245$\pm$ 0.003 \\
					$ \text{D/H} \times 10^5$ & 2.459$\pm 0.036$ & 2.547$\pm$ 0.025\\
					$^7 \rm {Li/H} \times 10^{10}$ & 5.62 $\pm 0.25$ & 1.6 $\pm$ 0.3 \\
					\hline
				\end{tabular}
				
			\end{center}
		\end{footnotesize}
		\caption{Primordial abundances of $^4$He,  D and $^7$Li:  theoretical predications and astrophysical measurements. Here we take the convention in~\cite{Zyla:2020zbs}. In particular, $Y_{\rm p}\equiv \rho(^4\text{He})/\rho_b$ is the primordial mass fraction of $^4$He and D($^7$Li)/H represents that the D($^7$Li) primordial abundances relative to that of H. The theoretical predictions are based on the CMB baryon-to-photon ratio $\eta=6.091\times 10^{-10}$~\cite{Pitrou:2018cgg,Fields:2019pfx}.}
		\label{tab:theory}
	\end{table}

	The shift of Higgs VEV from its current value $\sim 246\,{\rm GeV}$ impacts BBN mainly by modifying the following parameters in particle physics:
	\begin{itemize}
		\item Fermi constant $G_F \propto v^{-2}$, or equivalently $m_{W}\propto v$. The change to $m_W$ modifies all weak interactions, such as the $n \rightleftharpoons p$ conversion and the neutron lifetime. A larger $m_W$ leads to an earlier freeze out of the $n \rightleftharpoons p$ conversion and a longer neutron lifetime. It introduces a larger neutron density than that in the standard BBN picture and thus higher light-element abundances.
		
		\item Electron mass $m_e \propto v$. $m_e$ also plays an important role in weak interactions. A larger $m_e$ will reduce the rate of the $n  \rightleftharpoons p$ conversion and delay neutron decay. Additionally, it may reheat more the photon bath before BBN via electron-positron annihilation. 
		
		\item Mass difference between up and down quarks $\Delta m_q\equiv m_d-m_u \propto v$. The isospin-breaking $\Delta m_q$ effect contributes to the mass splitting between neutron and proton $\Delta m_{np}$~\cite{Walker-Loud:2014iea}, while the latter impacts the $n  \rightleftharpoons p$ conversion and neutron decays oppositely, relative to $m_W$ and $m_e$, as the Higgs VEV varies.     
		
		\item Averaged light quark mass $\bar{m}_q \equiv (m_u+m_d)/2 \propto v$. The change of $\bar{m}_q$ may significantly influence the rates of strong/nuclear interactions. Heuristically, the effect of increasing $\bar{m}_q$ is manifested an enlarged pion mass $m_\pi$. From chiral perturbation theory, we have the well-known relation $m_\pi^2 \simeq \bar{m}_q \left\langle q\bar{q} \right\rangle / f_{\pi}^2$,  where $f_\pi$ is the pion decay constant and $\left\langle q\bar{q} \right\rangle$ is the VEV of quark condensate. Since pions are the main mediators between nucleons, a larger $m_\pi$ makes nuclei less tightly bound. The nuclear-reaction rates thus may change substantially. Here we follow the discussions in~\cite{Flambaum:2007mj,Berengut:2009js,Cheoun:2011yn,Mori:2019cfo}. Note, nucleon mass also changes with $\bar{m}_q$. But this effect is subleading in this context, since the nucleon mass receives contributions mostly from QCD interaction. For $\delta v =1 \%$, $\delta m_W=\delta m_e =1\%$, while $\delta m_{\pi} \simeq 0.5\%$ and $\delta m_p=\delta m_n \simeq 0.1\%$.
		
	\end{itemize}
	Aside from these tree-level impacts, the variation of Higgs VEV can also shift the values of the coupling constants, such as $\alpha$ or $\alpha_s$, or some other physical quantities like $\Lambda_{\rm QCD}$ and neutrino mass.  But, these effects are either of the next-to-leading order or highly model-dependent (for relevant discussions, see, e.g.,~\cite{Hall:2014dfa}). So we will not consider them in this study.

	\begin{table}
		\footnotesize
		\centering
		\begin{tabular}{ |c|c|c|c|c|c| }
			\hline
			\diagbox{Y}{X} & $m_W$~\cite{Dent:2007zu} &$m_e$~\cite{Dent:2007zu} &  $\Delta m_q$~\cite{Dent:2007zu} & $\bar{m}_q$~\cite{Cheoun:2011yn,Mori:2019cfo} & $\eta$ \\
			\hline 
			$Y_{\rm p}$  & $2.9$ & $0.40$ & $-5.9$ & $-1.0$ & $0.039$  \\
			\hline
			D/H  & $1.6$ & $0.59$ & $-5.3$ & $10$ & $-1.6$  \\
			\hline
			$^7$Li/H & $1.7$ & $-0.04$ & $-5.3$ & $-60$ & $2.1$  \\
			\hline
		\end{tabular}
		\caption{Numerical values of $Y_{|X}\equiv \frac{\partial \ln Y}{\partial \ln X}$ for $Y_{\rm p}$,  D/H and $^7$Li/H. The $Y_{|\Delta m_q}$ values are calculated like~\cite{Dent:2007zu}, but using the lattice average in~\cite{Walker-Loud:2014iea} instead. This modification introduces a rescaling factor $\sim1.16$ to the numbers in~\cite{Dent:2007zu}. The $Y_{|\bar{m}_q}$ values are taken from~\cite{Cheoun:2011yn,Mori:2019cfo}, which are derived based on the $E_B$-$\bar{m}_q$ relation presented in~\cite{Flambaum:2007mj}. Here $E_B$ is nucleus binding energy.}
		\label{tab:dependence}
	\end{table}

	We summarize the $Y_{|X}$ values for $Y_{\rm p}$,  D/H and $^7$Li/H in Tab.~\ref{tab:dependence}. Building on the modified binding energy~\footnote{In this work, the binding energy of the excited states is  assumed to have the same shift relative to the corresponding ground states, for the elements involved in the BBN.} and evolution history of $n \rightleftharpoons p$, D, $^3$H, $^3$He, $^4$He, $^7$Be, $^6$Li and $^7$Li, these values measure the dependence of the $^4$He,  D and $^7$Li primordial abundances on the Higgs-VEV-mediated parameters discussed above and the baryon-to-photon ratio~\cite{Dent:2007zu}. As expected, the $Y_{|\Delta m_q}$ values are universally negative for $^4$He,  D and $^7$Li. This distinguishes them from most of the $Y_{|m_W}$ and $Y_{|m_e}$ values by a sign. The only exception is $(^7\text{Li/H})_{|m_e}$. Due to the extra impact of the variation of $m_e$ on the photon bath and hence the generation of $^7$Li through the photon-associated $^7$Be production~\cite{Pitrou:2018cgg}, its value turns out to be slightly negative. Another observation is that the $(^7\text{Li/H})_{|\bar{m}_q} $ value is highly negative. This effect can significantly reduce the predicted $^7$Li primordial abundance given a positive $\delta v_{\rm BBN}$ or an enlarged $\bar{m}_q$, and hence lays out the footstone of addressing the $^7$Li puzzle in this context. We also present the values of $Y_{|\eta}$ in this table. $\eta$ determines baryon number density during the BBN epoch, and influence BBN directly. It also determines $\omega_b$ at recombination time. As an outcome, we have~\cite{Dent:2007zu,Walker-Loud:2014iea}
	\begin{align} 
		Y_{\rm p}(\delta v_{\rm BBN},\delta \eta)&\simeq Y_{\rm p}(0,0)(1-3.6\delta v_{\rm BBN}+0.039  \delta \eta) \; , \label{eq:Yp} \\
		{\rm D/H} (\delta v_{\rm BBN},\delta \eta)&\simeq  {\rm D/H}(0,0)(1+6.9 \delta v_{\rm BBN} - 1.6  \delta \eta) \; , \label{eq:DH} \\
		{\rm ^7Li/H} (\delta v_{\rm BBN},\delta \eta)&\simeq {\rm ^7Li/H}(0,0)(1-64 \delta v_{\rm BBN}+2.1  \delta \eta) \label{eq:LiH} \; .
	\end{align}
	
As just mentioned, the very large negative coefficient of $\delta v_{\rm BBN}$ in ${\rm ^7Li/H} (\delta v_{\rm BBN},\delta \eta)$, namely $-64$, is the reason why $\delta v_{\rm BBN} \sim 1\%$ can decrease the theoretical value for the $^7$Li abundance by a factor of 3 and thus solve the $^7$Li puzzle. Let us review why this coefficient is so big, and whether this linearized approximation is valid for $\delta v_{\rm BBN} \sim 1\%$. 

A large variation in a nuclear reaction rate due to a small change in the pion mass $\delta m_{\pi} =\delta v/2$ is possible if (on/off) resonances play an important role, as indicated in Tab.~\ref{tab:dependence}. Among the main reactions in BBN, $^3$H(d,n)$^4$He, $^3$He(d,p)$^4$He and $^7$Be(n,p)$^7$Li are the only ones whose cross sections are governed by the resonances. There are some uncertainties on how the nuclear forces and binding energies vary as $\delta v_{\rm BBN}$ is turned on. Ref.\cite{Berengut:2009js} points out that the $^7$Li problem remains if  $^3$H(d,n)$^4$He and $^3$He(d,p)$^4$He are sensitive to $\delta v$. But, Ref.\cite{Cheoun:2011yn} argues that, on general grounds, the impacts of $\delta v_{\rm BBN}$ on the $^3$H(d,n)$^4$He and $^3$He(d,p)$^4$He reactions are small and hence can be neglected. With this background, it is shown that the $^7$Be(n,p)$^7$Li resonances move to higher energies. The $^7$Be(n,p)$^7$Li reaction happens a bit off the resonance now, resulting in a smaller production of $^7$Li (because of  the hindered Boltzmann factor ($E \lesssim 0.3$ MeV)). Ref.\cite{Mori:2019cfo} eventually finds that the theoretical prediction matches the observed $^7$Li abundance for $\delta v_{\rm BBN} \sim 0.4 \% - 0.8 \%$.  This result is consistent with that obtained in Ref.\cite{Li:2005km} and Ref.\cite{Dent:2007zu}.
 
	\begin{figure}[h!]
		\centering
		\includegraphics[width=0.7\textwidth]{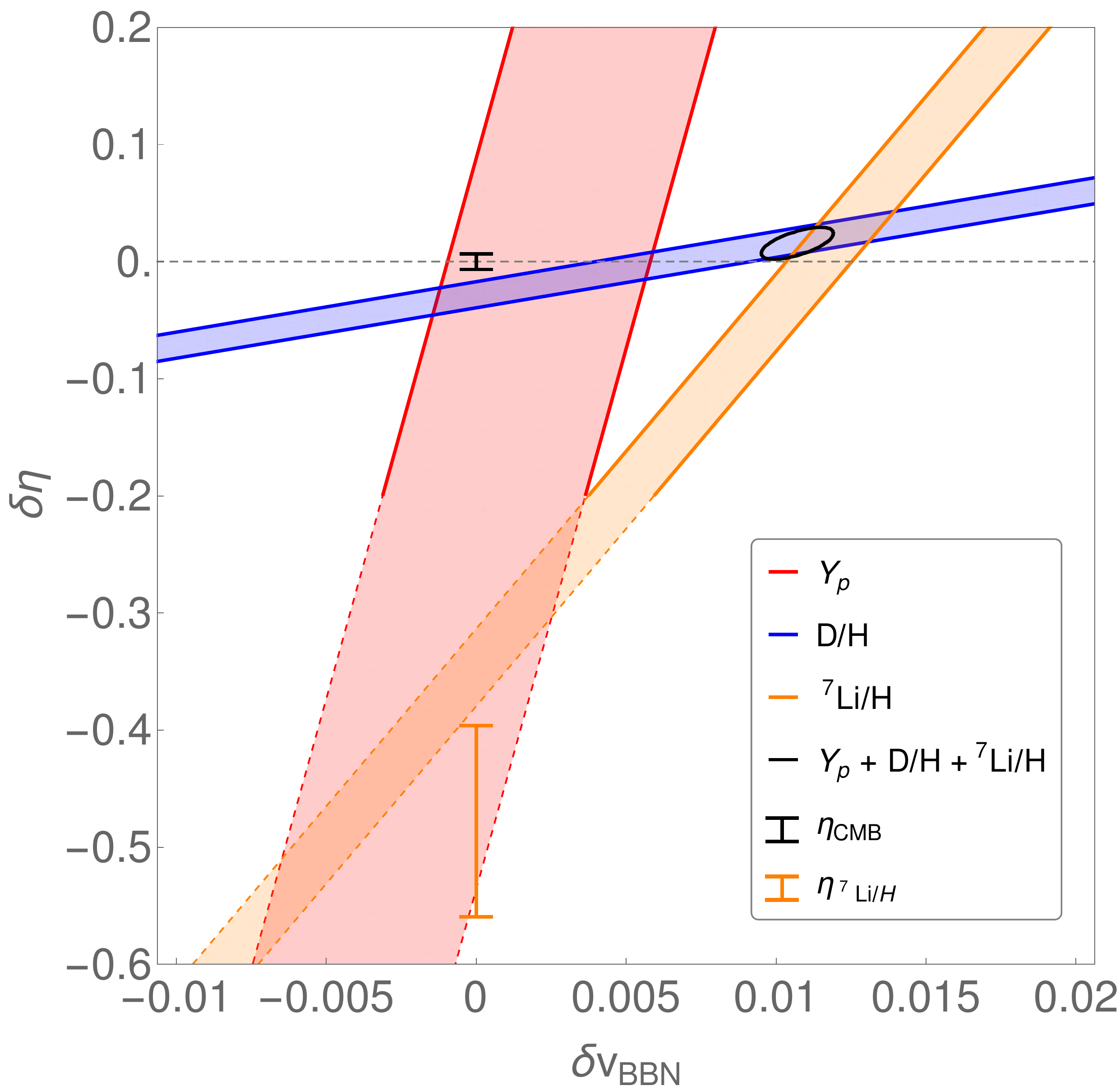}
		\caption{Constraints on $\delta v_{\rm BBN}$ and $\delta \eta$ at $1\sigma$ C.L. Here the black error bar ($\eta_{\rm CMB}=6.127 \pm \times 10^{-10}$ at $\delta v_{\rm BBN} = 0$), namely the interpretation of P18 data in the $\Lambda$CDM~\cite{Aghanim:2018eyx}, represents theoretical prediction from standard cosmology.}
		\label{fig:BBNcombine}
	\end{figure}

	  We present the linear-order constraints on $\delta v$ and $\delta \eta$ at $1\sigma$ C.L. in Fig.~\ref{fig:BBNcombine}. In the standard BBN scenario, the $^7$Li puzzle can be manifested as an $\eta$ value away from the CMB favored one with $\Lambda$CDM. We demonstrate this in this figure as a separation between the orange error bar and the point of $\delta \eta =0$ along the line with $\delta v_{\rm BBN} =0$. The story is dramatically changed in the model with varying Higgs VEV. As $\delta v_{\rm BBN}$ increases, the $\delta \eta$ value favored by $^7$Li/H  gets close to zero quickly. The black circle,
	which will be shown in Sec.~\ref{sec:cmb} fitting the CMB data well and hence can be approximately interpreted as new theoretical prediction, is within $1\sigma$ range of the observed $^7$Li/H! The best fit of $\delta v$ and $\delta \eta$ to $Y_{\rm p}$, D/H and $^7$Li/H now reads:
	\begin{equation} \label{BBN_linear}
		\delta v_{\rm BBN} = (1.1 \pm 0.1)\%,  \ \ \ \ \delta \eta = (1.7\pm 1.3)\%~. 
	\end{equation}
	The reduced $\chi^2$ value at this best-fit point is $\sim 7.0$, yielding a fit at $\sim 2.5\sigma$ level. As a comparison, the data can be fitted in the standard BBN scenario only with a $\chi^2$ value $\sim 42$ or at $\sim 8.8\sigma$ level. The $^7$Li puzzle is indeed greatly relieved in this new model. Notably we have not taken into account non-linear effects of $\delta \eta$ and $\delta v_{\rm BBN}$ in these discussions. While $|\delta \eta|$ being far from zero, its non-linear effects might not be negligible. We thus use dashed lines to represent the boundaries of the shaded regions with $|\delta \eta| > 0.2$ in this figure. This explains why the orange belt, obtained by fitting the observed $^7$Li/H, fails to pass the orange error bar at $\delta v_{\rm BBN} =0$. 
	
	At last, if non-linear corrections of $\mathcal{O}(\delta\bar{m}_q^2)$ from $^7$Li are incorporated, a slightly bigger value will be favored for  $\delta v_{\rm BBN}$~\cite{Mori:2019cfo} and hence for $\delta \eta$. In this case we have  
	\begin{equation} \label{BBN_nonlinear}
			\delta v_{\rm BBN} = (1.2 \pm 0.2)\%,  \ \ \ \ \delta \eta = (2.3 \pm 1.4)\%~. 
	\end{equation}
	So, the linearized approximation is quantitatively consistent with the non-linear treatment for $\delta v_{\rm BBN} \sim 1\%$. It is clear that a better understanding of the nuclear force (as $\delta v_{\rm BBN}  \ne 0$) is important for firmly establishing that $\delta v_{\rm BBN}  \ne 0$ fully solves the $^7$Li puzzle. On the other hand, assuming the validity of the axi-Higgs model in solving the $^7$Li puzzle, we learn something about the nuclear force, the nuclear binding energies and their impacts on the resonances and the nuclear reactions.

\section{Hubble Tension
	} \label{sec:cmb}
	
	 Our today's universe is well-described by Robertson-Walker metric, where its energy density is comprised of about $5\%$ baryons, $25\%$ CDM (be it weakly interacting massive particles or ultra-light axion) and $70\%$ dark energy $\Lambda$. However, today's cosmic expansion rate $H_0$ from CMB ($i.e.$, the early universe's prediction) is substantially smaller than the late-time determination, yielding a $\sim 4-6 \sigma$ discrepancy. We like to examine how a slightly larger Higgs VEV ($\delta v_{\rm rec}\sim 1\%$) at the recombination epoch impacts on the CMB prediction on $H_0$. Since $\delta v_{\rm rec}$ is small, we shall treat its effects on the Hubble parameter $H_0$, the matter density $\omega_m$ and the shift in the recombination redshift $z_*$ perturbatively, at a linear level. This allows us to study this problem analytically, so one can get a clearer picture than what a numerical multi-parameter fit provides. Feeding in the baryon density $\omega_b$ and $\delta v_{\rm BBN}$ determined from the BBN analysis and keeping unchanged the observed input data from P18 + BAO, we obtain an upward shift of $H_0$ relative to the P18 reference value. Our results are consistent with the numerical study by Hart and Chluba~\cite{Hart:2019dxi}. However, there is some subtlety related to how and what BAO data is applied. 

	\subsection{Standard $\Lambda$CDM Model} \label{LCDM_model}

		In the standard $\Lambda$CDM model, the dimensionless parameters are defined as 
	\begin{align}
		\omega_i = \Omega_i h^2, \quad \Omega_i = \dfrac{\rho_{i,0}}{\rho_{cr,0}}, \quad
		\rho_{cr,0} = \dfrac{3H_0^2}{8\pi G}, \quad h= \frac{H_0}{\left( 100 \text{ km}/s/\text{Mpc} \right)} ,
	\end{align}
for the universe today. We use the subscript $\gamma$, $\nu$, $b$, $c$, $m$ and $r$ to represent photon, neutrino, baryon, CDM, total matter and radiation, respectively. Then the radiation and total matter energy densities, $\Lambda$ and Hubble parameter evolve as
	\begin{align}
		&\rho_r(z) = \rho_{r,0}(1+z)^4, \quad \rho_m(z) = \rho_{m,0}(1+z)^3, \quad \rho_\Lambda = \text{const} \ , \\
		&H(z) = H_0 \sqrt{\Omega_r (1+z)^4 + \Omega_m (1+z)^3 + \Omega_\Lambda} \label{hubble_func} \ .
	\end{align}
	Here the number of relativistic D.O.F. is assumed to be a constant, since we are interested in the late-time universe.

	We define the reference model used in this paper as the baseline $\Lambda$CDM fitted with P18 data~\cite{Aghanim:2018eyx}. The cosmological parameters in this reference model then read~\footnote{Explicitly, we take the best-fit values of $\omega_b$, $\omega_c$ and $h$ from Planck 2018 TT,TE,EE+lowE+lensing and derive the values of other physical parameters ($i.e.$, $z_*$, $r_*$, $D_*$, $\theta_*$, $z_d$, $r_d$, $\eta$) from them. These reference values are denoted with a subscript ``P18" later on. Due to our simplified modeling of massive neutrino, slight difference exists in general between the reference value and the central value obtained from marginalization of Planck 2018 data~\cite{Aghanim:2018eyx}, for these parameters.}
	\begin{align}
		\omega_{b,\text{P18}} = 0.02238, \quad \omega_{c,\text{P18}} = 0.1201, \quad h_\text{P18} = 0.6732, \quad Y_{\rm P,\text{P18}} = 0.2454 \ . \label{P18_ref}
	\end{align}
	While $\omega_b$, $\omega_c$ and $h$ are subject to vary in the data fitting, we fix the radiation and neutrino sectors with 
	\begin{align}
		\omega_{\gamma,\text{P18}} = 2.47 \times 10^{-5}, \quad \omega_{\nu r,\text{P18}} = 1.15 \times 10^{-5}, \quad \omega_{\nu m, \text{P18}} = 0.64 \times 10^{-3} \, . \label{P18_ref_rad}
	\end{align}
These inputs can be inferred from the base-line $\Lambda$CDM setup: $T_0 = 2.7255$ K, $N_\text{eff} = 3.046$ and $\sum m_\nu = 0.06$~eV. $\omega_{\nu r}$ and $\omega_{\nu m}$ denote massless and massive neutrino densities here. The massive neutrino is modeled as radiation in the early universe and matter at late time~\cite{Lesgourgues:2012uu}. The redshift for its transition is determined by the condition $T_\nu (z_\nu) = \sum m_\nu$, which yields $z_\nu \simeq 356.91$.

	\subsection{$\Lambda$CDM Model with $\delta v_{\rm rec} \neq 0$} \label{MLCDM_model}

	In this subsection we will examine how a variation of Higgs VEV in the recombination epoch impacts the CMB prediction for the value of $H_0$ and some other cosmological parameters. We will treat its effects to be perturbative. This allows us to address the Hubble tension semi-analytically and postpone a comprehensive numerical analysis to a later time. We will assume that the variation of Higgs VEV steadily lasts from BBN to at least recombination and hence we have $\delta v_{\rm rec} \approx \delta v_{\rm BBN}$. Also, we will choose $\omega_b$, $\omega_c$, $h$ and $v$ as the free parameters.
	
	When the Higgs VEV increases, the mass of electron ($m_e$), proton ($m_p$) and Hydrogen atom ($m_H$) are all dragged up. However, $m_p$ and $m_H$ mainly arise from quark confinement. Their variations are hence relatively small compared to that of $m_e$ ($\delta m_{P/H} \lesssim \mathcal{O}(10^{-4})\delta m_e$). So we fix $m_p$ and $m_H$ here and assume $\delta m_e=\delta v$. In that case, the physical impacts of $\delta m_e > 0$ on the CMB spectrum enter mainly via the following quantities~\cite{Hart:2017ndk,Hart:2019dxi,Ade:2014zfo}:
	\begin{itemize}
		\item Thompson scattering cross-section: $\sigma_T \propto m_e^{-2}$;
		\item Atomic energy levels: $E_i \propto m_e$;
		\item Transition rate of the Lyman-$\alpha$ line: $A^*_{2p1s} = A_{2p1s} P_\text{esc} \propto m_e^3$, where $P_\text{esc}$ is the escape probability of emitting photons;
		\item ``Forbidden" two-photon decay rate: $A_{2s1s} \propto m_e$;
		\item Recombination coefficient: $\alpha_c \propto m_e^{-3/2}$;
		\item Photoionization coefficient: $\beta_c \propto \exp \left(-\frac{E_2 - E1}{T_M}\right)$, where $T_M \propto m_e^{-1}$ is the baryon temperature.
	\end{itemize}
A combination of these impacts eventually increases the redshifts of recombination and baryon drag, namely $z_*$ and $z_d$, for $\delta v_{\rm rec} > 0$. The baryon-photon sound horizon at $z_*$ and $z_d$ are then reduced accordingly. 
	
To see how this effect influences the standard cosmology, let us consider two most notable cosmological observables associated with CMB and BAO. The first observable is the angular sound horizon, defined as
	\begin{align} 
		\theta_* = \dfrac{r_*}{D_*} \ .
	\end{align}
	Here $r_*$ and $D_*$ are the sound horizon and the comoving diameter distance at the recombination (we use ``$_*$'' to denote quantities at the recombination in this paper). They are calculated respectively by 
	\begin{align} \label{rsDs}
		r_* &= \int^{\infty}_{z_*} dz \dfrac{c_s(z)}{H(z)} = \mathcal{D} \int_{z_*}^\infty \left. dz \middle/ \sqrt{3\left[ 1 + \dfrac{3\omega_b}{4\omega_\gamma}(1+z)^{-1} \right]\left[ \omega_r(1+z)^4 + \omega_m(1+z)^3 + \omega_\Lambda \right]} \right. , \\
		D_* &=  \int_{0}^{z_*} dz\dfrac{1}{H(z)}= \mathcal{D}  \int_0^{z_*} \left. dz \middle/ 
		\sqrt{\omega_r(1+z)^4 +  \omega_m(1+z)^3 + \omega_\Lambda} \right.,
	\end{align}
	with $c_s = 1 / \sqrt{3(1 + 3\rho_b/4\rho_\gamma)}$ and $\mathcal{D} = 2998 \text{ Mpc}$. 
	The angular sound horizon determines the separation of acoustic peaks and troughs of the CMB power spectrum. With the CMB anisotropies data~\cite{Aghanim:2018eyx}, its value has been exquisitely measured in the $\Lambda$CDM model with an extreme accuracy, as  
	\begin{align}
		(\theta_*)_{\rm CMB} = (1.04110 \pm 0.00031) \times 10^{-2}  \ . \label{cmb_ths}   
	\end{align}
Its reference value is calculated with Eq.~\eqref{P18_ref} as (with $z_{*,{\rm P18}} = 1089.87$). 
\begin{eqnarray}	
(\theta_*)_{\rm P18} = 1.04100 \times 10^{-2} \ .	
\end{eqnarray}

	The same scale can be observed via BAO peaks imprinted on the matter power spectrum at different redshifts. This feature has been measured directly with large-scale structure surveys~\footnote{\label{fot:BAO}This result is inferred from the BAO features of matter power spectrum by Ref.~\cite{Pogosian:2020ded} combining the high-redshift ($z > 0.6$) data~\cite{Alam:2020sor} including LRGs and ELGs~\cite{Zhao:2020tis, Wang:2020tje}, QSO~\cite{Hou:2020rse}, Lyman-$\alpha$ forest samples~\cite{duMasdesBourboux:2020pck} and the low red-shift galaxy data from 6dF~\cite{Beutler_2011} and MGS (SDSS DR7)~\cite{Ross:2014qpa}.} and indirectly with the CMB data~\cite{Aghanim:2018eyx}, constraining the following second observable
\begin{align}
	(r_{d}h)_{\rm BAO} = \left(99.95 \pm 1.20 \right) \; {\rm Mpc} \ , \ \ \ \   (r_d h)_\text{CMB} = (99.08 \pm 0.92) \; {\rm Mpc } \ ,  \label{bao_constraint}
\end{align}
where $r_d = \int^{\infty}_{z_d} dz \; c_s(z) / H(z)$ denotes sound horizon at the end of baryon drag epoch (find details on the computation of $z_*$ and $z_d$ in App.~\ref{rec_appendix}). The reference value for $r_d h$ is then computed to be (with $z_{d,{\rm P18}} = 1059.95$)
\begin{eqnarray}
 (r_d h)_\text{P18} = 99.01 \; {\rm Mpc } \ .
\end{eqnarray}	
	
	To quantitatively extract how the variation of Higgs VEV in the recombination epoch necessarily impacts the CMB predictions for cosmological parameters, let us consider the relation of $\theta_*$ and $r_{d}h$ with $\delta v_{\rm rec}$ via $\delta z_*$ and $\delta z_d$. By varying these two observables, we find 
	\begin{align}
		\label{constraint_eqs}
		\begin{split}
			&d \ln r_* - d \ln D_*  = d\ln \theta_* \ , \\
			&d \ln r_d + d \ln h = d\ln(r_d h) \ .
		\end{split}		
	\end{align}
Here we will treat $\theta_*$ as a fixed observable so $d\ln \theta_*=0$. The variations of $r_*$, $D_*$ and $r_d$ with respect to $v$ are given by 
(employing again the shorthand notation $r_{*||v} \equiv \frac{d\ln r_*}{d\ln v}$, $r_{*|v} \equiv \frac{\partial \ln r_*}{\partial \ln v}$, $r_{*|b}\equiv \frac{\partial \ln r_*}{\partial \ln \omega_b}$ et. al.)
	\begin{align}
		r_{*||v} &=  r_{*|b} \omega_{b||v}+  r_{*|c} \omega_{c||v}+  r_{*|h} h_{||v} +  r_{*|z_*} z_{*||v} \label{rstar_deriv}  \\
		&\simeq -0.135 \omega_{b||v} - 0.208 \omega_{c||v}-0.656 z_{*||v} \ , \\		
		D_{*||v} &=  D_{*|b} \omega_{b||v}+ D_{*|c} \omega_{c||v}+ D_{*|h} h_{||v} + D_{*|z_*} z_{*||v}  \\
		&\simeq  -0.062 \omega_{b||v}-0.335 \omega_{c||v}-0.193 h_{||v} +0.015 z_{*||v} \ , \\
		r_{d||v} &=  r_{d|b} \omega_{b||v}+ r_{d|c} \omega_{c||v}+ r_{d|h} h_{||v} + r_{d|z_d} z_{d||v}  \\
		&\simeq -0.137 \omega_{b||v}-0.210 \omega_{c||v}-0.652 z_{d||v}  \ ,
	\end{align}
with (for details, see App.~\ref{rec_appendix})
	\begin{align}
		\label{VEV_change}
		\begin{split}
			z_{*||v} &= z_{*|v} + z_{*|b} \omega_{b||v} + z_{*|c} \omega_{c||v} + z_{*|h} h_{||v}  \simeq 1.018  \ , \\
			z_{d||v} &= z_{d|v}  + z_{d|b} \omega_{b||v} + z_{d|c} \omega_{c||v} + z_{d|h} h_{||v} \simeq 0.945 \ .
		\end{split}
	\end{align}
With the explicit forms of $r_*$, $D_*$ and $r_d$, and the determination of $z_{*|v}$ and $z_{*||v}$, the relations in Eq.~(\ref{constraint_eqs}) are then reduced to 
	\begin{align}\label{nkeyeq1}
		-&0.055 \omega_{b||v} + 0.1204 \omega_{c||v} + 0.1934 h_{||v} - 0.6838 = 0, \\
		-&0.1687 \omega_{b||v} - 0.2154 \omega_{c||v} + h_{||v} - 0.6163 = (r_d h)_{||v} \ . \label{nkeyeq2}
	\end{align}

	\begin{table}
		\footnotesize
		\centering
		\begin{tabular}{ |c|c|c|c|c| }
			\hline
			\diagbox{Y}{X} & $\omega_b$ & $\omega_c$ &  $h$ & $z_{*/d}$ \\
			\hline
			$r_*$ & $-0.1351$ & $-0.2080$ & $-4 \times 10^{-10}$ & $-0.6563$ \\
			\hline
			$D_*$  & $-0.0624$ & $-0.3349$ & $-0.1934$ & $+0.0151$ \\
			\hline
			$r_d$ & $-0.1372$ & $-0.2100$ & $-4 \times 10^{-10}$ & $-0.6521$ \\
			\hline
		\end{tabular}
		\caption{Numerical values of $Y_{|X}$ for the cosmological parameters.}
		\label{derivatives}
	\end{table} 
	
	The system of Eq.\,(\ref{nkeyeq1}) and Eq.\,(\ref{nkeyeq2}) compactly encodes the correlation of the parameter variations, namely $\omega_{b||v}$, $\omega_{c||v}$, $h_{||v}$ and $(r_d h)_{||v}$, introduced by the CMB and BAO observations. Next we demonstrate how a bigger Hubble constant can be achieved in this context. Firstly,  among the four unknowns, $\omega_{b||v}$ can be inferred using the BBN fit in Sec~\ref{sec:bbn}, assuming that BBN analysis gives the best determination of $\omega_b$. From Eq.\,(\ref{BBN_linear}), one can see that $\delta v_{\rm BBN} \simeq 1.10\%$ yields a shift to $\eta_{\rm BBN}$ by $1.68\%$. This results in     
	\begin{equation}\label{wbv}
		\omega_{b||v} = \eta_{{\rm BBN}||v} = 1.68/1.10 \simeq 1.55 \ .
	\end{equation}
Note that this positive correlation for $\delta \omega_b$ and $\delta v$, motivated by solving the $^7$Li puzzle, is consistent with our anticipation, arising from addressing the Hubble tension with $\delta m_e > 0$~\cite{Hart:2019dxi}.  
Next, if $r_d h$ is also treated as a fixed observable, so $(r_d h)_{||v}=0$, then the remaining two unknowns in Eq.~(\ref{nkeyeq1}) and Eq.~(\ref{nkeyeq2}) can be solved out
	\begin{eqnarray}
		\omega_{c||v} \simeq 3.70, \ \ \ \ h_{||v} \simeq 1.67 \ .
	\end{eqnarray}
	This means that $\omega_c$ and $h$ increase roughly by 3.7\% and 1.7\% respectively for every percent increase in $v$.
	But, as indicated in Eq.~(\ref{bao_constraint}), there exists a mild discrepancy between the BAO and P18 central values of $r_dh$. So we need to include the uncertainty in $r_dh$ to determine the value of $(r_d h)_{||v}$. To incorporate the uncertainty in $r_dh$, we specifically adopt
	\begin{align}\label{eq:BAO2}
		(r_d h)_{||v} = \dfrac{[(r_d h)_{\rm BAO} - (r_d h)_\text{P18}]/(r_d h)_\text{P18}} { \delta v} \simeq 0.9 \pm  1.1   \ ,
	\end{align}
	by assuming that the discrepancy is caused by $\delta v_{\rm rec} = 0$ in interpreting the CMB data. Here the $1\sigma$ uncertainty for $(r_d h)_{||v}$ arises from that of $(r_d h)_{\rm BAO}$. Interestingly, at this significance level the value of $(r_d h)_{||v}$ is basically positive. This implies that the reduction of $r_d$ will push up the value of $h$. This alleviates the Hubble tension! Solving again $\omega_{c||v}$ and $h_{||v}$ but now with Eq.~(\ref{eq:BAO2}), one obtains 
	\begin{eqnarray}
	\label{eq:h-vev-scaling}
		\omega_{c||v} \simeq 2.67 \pm 1.31, \ \ \ \ h_{||v} \simeq 2.31 \pm 0.82 \ .
	\end{eqnarray}
They translates explicitly to
	\begin{eqnarray}
		\omega_c &=& \omega_{c,\text{P18}} (1 + \omega_{c||v} \delta v_{\rm rec}) = 0.1201 \left[1 + (2.67 \pm 1.31) \; 0.011 \right] \simeq 0.1236 \pm 0.017 \ , \\
		h &=& h_{\text{P18}} (1+ h_{||v}  \delta v_{\rm rec}) = 0.6732 \left[ 1 + (2.31 \pm 0.82) \; 0.011 \right] \simeq  0.6903 \pm 0.0061\ .
	\end{eqnarray}

	\begin{table}[htp]
		\centering
		\resizebox{\textwidth}{!}{%
			\begin{tabular}{ |c|c|c|c|c|c| }
				\hline
				\multirow{1}{2cm}{ Models} & $v/v_0$ & $\omega_b$ & $\omega_c$ & $h$ & $r_d$ \\
				\hline
				Ref & $1.000$ & $0.02238$ & $0.1201$ & $0.6732$ & $147.07$ \\
				\hline
				\multirow{2}{2.5cm}{$\Lambda$CDM+$m_e$ +P18+BAO} & $1.008 \pm 0.007$ & $0.02255 \pm 0.00017$ & $0.1208 \pm 0.0019$ & $0.6910 \pm 0.014$ & $146.0 \pm 1.3$ \\
				& ($0.8 \pm 0.7$)\% & ($0.8 \pm 0.8$)\% & $(0.6 \pm 1.6)\%$ & $(2.6 \pm 2.1)\%$ & $(-0.7 \pm 0.9)\%$ \\
				\hline
				\multirow{2}{2.5cm}{BM$_1$}
				& $ 1.011 $ & $ 0.02276 $ & $0.1236 \pm 0.0017$ & $0.6904 \pm 0.0061$ & $144.5 \pm 0.4$ \\
				& $1.1\%$ & $1.7\%$ & ($2.9 \pm 1.4$)\% & ($2.6 \pm 0.9$)\% & ($-1.8 \pm 0.3$)\% \\
				\hline
				\multirow{2}{2.5cm}{BM$_2$}
				& $1.010$ & $ 0.02238 $ & $0.1229 \pm 0.0017$ & $0.6872 \pm 0.0061$ & $145.4 \pm 0.5$ \\
				& $1.0\%$ & $0.0\%$ & ($2.3 \pm 1.4$)\% & ($2.1 \pm 0.9$)\% & ($-1.1 \pm 0.3$)\% \\
				\hline
				\multirow{2}{2.5cm}{BM$_3$}
				& $ 1.000 $ & $ 0.02260 $ & $0.1189 \pm 0.0017$ & $ 0.6793 \pm 0.0061$ & $147.1 \pm 0.5$ \\
				& $0.0\%$ & $1.0\%$ & ($- 1.0 \pm 1.4$)\% & ($0.9 \pm 0.9$)\% & ($0.1 \pm 0.3$)\% \\
				\hline
				\multirow{2}{2.5cm}{BM$_{\rm BBN}$} & $1.011 \pm 0.001$ & $ 0.02276\pm 0.00030$ & $0.1236 \pm 0.0019$ & $0.6904 \pm 0.0066$ & $144.8 \pm 0.7$ \\
				& $(1.1 \pm 0.1)\%$ & $ (1.7 \pm 1.3)\% $  & $(2.9 \pm 1.6)\%$ & $(2.6 \pm 1.0)\%$ & $(-1.6 \pm 0.5)\%$ \\
				\hline
				\multirow{2}{2.5cm}{BM$_{\rm BBN(NL)}$} & $1.012 \pm 0.002$ & $ 0.02290 \pm 0.00031$ & $0.1243 \pm 0.0019$ & $0.6924 \pm 0.0068$ & $144.3 \pm 0.8$ \\
				& $(1.2 \pm 0.2)\%$ & $(2.3 \pm 1.4)\%$ & $(3.5 \pm 1.6)\%$ & $(2.8 \pm 1.0)\%$ & $(-1.9 \pm 0.5)\%$ \\
				\hline
		\end{tabular}}
		\caption{Cosmological constraints ($1\sigma$ C.L.) on the parameters in different models. We define five benchmark models (BMs) in terms of $v/v_0$ and $\omega_b$ as the inputs. Their values are chosen to be fixed for BM$_{1,2,3}$, and BBN-favored (Eq.~(\ref{BBN_linear}) and Eq.~(\ref{BBN_nonlinear})) for BM$_{\rm BBN}$ and BM$_{\rm BBN(NL)}$. We also incorporate the results of P18 and CMB+BAO/$\Lambda$CDM+$m_e$ from~\cite{Hart:2019dxi}. The first one serves as the reference for demonstrating the impacts of $\delta v_{\rm rec}$ on the CMB physics, while the second one is used for their consistency check. In each box of these  analyses (except for P18 ones), the first line shows the favored parameter value, and the second line shows its relative shift to the reference point.}
		\label{cosmo_params_cons}
	\end{table}

	\begin{figure*}[!ht]
		\centering
		\includegraphics[width=0.8\textwidth]{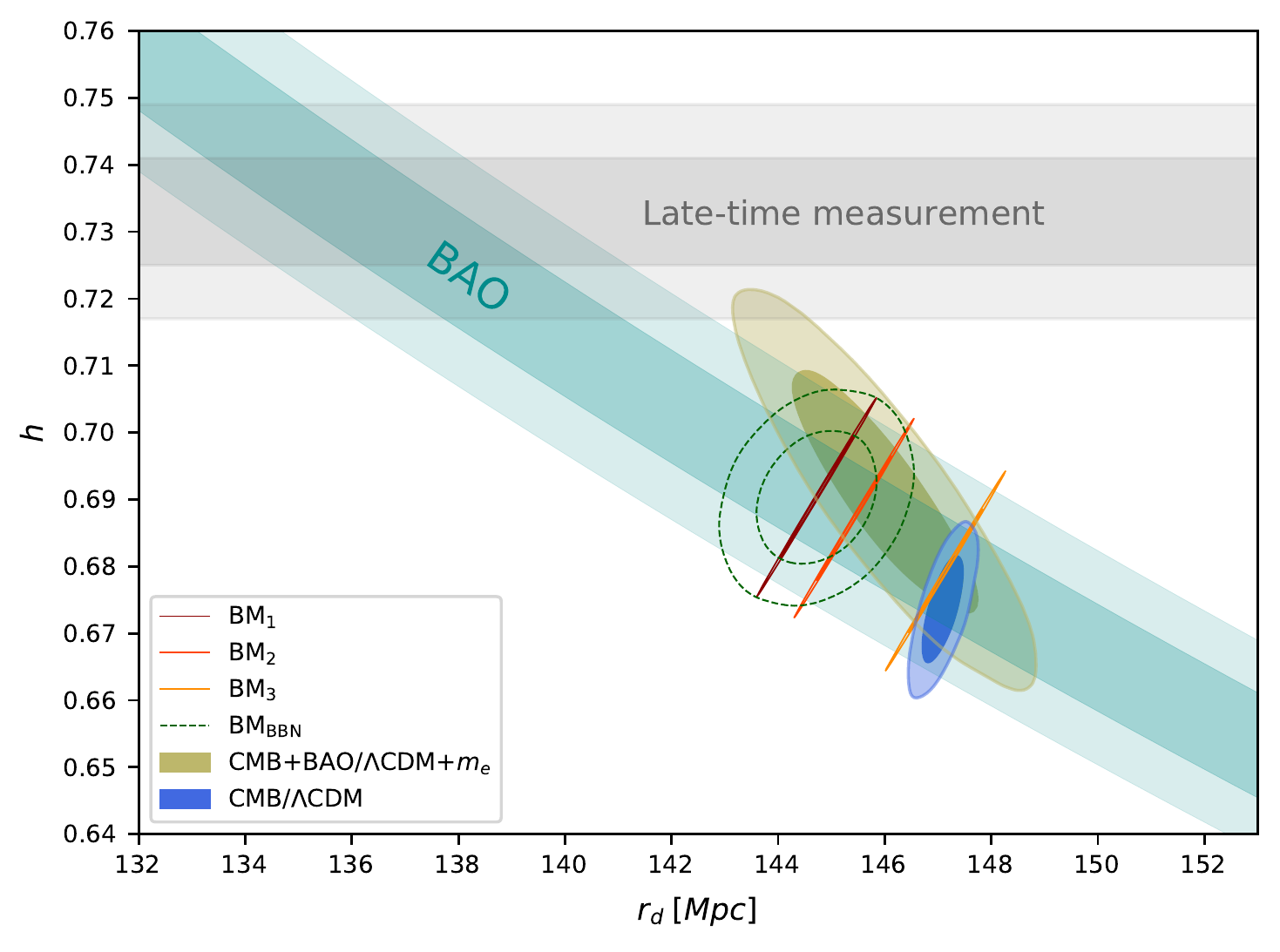}
		\caption{Cosmological constraints on $r_d$ and $h$ in different models. The late-time measurement of $H_0 = 73.3 \pm 0.8 \text{ km}/s/\text{Mpc}$ is taken from~\cite{Verde:2019ivm}, while the combined BAO constraint of $r_d h = 99.95 \pm 1.20$ is from \cite{Pogosian:2020ded}. The shaded blue and khaki regions are extracted from the MCMC chains publicly by Planck team~\cite{Aghanim:2018eyx} and privately provided by Hart \& Chluba~\cite{Hart:2019dxi}, respectively. The BMs are defined in the caption of Tab.~\ref{cosmo_params_cons}. The plot is generated with \textbf{GetDist} \cite{Lewis:2019xzd}.}
		\label{rdh_plot}
	\end{figure*}

The CMB-BAO predictions for more benchmark models (BMs) are presented in Tab.~\ref{cosmo_params_cons} and Fig.~\ref{rdh_plot}. The late-time measurement of $H_0$ shown as the gray band is taken from~\cite{Verde:2019ivm}, which combines the results of SH0ES~\cite{Riess:2019cxk}, H0LiCOW~\cite{Wong:2019kwg}, MCP~\cite{Reid:2008nm}, CCHP~\cite{Freedman:2019jwv}, SBF~\cite{Potter:2018} and MIRAS~\cite{Huang:2018dbn}. The BMs are chosen by various inputs of $\delta v$ and $\delta \omega_b$ to determine $\delta \omega_c$ and $\delta h$ from the aforementioned reference point.

	In the recombination epoch, the most significant effect of $\delta v_{\rm rec}$ is on electron mass, where  $\delta m_e=\delta v_{\rm rec}$ as mentioned. Therefore, our current consideration is similar to $\Lambda$CDM+$m_e$ previously studied by Ref.~\cite{Hart:2019dxi}. Indeed, one can see that a similar trend for the favored $r_d$ and $h$ values is shared for the BMs and the CMB+BAO/$\Lambda$CDM+$m_e$ analysis~\cite{Hart:2019dxi},  regardless of the difference existing in their analysis methods. In~\cite{Hart:2019dxi}, the relevant parameters are allowed to freely vary to fit CMB and BAO data. In our analysis, we emphasize that the two parameters, namely $\delta v_{\rm rec}$ and $\omega_b$, are treated as inputs from BBN, while the others are determined analytically. The uncertainties for the BMs thus mainly arise from the BBN and BAO data. This explains why the favored $r_d$ and $h$ values in BM$_{1,2,3,}$ (where the $\delta v_{\rm rec}$ and $\omega_b$ values are fixed) vary along the uncertainly direction of $(r_dh)_{\rm BAO}$, and in BM$_{\rm BBN}$ additionally along another one determined by the BBN data. This also explains, at least partly, why the uncertainty contours extend for the CMB/$\Lambda$CDM and CMB+BAO/$\Lambda$CDM+$m_e$, but do not for these BMs where $\delta v$ and $\delta \omega_b$ are correlated differently. 
	
	Our semi-analytical approach requires less numerical efforts but still illustrates to some extent the mechanism of how the BAO data help to break the pronounced degeneracy of $\Lambda$CDM+$m_e$ fitted with the CMB data alone~\cite{Hart:2019dxi}. As a consistency check, we plug in the $\Lambda$CDM+$m_e$ model fitted with the CMB+BAO data in~\cite{Hart:2019dxi} as the input, namely~\footnote{We extract the samples of $\omega_b$, $m_e/m_{e,0}$ and $r_d h$ from the corresponding \textbf{CosmoMC} chain kindly provided by the authors. Also, Hart \& Chluba used a combined set of only low-z BAO data: 6dF~\cite{Beutler_2011} + MGS~\cite{Ross:2014qpa} + BOSS DR12~\cite{Gil-Marin:2015nqa}, which leads to a different constraint on $r_d h$ compared to Eq.~(\ref{bao_constraint}).}
	\begin{align}
		\delta v = 0.0079 \pm 0.0071, \quad \delta \omega_b = 0.0076 \pm 0.0076, \quad (r_d h)_\text{BAO} = 100.9 \pm 1.4,
	\end{align}
	and find 
	\begin{align}
		\omega_c = 0.1208 \pm 0.002, \quad h = 0.691 \pm 0.013 \ .
	\end{align}
	in perfect agreement with the values quoted earlier in Tab.~\ref{cosmo_params_cons}. This check essentially validates our linear extrapolation approach. 
	
	In conclusion, in the paradigm where the Higgs VEV is allowed to vary by $\delta v_{\rm rec} = \delta v_{\rm BBN}$, the combination of the unchanged $\theta_*$ and varied $r_d h$ favors a higher value of $H_0$ than the prediction of standard $\Lambda$CDM. The Hubble tension is alleviated but not resolved.

	\subsection*{Remarks}

	The method of linear extrapolation introduced for the study above is quite general. It allows us to quantitatively analyze the leading-order variations of cosmological parameters in beyond-$\Lambda$CDM models, as a prior step to the full-fledged MCMC simulations, and well-complements the analysis of real data. To apply this method, a crucial step is to introduce a set of relevant (information-rich, readily inferred from real data, and sufficiently many) cosmological/astronomical observables, either direct or derived ones, and define the reference point. The parameter variations w.r.t. the reference point will be constrained by the equations such as Eq.~(\ref{constraint_eqs}) derived by varying these observables. In our analysis, we take $\theta_*$ and $r_d h$ to serve this purpose.

	$\theta_*$ is the angular sound horizon at $z_*$. As the distance measure of the CMB acoustic peaks, $\theta_*$ is believed to be one of the well-measured parameters, which is independent of any cosmological model. In contrast, $r_d h$ is inferred from 
		\begin{align}
			\alpha_\perp(z_\text{eff})^{-1} \propto \dfrac{r_d}{D(z_\text{eff})}, \ \ {\rm with} \  \ D(z_\text{eff}) \simeq  \frac{ \mathcal{D}}{h} \int_0^{z_\text{eff}} \dfrac{dz}{\sqrt{\Omega_m(1 + z)^3 + 1 - \Omega_m}}   \ .
		\end{align}
Here $\alpha_\perp(z_\text{eff})^{-1}$ is the angular sound horizon at some redshifts $z_{\rm eff}$ where BAO peaks are observed. In our analysis, we have implicitly assumed that $r_d h$ can be CMB-independently determined by BAO data. To be more general, we can replace $r_d h$ with $\alpha_\perp(z_\text{eff})^{-1}$ 		
	\begin{align}
		\d\ln D(z_\text{eff}) - \d\ln r_d = \d\ln\alpha_\perp(z_\text{eff}) \ .
	\end{align} 
As another test, we can separately apply this to the BAO data at different $z_{\rm eff}$~\cite{Pogosian:2020ded}, including 6dF at $z_\text{eff} = 0.106$~\cite{Beutler_2011}, MGS at $z_\text{eff} = 0.15$~\cite{Ross:2014qpa}~\footnote{Only the isotropic BAO scales, defined as $\alpha_V \propto \alpha_\perp^{2/3} \alpha_\parallel^{-1/3}$, are available for 6dF and MGS.}, LRG and ELG at $z_\text{eff} = 0.7$, $0.77$, $0.845$~\cite{Zhao:2020tis, Wang:2020tje}, QSO at $z_\text{eff} = 1.48$~\cite{Hou:2020rse}, Ly-$\alpha$ at $z_\text{eff} = 2.33$~\cite{duMasdesBourboux:2020pck}. With $\delta v = 1\%$ and $\delta \omega_b = 0$, we find
	\begin{gather*}
		H_0 (z_\text{eff} = 0.106) = 70.4 \pm 2.5, \quad H_0 (z_\text{eff} = 0.15) = 66.3 \pm 2.2, \\ 
		H_0 (z_\text{eff} = 0.7) = 66.5 \pm 2.3, \quad H_0 (z_\text{eff} = 0.77) = 68.7 \pm 1.6, \quad H_0 (z_\text{eff} = 0.845) = 74.5 \pm 3.3, \\
		H_0 (z_\text{eff} = 1.48) = 66.5 \pm 3.5, \quad H_0 (z_\text{eff} = 2.33) = 75.3 \pm 4.7.
	\end{gather*}
The combination of these $H_0$ values, weighted by their errors, eventually gives 
	\begin{align}
		H_{\text{0,com}} = 68.7 \pm 0.9 \;\;\text{km/s/Mpc} \ . 
	\end{align}
This result agrees with the central value of $H_0$ we have obtained in BM$_2$, where $(r_d h)_{\rm BAO}$ is derived based on a combination of the  BAO data sets said above (see footnote~\ref{fot:BAO}), therefore justifies our choice of $r_d h$ as an observable.

Finally, we are aware that the CMB spectrum contains more intrinsic features other than the peaks spacing. As pointed out by Hu et.al ~\cite{Hu:1995kot,Hu:2000ti,Hu:2001bc}, the angular sound horizon $\theta_*$ is just one of the four key parameters to characterize the spectrum. Another three are the particle horizon at matter-radiation equality $l_\text{eq} \equiv k_\text{eq} D_*$, the damping scale $l_d \equiv k_d D_*$ and th baryon-photon momentum density ratio $R_*$. Potentially, these observables can be also incorporated into this analysis, which will be elaborated more systematically in our coming work. But, developing such a comprehensive formalism is beyond the scope of this work.

	\section{Axi-Higgs Model}\label{sec:model}
	
	To address the $^7$Li puzzle and the Hubble tension, Higgs VEV $v$ has to stay $ \sim 1\%$ higher than its present value $v_0$ from the BBN epoch ($\sim$ 3 minutes after big bang) to the recombination epoch ($\sim$ 380,000 years) and then drops to $v_0$ afterwards, which is known to be stabler than a variation of $10^{-16}$ per year~\cite{Huntemann:2014dya,Godun:2014naa}. Here we will present a model of an axion coupled to the Higgs field to achieve this goal. The properties of this model help to resolve the discrepancies in the $\rm ^7Li$ abundance and the Hubble tension discussed in the above sections as well as provide a natural explanation to the $S_8/\sigma_8$ tension and the ICB anomaly which will  be discussed in the next sections. 
	
	When the electroweak scale $v_0$ and the SUSY breaking scale $m_s$ are $100$ GeV or larger, each of both (with its radiative corrections) will introduce a shift to the vacuum energy density $\Lambda$ by many orders of magnitude bigger than the observed value $\Lambda_{\rm obs}$ which is exponentially small. So a fine-tuning is needed to obtain the right $\Lambda_{\rm obs}$. To naturally generate such a $\Lambda_{\rm obs}$, in the SUGRA model, the SUSY-breaking and the electroweak-scale contributions to $\Lambda$ must shield each other precisely~\cite{Qiu:2020los}. Motivated by string theory, one can start with a SUGRA model as a low-energy effective theory. A natural SUSY breaking mechanism is to introduce anti-D3-branes~\cite{Kachru:2003aw}, where $m_s^4$ is the warped brane tension.  In the brane world scenario, the anti-D3-branes span our 3-dimensional observable universe, where all known SM particles (except the graviton) are open string modes living inside these anti-D3-branes.
	
	In flux-compacified Calabi-Yau orientifold in Type IIB string theory, an anti-D3-brane brings in a
	nilpotent superfield $X$ ($i.e.$, $X^2=0$, so the scalar degree of freedom in $X$ is absent)~\cite{Cribiori:2019hod,Parameswaran:2020ukp,Vercnocke:2016fbt}, to facilitate the SUSY breaking~\cite{Antoniadis:2010hs,Kallosh:2015nia,GarciadelMoral:2017vnz}. Applying $X$ as a projection operator~\cite{Komargodski:2009rz} to carry out the projection employed in~\cite{Li:2020rzo}, the two electroweak Higgs doublets $H_u, H_d$ in SUGRA are reduced to a single doublet $\phi$. The  superpotential $W$ contributes to the Higgs potential $V_{\phi}$ as
	\begin{equation}\label{modelVphi}
		W= X\big(m_s^2G(A) - \kappa K(A)H_uH_d\big) +\cdots\, \to\, V_{\phi}= \left|m_s^2G(a) - \kappa K(a)\phi^{\dagger}\phi \right|^2 \ ,
	\end{equation}
	where we have introduced coupling functions $G(A)$ and  $K(A)$. It has been shown that the anti-D3-branes couple to the closed string modes like complex-structure moduli and dilaton~\cite{GarciadelMoral:2017vnz,Cribiori:2019hod, Dudas:2019pls, Parameswaran:2020ukp}, collectively described here as superfield $A$. The coupling $G(A)$ is expected, as the warped throat in which the anti-D3-branes sit is described by the complex-structure moduli and the dilaton (as well as fluxes with discrete values). Since each of these modes contains a complex scalar boson, an axion field can come from either a complex-structure modulus or the dilaton, or some combination, as a partner of them. Because the Higgs fields are open string modes inside the anti-D3-branes, we expect a coupling between $a$ and $\phi$ also, which is mediated by $K(a)$. 
	
	In this model, because of the perfect square form of $V_{\phi}$ in Eq.~(\ref{modelVphi}), the SUSY breaking and the Higgs contribution to the vacuum energy density are arranged to precisely cancel each other, allowing a naturally small $\Lambda$, as proposed in the Racetrack K\"ahler Uplift (RKU) model~\cite{Sumitomo:2013vla}.  
	Here, all moduli are assumed to be stabilized except for the axion $a$ (or multiple fields in $A$) and the Higgs field $\phi$.
	For later convenience, we choose this particular form such that 
	\begin{equation}
		v_0=\frac{\sqrt{2}m_s}{\sqrt{\kappa}}=246\,{\rm GeV}, \qquad m_\phi =2\sqrt{\kappa} m_s =125\,{\rm GeV}\ .
	\end{equation}
	This implies $m_s=104.3\, {\rm GeV}$ and $\kappa=0.36$. 
	
	Since $W$ has mass dimension three, we have, to a leading-order approximation,
	\begin{equation}
		G(a)= 1+ \frac{g a^2}{M_{\rm Pl}^2}, \quad K(a)=1+\frac{k a^2}{M_{\rm Pl}^2}\; ,
	\end{equation}
	where $g$ and $k$ are parameters of order one. We have normalized the functions $G(a)$ and $K(a)$ so that at the locally stable minimum $a=0$, they take values $G_0=K_0=1$. 
	The axion $a$ naturally has its scale $f_a$ that appears as a dimensionless quantity $a/f_a$ in the axion potential. Here we have absorbed these scales into the coupling constants at the front like $g'a^2/f_a^2=g a^2/M_{\rm Pl}^2$. For the sake of simplicity, we do not introduce mixing between axions. Then the total scalar potential can be expressed as
	\begin{equation}
		V=V_a+V_\phi=V_a(a) + \left|K(a)\left(m_s^2 F(a)-\kappa \phi^\dagger \phi\right)\right|^2 \ ,
		\label{eq:v}
	\end{equation}
	with 
	\begin{equation}\label{GKF}
		F(a)=\frac{G(a)}{K(a)} \simeq 1+ \frac{C a^2}{M_{\rm Pl}^2}\;,
	\end{equation}
	where $C=g-k$ is a constant whose positivity is undetermined. Then we have
	\begin{equation}
		\braket{\phi^\dagger \phi}=\frac{v^2}{2}=\frac{m_s^2}{\kappa}F(a)\;.
		\label{eq:prof}
	\end{equation}
	Here the impact of the function $F(a)$ is screened by Higgs VEV. $K(a)$ plays no important role here, so we simply set $K(a)=1$. If $a$ is replaced with a scalar mode $\varphi$, we will have $F(\varphi) = 1 + d_1 \varphi/M_{\rm Pl} + d_2 (\varphi/M_{\rm Pl})^2 + \cdots $ instead. The oscillation of $\delta v$ then follows $\varphi(t)$ (instead of $\varphi(t)^2$) at leading order, and hence cannot be suppressed to a level allowed by observations today. So it is hard to find a  solution, unless $d_1$ is fine-tuned to be negligibly small while $d_2$ is kept $\sim 1$.
	
	The dilaton $S$ and the complex-structure moduli $U_j$ also enters the superpotential $W=W_0(U_j,S)+ \cdots$, where $V_a(a) \sim |DW_0|^2$, so $V_a(a)$ is proportional to a perfect square and vanishes at its minimum. As an axion enters $W$ as a phase, it is reasonable that $V_a(a) \sim |\sin (a/2f_a)|^2= 1-\cos(a/f_a)$. The evolution of $a$ essentially depends on the form of $V_a(a)$ which is typically given by 
	\begin{equation}\label{eq:Vae}
		V_a=m_a^2 f_a^2 \left(1-\cos{\frac{a}{f_a}}\right)=\frac{1}{2}m_a^2a^2-\frac{m_a^2 a^4}{24f_a^2}+\cdots\;.
	\end{equation}
	Here $f_a$ appears only as a next-order effect. But, it appears in the interaction with the electromagnetic (EM) field via
	\begin{equation}\label{eq:aFF}
		\mathcal{L} \sim \frac{c_\gamma}{32\pi^2} \frac{a}{f_a} F_{\mu\nu}\tilde{F}^{\mu \nu}
	\end{equation}
	where $F_{\mu\nu}$ is the EM field strength and $\tilde F_{\mu\nu}$ is its dual. $c_\gamma$ is the parameter introduced by hand to describe physics beyond. In this article, we fix it to unity, $c_\gamma=1$, a value usually viewed to be ``natural''.

	\subsection{Single-Axion Model}
	
	Let us consider single-axion model first. Because of their interaction in $V_\phi$ (see Eq.~(\ref{modelVphi}) and Eq.~(\ref{eq:v})), the axion and the Higgs field evolve as a coupled system in the early universe.  In general, the evolution of the heavier boson will significantly affect the evolution of the lighter one in such a system. But, this axi-Higgs model, originally motivated by string theory and the requirement of a naturally small $\Lambda$, demonstrates an opposite but desirable behavior.

With $\phi=v/\sqrt{2}$, the potential is 
\begin{equation}\label{eq:Vaphi}
V=V_a+V_\phi \simeq \frac{m_a^2}{2}a^2+ |B(a,v)|^2 
\end{equation}
where
\begin{equation}\label{eq:B}
B(a, v) = m_s^2 \left(1+\frac{C a^2}{M_{\rm Pl}^2}\right) -\kappa \frac{v^2}{2} \ .
\end{equation}
Adopting canonical kinetic terms for the fields, the equations of motion for $a(t)$ and $v(t)$ are
\begin{align}
\ddot{a}+3H\dot{a}+\left[m_a^2+\frac{4Cm_s^2}{M_{\rm Pl}^2}B \right]a&\simeq 0 \label{eq:evoa} \ , \\
\ddot{v}+\left(3H+\Gamma_\phi\right)\dot{v}-2\kappa B v &=0 \ . \label{eq:evov}
\end{align}
Here the scale of $B(a,v)$ is $m_s^2 \simeq (100\, {\rm GeV})^2$.
At first sight, the term $4Cm_s^4/M_{\rm Pl}^2$ is $\sim 10^{50}\, m_a^2$, and hence may have a huge impact on the evolution of $a$. Fortunately, this is not the case, thanks to the perfect square form of $V_{\phi}$ and the large decay width $\Gamma_\phi$ of the Higgs boson. To show this point explicitly, let us assume an initial profile of $B(a, v)_\text{min} = 0$ for the axion and Higgs fields, and determine, while the axion field slightly evolves from $a$ to $a'$, the time scale for the Higgs field to reach its new stable profile $v'$, where $B(a', v')_\text{min} = 0$. For this purpose, we introduce $\Delta v (t)$ to denote the deviation of the Higgs VEV from $v'$, with its initial value being $\Delta v (t=0) = \Delta v_{\rm ini} = v - v'$. Then we have  
\begin{eqnarray}
B(a',v' + \Delta v) &=&  \frac{\partial B}{\partial v'} \Delta v + \mathcal{O}(\Delta v^2) \simeq - \kappa v' \Delta v  \ , \nonumber \\
-2\kappa B (v' + \Delta v)  &\simeq& 2 \kappa^2 v'^2 \Delta v  \simeq  m^2_\phi \Delta v \; .
\end{eqnarray}
With these relations, Eq.~(\ref{eq:evov}) can be recast as the equation of motion for $\Delta v (t)$, namely
\begin{equation}
	\ddot{(\Delta v)} + \Gamma_\phi \dot{(\Delta v)} + m^2_\phi \Delta v \simeq 0 \label{eq:H_per_eq}  \ ,
\end{equation}
where the Hubble factor has been dropped because $H(z) \ll \Gamma_\phi \sim 4 \text{ MeV}$ for $z < z_\text{BBN}$. Solving this equation finally gives 
\begin{equation}
	\Delta v (t) \simeq\Delta v_{\rm ini} e^{-\Gamma_\phi t / 2}\sin (m_\phi t) \; .  
\end{equation}
Here the amplitude of $\Delta v (t)$ is exponentially damped, with a time scale $\Gamma_\phi^{-1} \sim \mathcal{O}(10^{-22})$~s much smaller than that for axion evolution, namely $H(t)^{-1} \sim m_a^{-1} \sim \mathcal{O}(10^{6})$~yr. So, the Higgs field is stabilized to the axion-driven profile (see Eq.~(\ref{eq:prof})) instantly, as we state above, while the axion evolution is approximately described by physics for a dynamically damped harmonic oscillator: 
\begin{equation}
	\ddot{a}+3H(t)\dot{a}+\frac{\partial V_a}{\partial a}= 0 \ , \label{eq:aEVO}
\end{equation}
where the full cosine form of $V_a$ in Eq.~(\ref{eq:Vae}) is used.
%Eq.~(\ref{eq:evov}) implies that $\phi$ (or $v$) would stabilize at the value that makes the last term vanish, which is $B_{\rm min}=0$ (also the minimum of its potential). Due to the presence of $\Gamma_\phi\simeq 4$ MeV, this process happens rapidly. As the evolution of $a$ is much slower, this implies that we can simply treat $B=0$ in Eq.~(\ref{eq:evoa}) for all  time. 
%In short, due to the strong dissipation caused by a large decay width $\Gamma_\phi \sim 4 \text{ MeV} \gg H (t > t_{\rm EW})$, any deviation of the Higgs evolution from its stable point is instantly damped. Then thanks to $V_{\phi}$'s perfect square form, the $C$ term in Eq.~(\ref{eq:evoa}) drops out. 

At early cosmic time, the large $H(t)$ freezes the axion field $a(t)$ to an initial value $a_{\rm ini}$. This value is determined by Eq.~(\ref{eq:prof}) to be
	\begin{equation}
		a_{\rm ini}= \left( \frac{2\delta v_{\rm ini}M_{\rm Pl}^2}{C} \right)^{1/2}\; ,
		\label{eq:dvi}
	\end{equation}
	together with an assumption of $\delta v_{\rm ini} = \delta v_{\rm rec} = \delta v_{\rm BBN}$. The axion field $a(t)$ will not roll down to its potential minimum until $H(t) \lesssim m_a$. Then it starts to oscillate around the minimal point in an underdamped manner, yielding   
	\begin{equation}
		a(t)\simeq \mathcal{A}_m(t)a_{\rm ini}\cos{\left(m_a t\right)}\; .  \label{eq:aevo}
	\end{equation}
	Here the oscillation period is dictated by $m_a$. The dimensionless amplitude $\mathcal{A}_m(t)$ decreases exponentially with a characteristic time scale $\sim H(t)^{-1}$.
	
	\begin{figure}
		\centering
		\includegraphics[width=0.7\textwidth]{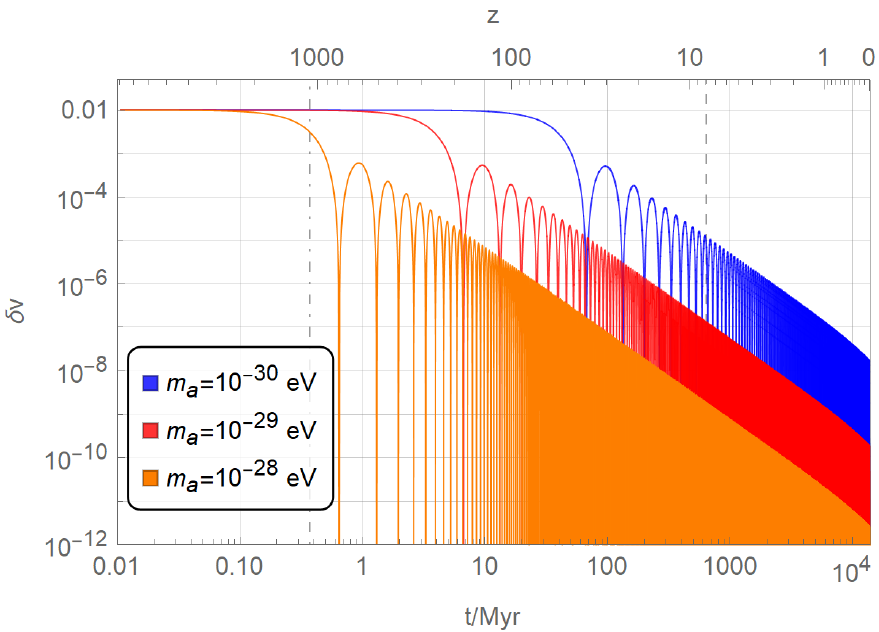}
		\caption{Cosmic evolution of the fractional deviation $\delta v=\Delta v/v_0$ of Higgs VEV, for different axion masses. The recombination time, $t_{\rm rec}\sim 0.34\;{\rm Myrs}$, is labelled by the vertical dot-dashed line on the left. The redshift $z=8$, which sets the earliest possible QS probe for $\delta v$, is labelled by a vertical dashed line on the right.}
		\label{fig:roll}
	\end{figure}
	
	In the axi-Higgs model, we are interested in the mass range such that the axion field starts to roll down near or after the recombination and oscillates with a highly-suppressed amplitude at low redshift and today. The former requirement ensures that the assumption of $\delta v_{\rm BBN} = \delta v_{\rm rec}$, which lays out our discussions so far, is not broken, as the Higgs VEV evolves following  
	\begin{equation}
		\delta v(t)=F(a(t))^{1/2}-1\simeq\frac{Ca(t)^2}{2M_{\rm Pl}^2} \ .
		\label{eq:yt}
	\end{equation} 
	This sets the upper limit of $m_a$ to be $\lesssim 3.3\times 10^{-29}\, {\rm eV}$. The latter requirement ensures that this model can survive the existing constraints for the variation of Higgs VEV, from both astronomical observations, e.g., the QSs, and local laboratory experiments such as ACs~\cite{Uzan:2010pm,Safronova:2017xyt}. Given that a smaller  axion mass yields a later rolling down of the axion field, which in turn yields a bigger axion amplitude and a bigger Higgs VEV oscillation amplitude today, this requirement puts a lower bound on $m_a$. Explicitly, the time variation of $\delta v$ is given by 
	\begin{equation}
		\left.\frac{\d (\delta v)}{\d t}\right|_{t_0} \simeq \frac{\d}{\d t}\left[\frac{C\mathcal{A}_{m,0}^2a_{\rm ini}^2 \cos^2{(m_a t)}}{2
			M_{\rm Pl}^2}\right]=-\delta v_{\rm ini} m_a \mathcal{A}_{m,0}^2\sin{(2m_a t)}\;,
		\label{eq:vdot}
	\end{equation}
where the condition that the current characteristic time scale of $\mathcal{A}_{m}$ decay is much longer than $m_a^{-1}$ is applied, and a shorthand notation of $\mathcal{A}_{m,0} = \mathcal{A}_{m} (t_0)$ is taken. The AC measurements~\cite{Huntemann:2014dya,Godun:2014naa} put a strong bound on the variation rate  in electron-to-proton mass ratio $\mu =m_e/m_p$, yielding $|\d (\delta v)/\d t|_{t_0}\lesssim 10^{-16}\;{\rm yr}^{-1}$. Then the lower bound on $m_a$ can be found by marginalizing the axion oscillation phase in Eq.~(\ref{eq:vdot}). Eventually, the AC measurements results in 
\begin{equation}
		m_a  \in [1.0,3.3] \times 10^{-29}\,{\rm eV}\;.
		\label{eq:mass}
	\end{equation}
at $68\%$ C.L., with the lower bound being extended to $1.6 \times 10^{-30}$~eV at $95\%$ C.L. Such an ultralight axion theoretically is quite acceptable in string theory~\cite{Hui:2016ltb,Tye:2016jzi}. The variation of the Higgs VEV with $z<8$ can be also probed by measuring the molecular absorption spectra of the QSs. The details of these analyses are presented in Sec.~\ref{sec:test}. 
		
	The cosmic evolution of $\delta v(t)$ for various axion masses is shown in Fig.~\ref{fig:roll} with $\delta v_\text{ini} = 1\%$.  As expected, $\delta v(t)$ starts to roll down before the recombination for $m_a \sim 10^{-28}$ eV, yielding a $\delta v_{\rm rec}$ too small to help in resolving the Hubble tension. In contrast, for $m_a \sim  10^{-30}$ eV, $\delta v(t)$ still oscillates with a relatively large amplitude at low redshift. It is ready to be confirmed or disproved at $2\sigma$ C.L. by the ongoing measurements. These tests could be extended to the scenario with $m_a \sim 10^{-29}$~eV in the near future.

	Now let us take a look how the parameters in this model are determined. We take $m_a=10^{-29}\,{\rm eV}$ as the benchmark and assume $\delta v_{\rm ini}=\delta v_{\rm BBN}$ at the initial moment. From the ICB analysis~\cite{Minami:2020odp,Fujita:2020ecn}, the ratio between $a_{\rm ini}$ and $f_a$ is  determined by ($\beta$ is defined in Sec.~\ref{sec:icb})
	\begin{equation}
		\frac{a_{\rm ini}}{f_a} = 16 \pi^2 \beta \simeq 0.97 \; .
		\label{eq:bif}
	\end{equation}
Assuming that this axion contributes a small fraction $x$ of total matter density today, we have 
	\begin{equation}
		\left(\frac{1}{1+z_a}\right)^3m_a^2f_a^2\left(1-\cos{\frac{a_{\rm ini}}{f_a}}\right)\simeq \omega_a \left(0.0030\,{\rm eV}\right)^4\;.
		\label{eq:mfa}
	\end{equation}
Here $\omega_a$ is the axion physical density and $z_a$ is the redshift when this axion field starts to roll down. Explicitly, $z_a$ solves the equation $\xi H(z_a)=m_a$, with $1 \le \xi \le 3$.  We are talking about the matter-dominated epoch, where $H(z)\propto (1+z)^{3/2}$, so we have $(1+z_a)\propto m_a^{2/3}$. This means that $z_a$ and $m_a$ essentially have no impacts on $f_a$ and $a_{\rm ini}$ due to a cancellation, eventually yielding 
	\begin{align}
		a_{\rm ini}&\simeq 3.7\times 10^{17}\,{\rm GeV}\,\left(\frac{x}{0.01}\right)^{1/2}\left(\frac{\xi}{1.5}\right)^{-1} \ , 
		\nonumber\\
		f_a&\simeq 3.8 \times 10^{17}\,{\rm GeV}\,\left(\frac{x}{0.01}\right)^{1/2}\left(\frac{\xi}{1.5}\right)^{-1}
		\label{eq:ai} \ .
	\end{align}
	According to Eq.~(\ref{eq:dvi}), one finds	
	\begin{equation}
		C\simeq 0.84\left(\frac{\delta v_{\rm ini}}{0.01}\right)\left(\frac{x}{0.01}\right)^{-1}\left(\frac{\xi}{1.5}\right)^{2}
		\;.
		\label{eq:c}
	\end{equation}
	We have four parameters in this single-axion model: $m_a$, $\delta v_{\rm ini}$, $a_{\rm ini}$, $f_a$. The amazing thing is that they are all reasonably constrained (see Fig.~\ref{fig:overall}). $\delta v_{\rm ini}=\delta v_{\rm BBN}=\delta v_{\rm rec}$ is imposed to resolve the BBN and Hubble tensions. The ICB measurement puts constraint on $a_{\rm ini}/f_a$. Together with the constraint from $S_8/\sigma_8$, we obtain the values of $a_{\rm ini}$ and hence of $f_a$. If the fraction $x$ is too small, the coupling constant $C$ in Eq.~(\ref{eq:c}) would be unreasonably large. Therefore, the parameters of this model are well-determined.

	\subsection{Two-Axion Model}
	
	In the single-axion model, $\delta v(t)$ drops over time. So the Higgs VEV at the BBN epoch is larger than that at the recombination time, $i.e.$, $\delta v_{\rm BBN} \gtrsim \delta v_{\rm rec}$. But, according to the discussions above, a $\delta v_{\rm rec}$ bigger than $1.1\%$, namely the value favored by the BBN, may address better the Hubble tension. We argue that this behavior could be achieved with the introduction of a second axion.
	
	In fact, in the FDM scenario~\cite{Hu:2000ke,Schive:2014dra,Marsh:2015xka,Hui:2016ltb}, an axion with mass $\sim 10^{-22}\;{\rm eV}$ as CDM can resolve galactic small-scale problems which are challenging the paradigm of weakly interacting massive particle. Thus we are naturally led to consider a model with two axions, denoted as $a_1$ and $a_2$: one with a mass $m_1 \simeq 10^{-29}\;{\rm eV}$ (here, $m_1$ can be relaxed from the mass range given in Eq.\,(\ref{eq:mass})) responsible for $\delta v_{\rm rec} > 0$ and another one with a mass $m_2 \simeq 10^{-22}\;{\rm eV}$ serving as FDM, extending $F(a)$ in Eq.~(\ref{GKF}) to 
	\begin{eqnarray}
	F(a_1, a_2) = 1+ \frac{C_1 a_1^2}{M_{\rm Pl}^2} + \frac{C_2 a_2^2}{M_{\rm Pl}^2}
	\end{eqnarray}
	together with a corresponding potential $V(a_2)$ for $a_2$.
	The formula for $\delta v(t)$ is then extended by including one more axion in the function $F$. To the leading order, it is given by
	\begin{equation}
		\delta v(t)=F(a_1,a_2)^{1/2}-1\simeq \frac{C_1 a_1^2}{2M_{\rm Pl}^2}+\frac{C_2 a_2^2}{2M_{\rm Pl}^2}\;.
		\label{eq:dv2}
	\end{equation}
	
The contributions of $a_1$ and $a_2$ to total matter density today, $i.e.$, $x_{1,2} = \frac{\omega_{1,2}}{\omega_m}$, are related by $x_2 \sim  100 x_1$.  Here the mixing between these two axions has been neglected. To study how $a_1$ and $a_2$ evolve, we simply assume the potential of $a_2$ to be $V_2\approx \frac{1}{2}m_2^2a_2^2$ and take a value of 1.5 for $\xi$. Here,  $a_2$ starts at $a_2=a_{2, {\rm ini}}$ and begins to roll down at a redshift $z_{\rm rec} \ll z_2\simeq 2.0 \times 10^{6} \ll z_{\rm BBN}$, where the universe is still dominated by radiation. While applying Eq.~(\ref{eq:mfa}) to derive $a_{2,{\rm ini}}$, note that no cancellation happens between the $(1+z_2)^{-3}$ and $m_2^2$ factors. We find (with Eq.\,(\ref{eq:ai}))
\begin{equation}
		a_{1,\rm ini}\simeq 3.7\times 10^{17}\,{\rm GeV}\;,\quad
		a_{2,\rm ini}\simeq 1.5\times 10^{17}\,{\rm GeV}\;.
\end{equation}
	The fact that $a_{1,\rm ini}$ and $a_{2,\rm ini}$ are comparable indicates that both can be important in $F(a_1, a_2)$. At $z_{\rm BBN}\sim 10^9$, if we have $C_1 C_2< 0$, the contributions of $a_1$ and $a_2$ to $\delta v_{\rm BBN}$ can be cancelled to some extent. At $z_{\rm rec}$, since the oscillation amplitude for $a_2$ is already highly suppressed, $\delta v_{\rm rec}$ will be determined by $a_1$ only. A scenario with $\delta v_{\rm rec}> \delta v_{\rm ini}=\delta v_{\rm BBN}$ thus can be easily achieved. Explicitly,  by solving  
		\begin{equation}
		\frac{C_1 a_{1,\rm ini}^2}{2M_{\rm Pl}^2}+ \frac{C_2a_{2,\rm ini}^2}{2M_{\rm Pl}^2}=\delta v_{\rm BBN}\;,\quad \frac{C_1 a_1^2(t_{\rm rec})}{2M_{\rm Pl}^2}=\delta v_{\rm rec}\;. 
	\end{equation}
For example, for $\delta v_{\rm BBN}=\delta v_{\rm ini}=1\%$ and $\delta v_{\rm rec}=2\%$,
	\begin{equation}
		C_1\simeq 1.7\;,\quad C_2\simeq -5.1\; ,
	\end{equation}
 a scenario more favored in addressing the Hubble tension. Here  	
$C_1$ and $C_2$ are of the same order and hence no fine-tuning is involved. 

	In summary, the FDM axion, namely $a_2$ with $m_2\sim 10^{-22}\,{\rm eV}$ and $\omega_2=\omega_c$, can be easily incorporated into the axi-Higgs model.
	We can choose the 5 parameters: $m_1 \simeq m_a$, $f_1 \simeq f_a$, $\delta v_{\rm BBN}$, $\delta v_{\rm rec}$ and $x$ (or $\omega_a$) to fix the model. Thus, with one extra parameter beyond the single axion model, the Hubble tension could be better addressed.

	\subsection*{Remarks}
	
	\begin{itemize}
		\item Here we point out that the axion mass, although being much smaller than the EW scale, is natural: the axion coupling to the Higgs VEV does not shift the axion mass significantly and hence no fine-tuning needs to be assumed. 
		Consider the potential $V$ in Eq.~(\ref{eq:Vaphi}), which can be simplified to (after dropping the order-one parameters $\kappa$ and $C$)
		\begin{equation}
			V=\frac{1}{2}m_a^2 a^2+B(a,v)^2\;,\quad B=m_s^2\left(1+\frac{a^2}{M_{\rm Pl}^2}\right)-\frac{v^2}{2}\;.
		\end{equation}
		There exists a valley in the $a$-$\phi$ field space, along which the Higgs VEV is (nearly always) stabilized at the bottom of its $B^2$ term. We are interested in the field evolution along this trajectory. It is determined by $\partial_v V=0 \Rightarrow B(a,v)=0$.
		The Hessian (mass-squared matrix) is given by
		\begin{equation}
			\mathbf{M}=
			\begin{pmatrix}
				m_a^2+8m_s^4a^2/M_{\rm Pl}^4 &  - 2\sqrt{2}m_s^2av/M_{\rm Pl}^2\\[0.6em]
				- 2\sqrt{2}m_s^2av/M_{\rm Pl}^2 & v^2
			\end{pmatrix}\; .
		\end{equation}
		Although the second term in $\mathbf{M}_{aa} \equiv \partial_a^2V$ is much bigger than $m_a^2$ in general, 
		we have
		\begin{equation}
			\lim_{m_a\to 0}\det{\mathbf{M}}=0\;.
		\end{equation}
		This indicates that the axion mass is slightly shifted only by its interaction with the Higgs field. 
		If we expand them in the power series of $m_a/m_s$, to leading order we have
		\begin{align}
			\left(m_\phi^{\rm phy}\right)^2 &\simeq 4m_s^2\left(1+\frac{a^2}{M_{\rm Pl}^2}\right)+\mathcal{O}\left(m_a^2\right)   \ , \\
			\left(m_a^{\rm phy}\right)^2 &\simeq m_a^2+\mathcal{O}\left(m_a^4\right)  \ .
		\end{align}
		Even though $m_a^{\rm phy}$ is evolving, its variation from $m_a$ is negligibly small. So, for the sake of convenience we shall not distinguish the $a$ field and the axion mass eigenstate in our discussions, unless otherwise specified.
		
		\item Next, we consider the radiative corrections to the axion mass. {\it A priori}, the $B^2$ term allows a Higgs-loop correction to shift the axion mass. Recall that, the axion mass is technically natural, as the shift symmetry protects the axion potential. (In the limit of zero axion mass, the shift symmetry is exact and there is no radiative correction to the axion mass.) All radiative corrections contributing to the axion mass term will only introduce terms proportional to its tree level mass. In particular, the axion potential in Eq.~(\ref{model}) has energy density $\sim m_a^2f_a^2$. Above the scale $\Lambda_a \simeq \sqrt{m_a f_a} $, the shift symmetry is expected to be unbroken. So the radiative corrections involving the Higgs and other SM particles largely vanish for the loop momenta above this scale. This means that 
		\begin{eqnarray}
			\Delta m_a^2 \sim \frac{1}{\pi^2} \left(\frac{m_s^2}{M_{\rm Pl}^2}\right) \left(\frac{m_a^2f_a^2}{m_{\phi}^2}\right) \lesssim m_a^2 \ , 
		\end{eqnarray}
		where the first factor comes from the $a^2\phi^{\dagger}\phi$ coupling and $m_a^2f_a^2$ is the momentum cut-off.  	So the radiative corrections remain small and our axion remains light.
		
		\item So far, we have not mentioned the K\"ahler modulus $T=t + i\tau$.  Its inclusion will change $K(a) \to K(a,t, \tau)$  in $V_{\phi}$ in Eq.\,(\ref{eq:v}). But, after a rescaling, it does not come into $F(a)$~\cite{Qiu:2020los,Li:2020rzo}, so our axi-Higgs model is not affected. To be specific, let us consider the RKU model~\cite{Sumitomo:2013vla,Qiu:2020los}. There, when one scans over the string landscape, the probability distribution for $\Lambda$ peaks sharply at $\Lambda=0$, so one can obtain a naturally small $\Lambda$. Matching it to the observed $\Lambda_{\rm obs}$, one finds that $m_t \sim m_{\tau} \sim 10^{-33}$ eV.  For $m \lesssim 10^{-33}\; {\rm eV} \sim H_0$, the field has not yet or just started to roll down, so it may contribute more to the dark energy density than to the dark matter density today. The undissipated vacuum energy may dictate the cosmic acceleration today, as it is in the ``quintessence'' mechanism.
		
		\item It is interesting to note that, the (dimension 3) superpotential takes the form
		$$W = X\big(m_s^2G(A) - \kappa K(A)H_uH_d\big) + \mu H_u H_d + \cdots$$
		Although the $\mu$ term does not appear in $V_{\phi}$ due to the removal of the Higgsinos and the corresponding auxiliary fields\,\cite{Li:2020rzo}, it does help to determine the magnitude of $W$~\cite{Andriolo:2018dee,Qiu:2020los}, which leads to
		$$ \Lambda_{\rm obs}= 10^{-120} \, M_{\rm Pl}^4 \ \  \Rightarrow \ \ \mu v^2 \sim \left(100 \,{\rm GeV}\right)^3 \ .$$ 
		So the electroweak scale emerges naturally without fine-tuning.
		
	\end{itemize}

	\section{$S_8/\sigma_8$ Tension}\label{sec:s8}
	
	\subsection{Matter clustering in $\Lambda$CDM and observations}
	
	The matter clustering amplitude $\sigma_8$ is the root mean square of matter density fluctuations on the scale of $R_8=8 h^{-1}$ Mpc. Intuitively, a sphere with a radius $R_8$ encloses a mass $\sim 10^{14} h^{-1} \text{ M}_\odot$, a typical value for galactic clusters. In the Fourier momentum space, $\sigma_8$ is defined as 
	\begin{eqnarray}
		\sigma_8^2= \sigma^2(r = R_8) =  \int_0^\infty \mathcal {P}_m (k) W^2(k R_8)   dk  \ ,  \label{eq:sig8}
	\end{eqnarray}
	with $W^2(kr)$ being a window function to exclude the contributions from the scales away from $r$. 

	Similar to many other cosmological parameters, $\sigma_8$ can be constrained by both lensing and CMB measurements. Yet, the  effect of $\sigma_8$ is generically inseparable from the growth rate of structure, in the galaxy-clustering observations. The direct observable is instead 	
	\begin{equation}
		S_8 \equiv \sigma_8 \left(\frac{\Omega_m}{0.3} \right)^{\gamma} \ , \label{eq:s8}
	\end{equation}
which can be measured by counting the number of galaxies in the redshift space or via weak lensing. Here the power-law index $\gamma$ depends on the observed redshift and the details of the gravity model. Conventionally, for the low-redshift universe and in the Newtonian limit, its value is fixed to $\gamma = 0.5$. 	
	
	The so-called $S_8$/$\sigma_8$ tension~\cite{DiValentino:2018gcu, Handley:2019wlz} arises from a $2-3\sigma$ discrepancy between the inferred $S_8$ value from the CMB data assuming the $\Lambda$CDM model~\cite{Aghanim:2018eyx} 
	\begin{equation}
		S_{8,\text{CMB}}= 0.832\pm 0.013 \ , 
	\end{equation}
and its value obtained from direct measurements in the late-time universe. In particular, Dark Energy Survey (DES)~\cite{Troxel:2017xyo} and Kilo-Degree Survey (KiDS-1000)~\cite{Heymans_2021} give~\footnote{The late-time value for $S_8$ reported by DES is a combined constraint utilizing a variety of independent measurements, mostly relying on weak lensing. In fact, these weak-lensing-based measurements are consistent with other late-time measurements of cluster abundance~\cite{Troxel:2017xyo,Hildebrandt:2016iqg,Rykoff:2013ovv}. All the late-time measurements consistently converge on the late-time value for $S_8$, and thus the discrepancy between the early time and the late-time measurements is less induced by calibration error only.}
	\begin{equation}
		S_{8,\text{DES}}= 0.773 ^{+0.026}_{-0.020} \ , \ \ \ \ \ S_{8,\text{KiDS}-1000}=0.766^{+0.020}_{-0.014} \  . 
	\end{equation}
Below, we will examine how this discrepancy can be addressed in the axi-Higgs model, using 
\begin{eqnarray}	
 S_{8,\text{P18}} = 0.8325 
\end{eqnarray}	
as the reference point for our linear extrapolation.
   
\subsection{Variations of $S_8/\sigma_8$}

To extract out the $S_8$ physics in the axi-Higgs model with an additional axion of mass $\sim 10^{-29}$~eV, let us start with its variation
		\begin{equation} \label{eq:dS8}
	    \delta S_8 = 0.5\delta \Omega_m + \delta \sigma_8 \ .
	\end{equation}
With the relation $\Omega_m \equiv (\omega_b + \omega_c + \omega_a + \omega_\nu)/h^2$, the derivation of $\delta \Omega_m$ is straightforward, yielding
\begin{equation} \label{eq:dOm}
    \delta \Omega_m = \frac{\omega_b}{\omega_m} \delta \omega_b + \frac{\omega_c}{\omega_m} \delta \omega_c + x - 2\delta h \ .
\end{equation}
In the base-line $\Lambda$CDM, $\omega_\nu$ is fixed to be its lower bound set by neutrino oscillation experiments, thus we do not vary $\omega_\nu$ here. 
	The shift of $\delta v_{\rm rec} \neq 0$ also impacts on $S_8$ via $\sigma_8$. Using the numerical Boltzmann solver code \textbf{axionCAMB}~\cite{Hlozek:2014lca, Hlozek:2017zzf}, we approximately find 
\begin{equation}\label{eq:sigma8-vev}
		\begin{split}
			\delta \sigma_8 &= \sigma_{8| b} \delta \omega_b + \sigma_{8|c} \delta \omega_c + \sigma_{8| h} \delta h + \frac{\partial\ln \sigma_8 }{ \partial x} x \\
			& \simeq - 0.179 \delta \omega_{b} + 0.635 \delta \omega_{c} + 0.232 \delta h - 3.22 x  \ ,
		\end{split}
	\end{equation}
where $\sigma_{8|v}$ has been neglected~\footnote{We take a check indirectly using the MCMC chain from~\cite{Hart:2019dxi}, and find $\sigma_{8|v} \sim \mathcal{O}(10^{-2})$. The other derivatives are calculated using second-order numerical formulae as shown in App.~\ref{rec_appendix}.}. Then with (the counterparts of Eq.~\eqref{nkeyeq1} and \eqref{nkeyeq2} in this context, with $\partial\ln\theta_*/\partial x = 0.3905$ and $\partial\ln r_d/\partial x \simeq 0$)
\begin{align}
	\delta \omega_c = 3.484 \delta v_\text{rec} + 0.138 \delta \omega_b - 1.193 \delta (r_d h) - 2.409 x \ , \label{eq:dwc} \\ 
	\delta h = 1.367 \delta v_\text{rec} + 0.198 \delta \omega_b + 0.743 \delta (r_d h) - 0.519 x  \ , \label{eq:dh}
\end{align}
we eventually find 
\begin{eqnarray} \label{eq:sig8n}
	\delta \sigma_8  &=& - 0.0454 \delta \omega_b - 0.585 \delta (r_d h) + 2.53 \delta v_\text{rec} - 4.87 x \ , \\
	\delta S_8 &=& -0.101 \delta \omega_b + 1.05 \delta \omega_c - 0.768 \delta h - 2.72 x   \ ,  \nonumber \\
	&=& - 0.108 \delta \omega_b - 1.83 \delta (r_d h) + 2.62 \delta v_\text{rec} - 4.86 x \ . \label{eq:s8n}
\end{eqnarray}
For the four variables in this formula, $\delta (r_d h)$ is determined by BAO data while $\delta \omega_b$ and $\delta v_\text{rec}$ have coefficients either too small to be useful or with a inappropriate sign for obtaining a negative $\delta S_8$. Differently, the coefficient of $x$ has a relatively big magnitude, with the right sign. The potential to resolve the $S_8/\sigma_8$ tension thus arises from $x$, namely the matter density of the axion.

	\begin{figure}
		\centering
		\includegraphics[width=0.75\textwidth]{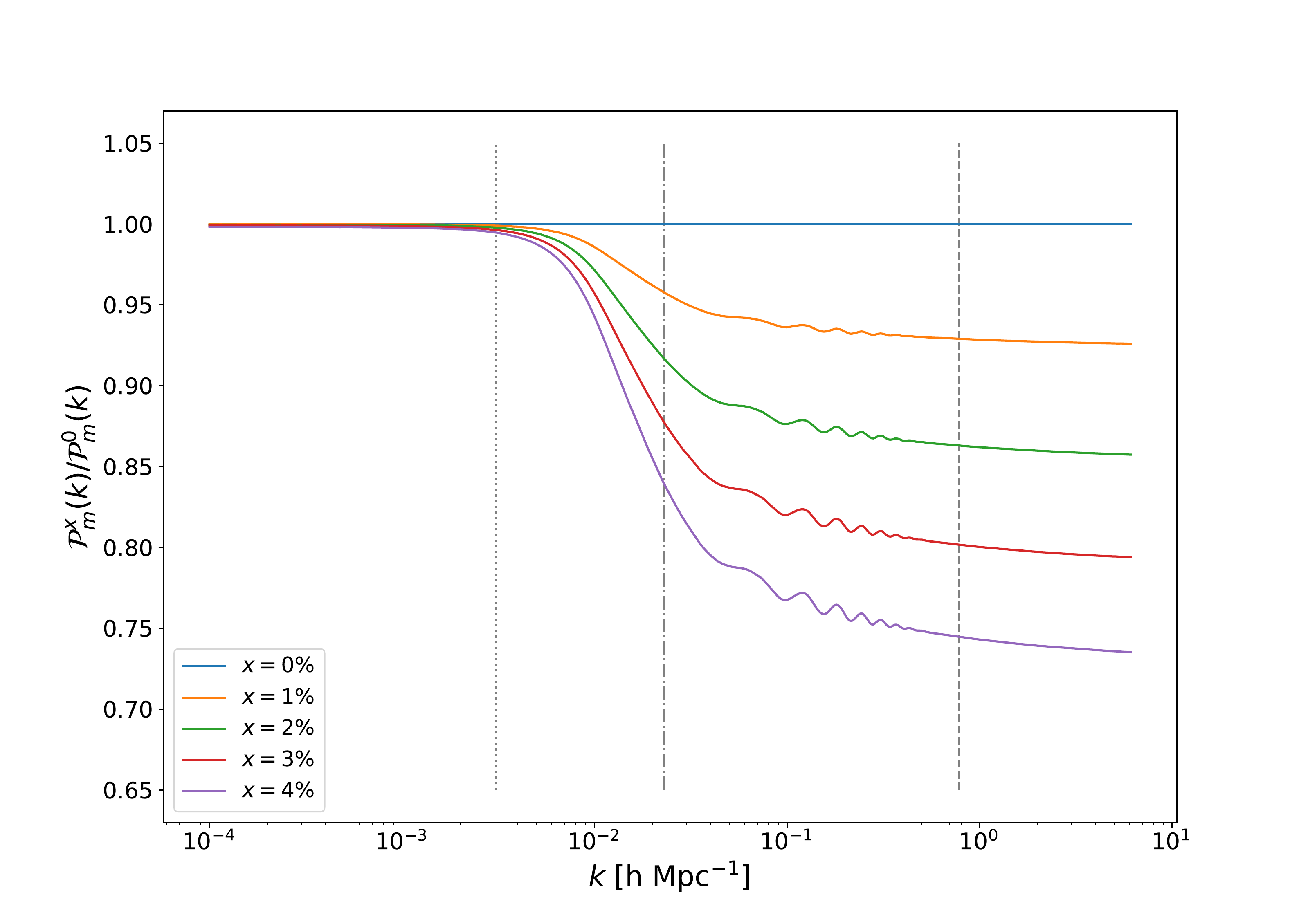}
		\caption{Suppression of the matter power spectrum for $m_a = 10^{-29}$ eV and different $x$ values (replotting of Fig.~7 in~\cite{Kobayashi:2017jcf}, with different $m_a$). Here the dotted, dot-dashed and dashed vertical lines represent $k_{\rm J} (z_{\rm eq})$, $k_{\rm J} (z_0)$ and $k(R_8)$, respectively. }
		\label{fig:mps}
	\end{figure}

In the axi-Higgs model, this axion is ultralight. Due to its astronomically long de Broglie wavelength, this axion field can scale-dependently suppress structure formation in the universe. Here the suppression scale is determined by its mass $m_a$ and redshift-dependent, while the suppression strength is determined by its matter density $x$. This impact has been verified by various theoretical considerations~\cite{scalar-field-structure-formation} and numerical work~\cite{Marsh:2010wq}, despite that most of existing works focus on an axion field with $m_a \sim 10^{-22}$ eV. As discussed in details in~\cite{Marsh:2010wq, Kobayashi:2017jcf}, the density fluctuations of such an axion field and the CDM evolve as a coupled system in this context. Particularly, the axion quantum pressure and its potential force jointly defines a $z$-dependent Jeans scale   
	\begin{eqnarray}
		k_{\rm J}(z) = \frac{\sqrt{m_a H(z)}}{(1+z)}   
	\end{eqnarray}
	for the density perturbation evolution, with $k_{\rm J}(z_0) \simeq 7.4 k_{\rm J}(z_{\text{eq}}) = 0.015 {\rm \; Mpc}^{-1}$ for $m_a = 10^{-29}$ eV. Here $z_0 \equiv 0$ denotes the redshift today. CDM and hence baryon fluctuations would feel this scale-dependent impact, 
	yielding $k$-dependent growing modes. For the modes with $k < k_{\rm J}(z_{\text{eq}})$ and $k>  k_{\rm J}(z_0)$, they stay super-Jeans and sub-Jeans respectively throughout the matter-dominated epoch. The former grows like CDM while the latter oscillates, hence being suppressed. As for the modes with $k_{\rm J}(z_{\text{eq}}) < k < k_{\rm J}(z_0)$, they do not grow until they cross the axion Jeans scale at some moment $z_k$ with $0 < z_k < z_{\text{eq}}$. Then under the assumption of the aligned total matter and CDM perturbations, one finds that the matter power spectrum grows as~\cite{Kobayashi:2017jcf}
	\begin{eqnarray}
		\frac{\mathcal {P}^x_m(k)}{\mathcal {P}^0_m(k)} =
		\begin{cases}
			1       & \quad \text{for } k < k_{\rm J} (z_{\rm eq}) \ ,  \\
			\left (\frac{k_{\rm J} (z_{\rm eq})}{k} \right)^{10-2\sqrt{25-24x}}   & \quad \text{for } k_{\rm J}(z_{\text{eq}}) < k < k_{\rm J}(z_0) \ , \\
			\left (\frac{k_{\rm J} (z_{\rm eq})}{k_{\rm J} (z_0) } \right)^{10-2\sqrt{25-24x}}     &  \quad \text{for } k > k_{\rm J} (z_0)  \ .
		\end{cases}    \label{eq: sol}
	\end{eqnarray}
	We demonstrate the suppression of the matter power spectrum for $m_a = 10^{-29}$ eV and different axion fractions $x$ in Fig.~\ref{fig:mps}. Indeed, the super-Jeans modes throughout the matter-dominated epoch are not suppressed. The modes with $k_{\rm J}(z_{\text{eq}}) < k < k_{\rm J}(z_0)$  are suppressed to some extent which  depends on when they cross the axion Jeans scale. As for the sub-Jeans modes today, to which the $\sigma_8$ mode belongs, are suppressed most.

Given $\sigma_8 \propto \mathcal {P}_m(k(R_8))^{1/2}$, the dependence of $\sigma_8$ on $x$ can be calculated straightforwardly, which yields   
	\begin{align}
		\dfrac{\partial\ln\sigma_8}{\partial x} = \dfrac{1}{2} \dfrac{\partial\ln\left[ \frac{\mathcal{P}^x_m(k(R_8))}{\mathcal{P}^0_m(k(R_8))} \right] }{\partial x}  = \frac{12}{\sqrt{25-24x}} \ln \left( \frac{k_{\rm J} (z_{\rm eq})}{k_{\rm J} (z_0) }  \right) 
		\ ,
	\end{align}
At the reference point, where $x=0$, we then have  	
  \begin{eqnarray} \label{eq:sig8-x-derivative}
 \left. 	\frac{ \partial \ln \sigma_8 }{\partial x} \right|_{x=0} \simeq -4.8  \ .
  \end{eqnarray}
This outcome is consistent with the result obtained from the linear extrapolation in Eq.~(\ref{eq:sig8n}).

	\begin{figure}
		\centering
		\includegraphics[width=0.75\textwidth]{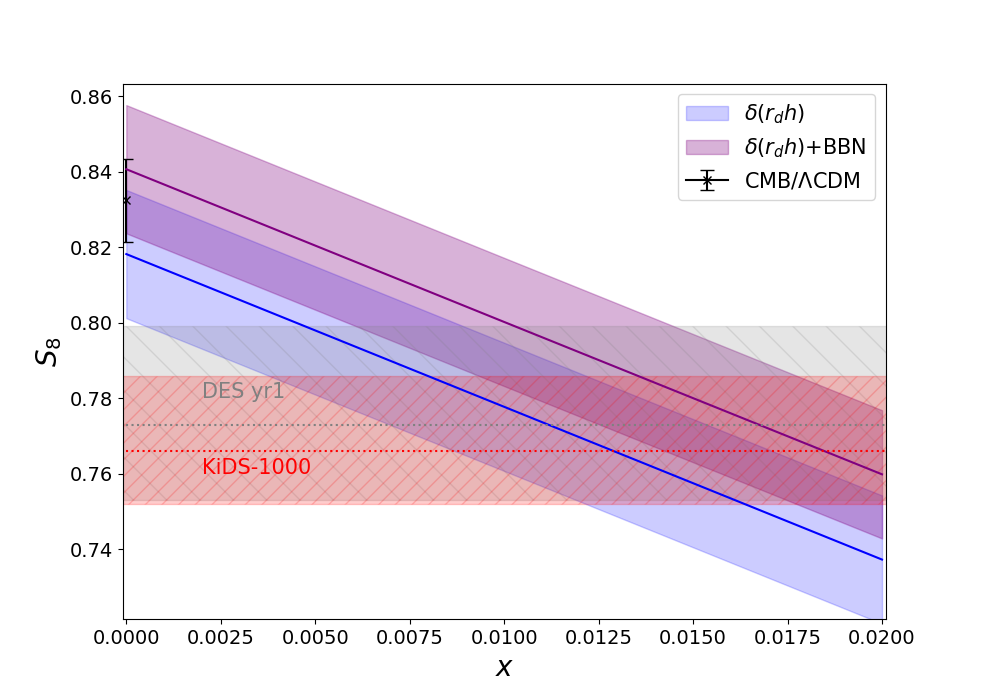}
		\caption{Dependence of $S_8$ on $x$ in the axi-Higgs model. The CMB favored value in the $\Lambda$CDM model is displayed as an error bar. $\delta (r_d h)$ is turned on for both blue and purple bands. But, $\delta x_{\rm rec}$ and $\delta \omega_b$ are set to zero in the former case while the BBN inputs for them are applied in the latter one. The grey horizontal band is from DES~\cite{Troxel:2017xyo} and the orange band is from KiDS-1000~\cite{Heymans_2021}. All uncertainties are shown at $1 \sigma$ C.L. }
 	\label{fig: x-s8}
	\end{figure}

As $\sigma_{8||v}$ (recall $\sigma_{8||v} \equiv (\sigma_{8|b} \delta\omega_b + \sigma_{8|c} \delta\omega_c + \sigma_{8|h} \delta h) / \delta v$) arises from the base of the quantity in the last-row of Eq.~(\ref{eq: sol}) while $\frac{ \partial \ln \sigma_8 }{\partial x}$ is generated from its exponent, one would expect $\sigma_8$ to be more sensitive to $x$ and hence $|\frac{ \partial \ln \sigma_8 }{\partial x}| \gg |\sigma_{8||v} |$. This analytically-derived value of $\frac{ \partial \ln \sigma_8 }{\partial x}$ is partly justified by our numerical calculations in Eq.~(\ref{eq:sigma8-vev}), and strongly indicates that $x$ may play a crucial role in solving the $S_8/\sigma_8$ tension. Finally, we show the dependence of $S_8$ on $x$ in the axi-Higgs model in Fig.~\ref{fig: x-s8}. The blue and purple bands are slightly different in their heights. But in general, an $x$ value of $\sim 1-2\%$ will greatly mitigate the tension, suppressing the discrepancy from $\sim 2-3 \sigma$ to $< 1\sigma$. The request of addressing this tension will eventually fix $x$ and hence determine the values of $a_{\rm init}$ and $f_a$ via Eq.~(\ref{eq:ai}).

\section{Hubble Tension versus $S_8/\sigma_8$ Tension}\label{sec:HS8}

In Sec.~\ref{sec:bbn} and \ref{sec:cmb}, we take an uplift for the Higgs VEV, namely $\delta v_\text{BBN} = \delta v_\text{rec} >0$, to explore its impacts on BBN and the $H_0$ value,  while in the previous section we introduce the axion ($m_a \sim 10^{-29}$\,eV) with a matter density $\omega_a$ to study its effect on $S_8/\sigma_8$. A priori,  $\delta v$ and $\omega_a$ are independent effects. In the axi-Higgs model, they are intimately connected. Here, we would like to discuss one intriguing feature of how the $H_0$ tension and the  $S_8/\sigma_8$ tension are correlated through a combination of $\delta v$ and $\omega_a$.

We have demonstrated how $\delta v_{\rm rec} > 0$ can reduce the Hubble tension. Turning on $\delta v_{\rm rec}$ alone however exacerbates the $S_8/\sigma_8$ tension, as shown in Fig.~\ref{fig: x-s8} with the purple band. Similarly, the $S_8/\sigma_8$ tension is largely resolved by introducing axion matter abundance with $x\sim 1\%$. But, it slightly downgrades the $H_0$ value, as indicated in~\cite{Hlozek:2017zzf}. So there exists some trade off in addressing these two problems in the axi-Higgs model.  The analysis in Sec.\,\ref{sec:cmb} focuses on $\delta v_{\rm rec} \simeq \delta v_{\rm BBN}$, here we will extend the analysis to allow $\delta v_{\rm rec} > \delta v_{\rm BBN}$.

In the single-axion model, $\delta v_\text{BBN} = \delta v_\text{rec}$, and $\delta \omega_b$ is fixed by $\delta \omega_b = 1.53\delta v_\text{BBN}$, see Eq.~\eqref{wbv}. However, in the two-axion model, $\delta v_{\rm rec}$ is decoupled from $\delta v_{\rm BBN}$, where we trade the two coefficients $C_1$ and $C_2$ in $F(a_1, a_2)$, see Eq.~(\ref{eq:dv2}), for $\delta v_{\rm rec}$ and $\delta v_{\rm BBN}$. So we are free to vary $\delta v_{\rm rec}$ to a larger value to fit the data, while maintaining $\delta v_{\rm BBN} =1.1\%$. The resulting scaling of $H_0$ and $S_8$ then reads:
\begin{align}\label{eq:H0}
	H_0 &\simeq H_{0,\text{P18}} \left(1.01 \pm 0.01 + 1.37 \delta v_\text{rec} - 0.52 x \right) \ , \\
	S_8 &\simeq S_{8,\text{P18}} \left(0.98 \pm 0.02  + 2.62 \delta v_\text{rec} - 4.86 x \right) \ ,
	\label{eq:S8}
\end{align}
with the inputs of $\delta \omega_b$ from BBN and $\delta (r_d h) \simeq (0.9 \pm 1.2)\%$ from Eq.~(\ref{bao_constraint}). The signs of the $\delta v_{\rm rec}$ and $x$ terms in these two equations manifest the trade-off effect mentioned above.

{\begin{table}[htp]
\centering
\resizebox{\textwidth}{!}{
\begin{tabular}{c|c|c|c|c|c|c|c}
\hline
 Model & $x$& $\delta v_{\text{\rm rec}}$  & $\omega_b$ & $\Omega_m$  & $h$ & $\sigma_8$ & $S_8$ \\ 
\hline
Ref & \multirow{2}{*}{$0\%$} & \multirow{3}{*}{$0\%$} & \multirow{3}{*}{$0.02238$} & $0.3158$ & $0.6732$ & $0.8114$& $0.8325$\\
\cline{1-1}\cline{5-8} 
BM$_4$ & & & & $0.3094\pm0.0086$  & $0.6771 \pm 0.0055$ & $0.8076 \pm 0.0052$ & $0.8204\pm0.0166$\\
\cline{1-2}\cline{5-8}
BM$_5$ & $1\%$ & & & $0.3123\pm0.0086$  & $0.6737 \pm 0.0055$ & $0.7681 \pm 0.0052$ & $0.7795\pm0.0166$\\
\hline
BM$_{\text{BBN}}^0$ & $0\%$ & \multirow{2}{*}{$(1.1\pm0.1)\%$} & \multirow{4}{*}{\makecell{$0.02276$ \\ $\pm 0.00030$}} & $0.3083\pm0.0087$ & $0.6903\pm0.0058$ & $0.8289 \pm 0.0057$ & $0.8430 \pm 0.0169$ \\
\cline{1-2}\cline{5-8}
BM$_{\text{BBN}}^x$ & $1\%$ & & & $0.3083\pm0.0087$ & $0.6868\pm0.0058$ & $0.7893 \pm 0.0057$ & $0.7999 \pm 0.0169$ \\
\cline{1-3}\cline{5-8}
BM$_6$ & \multirow{2}{*}{$2\%$} & \multirow{2}{*}{$4\%$} & & $0.3100\pm0.0087$ & $0.7099\pm0.0058$ & $0.8093 \pm 0.0057$ & $0.8232 \pm 0.0169$ \\
\cline{1-1}\cline{5-8}
BM$_{6}^{\text{HC}}$ &  &  & & $0.3089\pm0.0110$ & $0.7108\pm0.0072$ & $0.8086 \pm 0.0067$ & $0.8203 \pm 0.0212$ \\
\hline
\end{tabular}}
		\caption{ $S_8/\sigma_8$ values in the axi-Higgs model. Here $\delta(r_d h)$ is turned on for all BMs w.r.t. the reference point. All BMs assume $\delta (r_d h) \simeq (0.9 \pm 1.2)\%$ given by Eq.~(\ref{bao_constraint}) except for BM$_6^{\text{HC}}$ where a different value of $\delta (r_dh) = (1.1 \pm 1.4 )\%$ from~\cite{Hart:2019dxi} is applied. BM$_{\text{BBN}}^{0}$ and BM$_{\text{BBN}}^{x}$ take inputs for $\delta v_{\text{\rm rec}}$ and $\omega_b$  from the BBN data fitting, while BM$_6$ and BM$_{6}^{\text{HC}}$ are motivated by the two-axion model.}
		\label{tab: s8-table}
	\end{table}

	\begin{figure}
		\centering
		\includegraphics[width=0.75\textwidth]{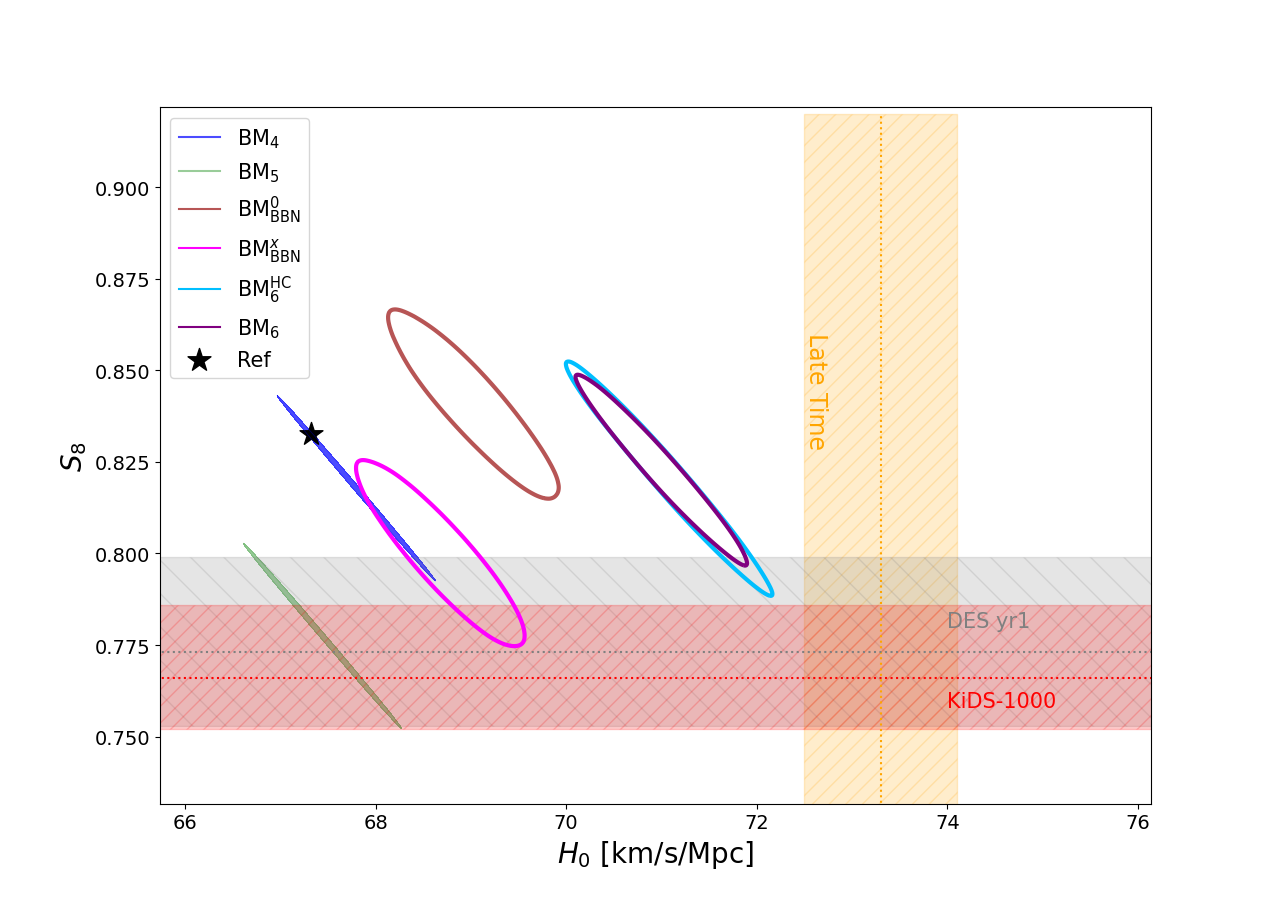}
		\caption{Correlated impacts of $\delta v_{\rm rec}$ and $x$ on $H_0$ and $S_8$. The orange band is the late-time determination of $H_0$~\cite{Verde:2019ivm}. The grey and red bands denote the $S_8$ values from DES~\cite{Troxel:2017xyo} and KiDS-1000~\cite{Heymans_2021}, respectively. 
		The BMs are defined in Tab.~\ref{tab: s8-table}.
		All uncertainties are shown at $1 \sigma$~C.L.}
			\label{fig: xv-s8} 
	\end{figure}}

The correlated impacts of $\delta v_{\rm rec}$ and $x$ on $H_0$ and $S_8$ are demonstrated in Fig.~\ref{fig:   xv-s8} (also see Tab.~\ref{tab: s8-table}). By varying their values, we can see how the $H_0$ and $S_8$ tensions could get resolved. In this figure, the circles are all stretched from left-upper to right-bottom by the uncertainty of $\delta(r_d h)$. According to Eq.~(\ref{eq:dh}) and Eq.~(\ref{eq:s8n}), a bigger $r_d h$ value increases the $H_0$ value and decreases the $S_8$ value, and hence reduces both Hubble and $S_8$ tensions from the data side. In terms of the characteristic parameters in the axi-Higgs model, $i.e.$, $\delta v_{\rm rec}$ and $x$, their impacts are demonstrated using a set of BMs. BM$_5$ and BM$^0_{\rm BBN}$ represent the scenarios where one of them is turned on. They are shifted away from BM$_4$, where $\delta v_{\rm rec}=x=0$, along the direction from left-bottom to right-upper or  its opposite. This feature reflects the said trade-off effect. But, one can see that BM$^x_{\rm BBN}$ does bring the favored values toward the intersection region of the late time $H_0$ (vertical) band and the $S_8$ (horizontal) bands. Hence both $H_0$ and $S_8$ tensions get alleviated to some extent. To get idea of the effect, the (approximate) choice of  
\begin{equation}   
	\delta v_{\rm rec} \sim 4\%\,, \quad \quad x \sim 2\% \ , 
\end{equation}    
namely BM$_6$ and BM$_6^{\rm HC}$ (characterized by different $r_d h$ values from BAO) which are motivated by the two-axion model, results in a slight overlap between the $H_0$ and the $S_8$ data sets, as shown in Fig.~\ref{fig: xv-s8}. A more precise determination of $\delta v_{\rm rec}$ and $x$ is forthcoming.

Notably, turning on $x$ or the axion matter density can have non-trivial impacts on the CMB data fitting. While incorporating an ultralight axion in $\Lambda$CDM alone constrains its abundance to a few percents, $i.e.$, $x \lesssim (2.2- 3.0)\%$ for $m_a \sim 10^{-30} - 10^{-29}$~eV at 95\% C.L~\cite{Hlozek:2014lca, Hlozek:2017zzf}, this bound could be relaxed in the axi-Higgs model due to the presence of $\delta v > 0$. On the other hand, an indication of $x \sim (1 - 2)\%$ in the two-axion model may better resolve the Hubble and $S_8/\sigma_8$ tensions. Yet, as $\delta v \equiv 0$ in~\cite{Hlozek:2017zzf}, a full-data analysis in the axi-Higgs model with $\delta v \ne 0$ and $x \ne 0$ simultaneously is required before more precise statements can be made.

	\section{Isotropic Cosmic Birefringence}
	\label{sec:icb}

	Most of the ongoing or proposed axion detections are based on the axionic Chern-Simon interaction with photons defined in Eq.~(\ref{eq:aFF}).  The magnitude of their coupling $g$ is model-dependent. This interaction violates parity in an axion background, correcting dispersion relation differently for left- and right-circularly-polarized photons. It thus yields an effect of cosmic birefringence when photons, if being linearly polarized,  travel over an axion background in the universe~\cite{Carroll:1989vb,Carroll:1991zs,Harari:1992ea}. 
	
	Cosmic birefringence opens an avenue to explore axion physics. In last decades a series of cosmological and astrophysical observations such as CMB~\cite{Harari:1992ea,Lepora:1998ix,Lue:1998mq}, radio galaxy and active galactic nucleus~\cite{Carroll:1989vb,Antonucci:1993sg}, pulsar~\cite{Liu:2019brz,Caputo:2019tms}, protoplanetary disk~\cite{Fujita:2018zaj}, blackhole~\cite{Chen:2019fsq}, etc. have been proposed to detect this effect. Recently, by reanalyzing P18 polarization data with an improved estimation on miscalibration in the polarization angle at its detectors, the authors of~\cite{Minami:2020odp} report that an ICB effect in the CMB, namely $\beta=0.35 \pm 0.14\,\deg$, has been observed with a statistical significance of $2.4\sigma$. Here $\beta$ is the net rotation made by cosmic birefringence in the linear polarization angle of CMB. If being confirmed later, this observation will be an unambiguous evidence for physics beyond the SM.

	 This ICB analysis is based on the $C_\ell^{EB}$ spectrum, a CMB observable known to be sensitive to parity-violating physics~\cite{Lue:1998mq}. Cosmic birefringence rotates the linear polarization of the CMB photons by an angle~\cite{Harari:1992ea} 
	\begin{eqnarray}
		\beta=\frac{1}{16\pi^2 f_a} \int^{t_0}_{t_{\rm LSS}} dt ~\dot{a} =  \frac{1}{16\pi^2 f_a} \Big[ a(t_0) - a(t_\mathrm{LSS})\Big] \ , \label{eq:cb}
	\end{eqnarray}
	and yields a contribution, namely  
	\begin{eqnarray}
		C_\ell^{EB, {\rm obs}}=\frac12\sin(4\beta)(C_\ell^{EE}-C_\ell^{BB})  \ ,
	\end{eqnarray}
	to the $C_\ell^{EB}$ spectrum observed today~\cite{Lue:1998mq,Feng:2004mq,Liu:2006uh}. Here $C_\ell^{EE}$ and $C_\ell^{BB}$ are the intrinsic $EE$ and $BB$ spectra at last scattering surface (LSS). $a(t_0)$ and $a(t_{\rm LSS})$ represent the axion profiles at present and LSS. Their fluctuations, which are anisotropic and hence not relevant here, have been left out. Generally, the calculation of $\beta$ in statistics is involved, as the value of $a(t_\mathrm{LSS})$ for the CMB photons may vary a lot. But, for $H_0 \lesssim m_a \lesssim H (t_{\mathrm{LSS}})$, the scenario that we are interested in, the axion field starts to roll down and oscillate after the last scattering of the CMB photons. $a(t_{\rm LSS})$ thus can be naturally approximated as a constant, $i.e.$, $a_{\rm ini}$,  for all CMB photons. Eq.~(\ref{eq:cb}) is then reduced to 
	\begin{eqnarray}
		\beta \sim  - \frac{1}{16\pi^2}  \frac{a_{\rm ini}}{f_a},  \quad \Rightarrow \quad  \frac{a_{\rm ini}}{f_a} \simeq 1.0 \pm 0.3 \ .
		\label{eq:new}
	\end{eqnarray}
Note that a minus sign has been dropped here for $\frac{a_{\rm ini}}{f_a}$ for convenience (see footnote~\ref{ft:z2}).

	Now we are able to draw an overall picture on the axi-Higgs cosmology in the single axion version of the model. (Note that, with $m_2\sim 10^{-22}$ eV in the two-axion version, the $a_2$ oscillations are rapid during recombination time so the time averaging of its fast oscillations renders negligible its contribution to ICB.)
 As discussed in Sec.~\ref{sec:model}, this axi-Higgs model is parametrized by four free parameters. For the convenience of presentation, we choose them to be: $m_a$, $C'$, $\frac{a_{\rm ini}}{f_a}$ and $f_a$. Here $C'$ is defined in Eq.~(\ref{eq:Cp}). These parameters are then applied to address five classes of astronomical/cosmological observations and measurements: (1)~AC/QS; (2)~CMB+BAO; (3)~BBN; (4)~$S_8/\sigma_8$ and (5)~ICB. This picture is demonstrated in Fig.~\ref{fig:overall}. In this figure, the right edge of the shaded green region shows the upper bound of the axion mass. It is determined by the requirement that the axion does not roll down until near or after the recombination. The lower limit of $m_a$ is set by the AC measurement of the $\mu$ drift rate~\cite{Lange:2020cul}. The projected lower limits from astronomical  observations of molecular absorption spectra, based on the present and the two-order improved precisions for eighteen known QSs~\cite{Levshakov:2020ule}, are also presented. The shaded purple region represents a recast of the CMB+BAO data interpretation in $\Lambda$CDM+$m_e$ (previously proposed to address the Hubble tension in~\cite{Hart:2019dxi}) in this axi-Higgs model. The shaded orange region is responsible for addressing the $^7$Li problem. At leading order, only $C'\left(\frac{a_{\rm ini}}{f_a}\right)^2$ matters for both. This quantity induces the shift in the Higgs VEV, namely $\delta v_{\rm BBN}$ and $\delta v_{\rm rec}$, according to the axion-Higgs coupling. As shown in Fig.~\ref{fig:overall}, the $\frac{a_{\rm ini}}{f_a}$ values (and hence the $C'\left(\frac{a_{\rm ini}}{f_a}\right)^2$ values) favored by the CMB+BAO and the BBN data are fully overlapped at $1\sigma$ level! The ICB is determined by $\frac{a_{\rm ini}}{f_a}$ only. A choice of $C' \sim \mathcal O(0.01)$ allows these three puzzles to be addressed simultaneously! At last, the $S_8/\sigma_8$ tension can be mitigated with a percent-level contribution from this axion to dark matter energy density. We present the $f_a$ contours in this figure, assuming $x=0.01$, with $a_{\rm ini}$ being approximately determined by 
\begin{eqnarray}	
	\frac{1 }{2} m_a^2 a_{\rm ini}^2 = x \rho_{m} (z_a+1)^3 \  \ . 
\end{eqnarray}
Here $\rho_{m}$ is the total matter energy density today. In the intersection region of all, $f_a$ is favored to be $\sim 10^{17}-10^{18}$ GeV. In summary, it is worthwhile to point out that an axion with $m_a \sim 10^{-30} - 10^{-29}$~eV, as favored in the axi-Higgs model, falls into the ``vanilla'' region to explain this ICB anomaly. Heavier axions such as the FDM axion tend to start oscillating earlier and hence to contribute less with a suppressed $a(t_{\rm LSS})$ (see, e.g.,~\cite{Fujita:2020ecn}).

	\section{Testing the Axi-Higgs Model}\label{sec:test}
	
In the axi-Higgs model, the axion (or the lighter axion in the two-axion verison) rolls down near or after the recombination and oscillates with a highly-suppressed amplitude at low redshifts and today. As discussed in Sec.~\ref{sec:model}, this expectation well-determines the mass range allowed for this axion. It also lays out the foundation to test this model in the near future. Here observations involved are the AC and QS measurements. 
	
	\begin{figure}
		\centering
		\includegraphics[width=0.45\textwidth]{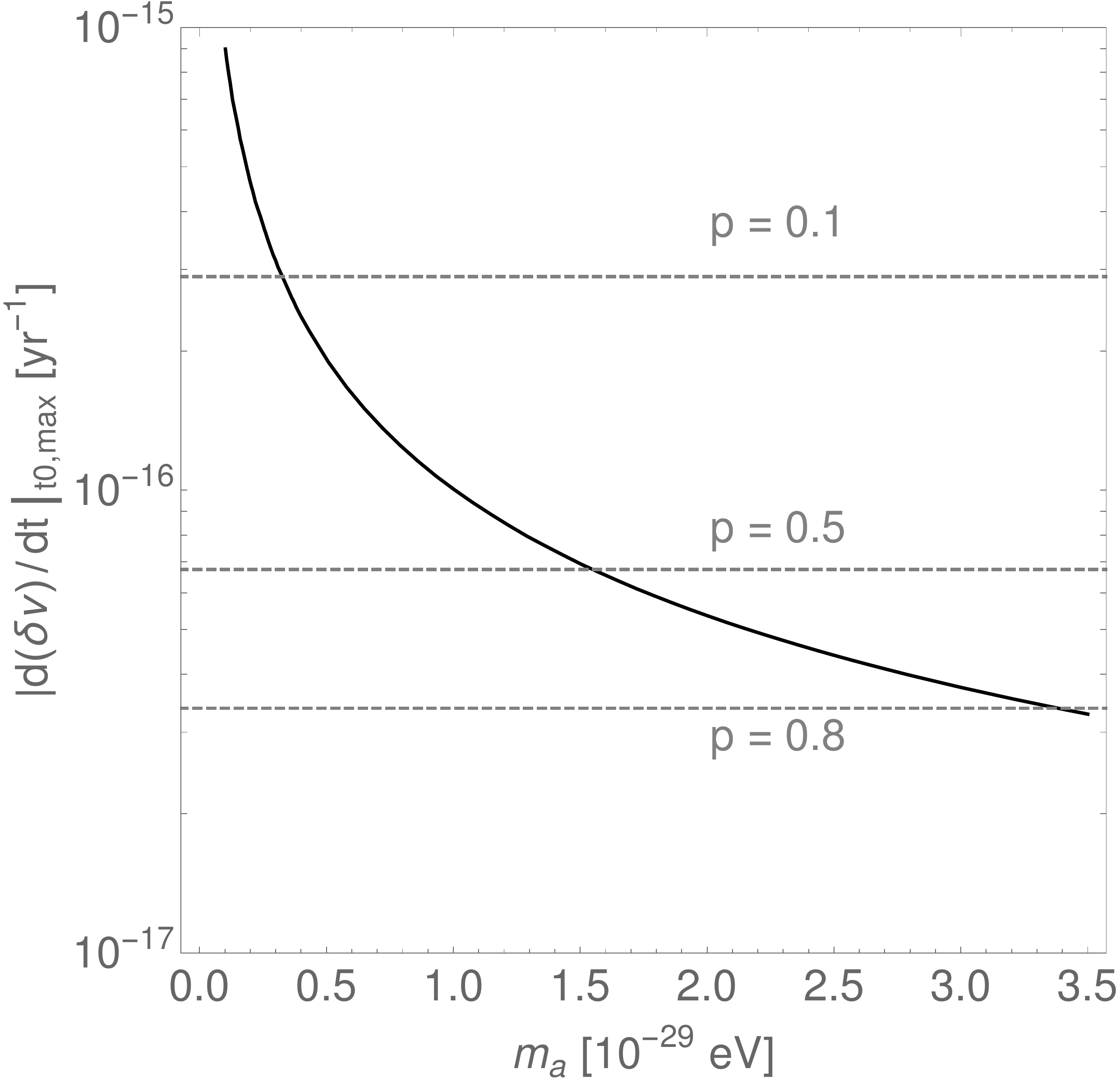}  \ \ \ \ 
		\includegraphics[width=0.45\textwidth]{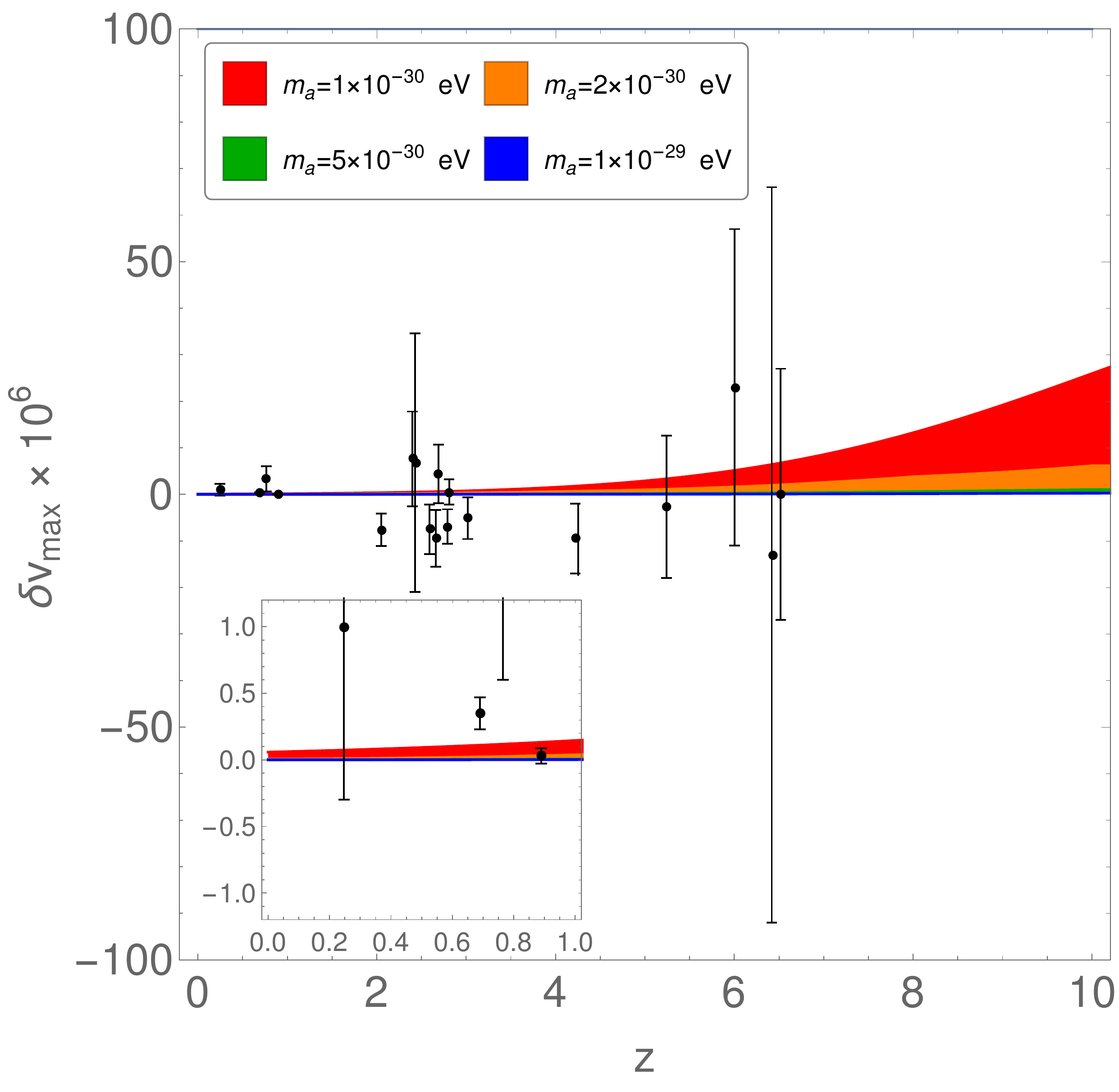}
		\caption{Late-time constraints on $\delta v(t)$. Left: Maximal drift rate $|d(\delta v)/d t|_{t_0}$ as a function of $m_a$, versus the $p$ values after  marginalizing the axion oscillation phase, with the AC data~\cite{Lange:2020cul}. Right: Maximal $\delta v$ as a function of $z$, versus the QS data at different redshifts~\cite{Levshakov:2020ule}. In both panels, we take $1.1\%$ as the $\delta v_{\rm ini}$ benchmark value.}
		\label{fig:late}
	\end{figure}
	
	The AC measurements are sensitive to the temporal drift rate of $\delta v(t)$ since atomic frequencies depend on the parameters such as $\mu= m_e/m_p$~\footnote{AC can also put limits on the drift rates of light quark masses. However, the corresponding sensitivities are at least one order of magnitude lower than that of $\mu$~\cite{Guena:2012zz}.}. By measuring the time dependence of various types of ACs, these experiments are able to limit the local drift rate $d (\delta v) / dt$ to a level of $\lesssim \mathcal{O}(10^{-16})\, {\rm yr}^{-1}$~\cite{Huntemann:2014dya,Godun:2014naa,Lange:2020cul}. Let us consider the latest limit reported in~\cite{Lange:2020cul}. This measurement is based on the observation of $^{171}$Yb$^+$ electric quarduple/octuple frequencies for several years, yielding 
	\begin{equation}
		\frac{d(\delta v)}{d t} \bigg|_{t_0} \simeq \frac{d(\delta \mu)}{d t}\bigg|_{t_0} = -(0.08\pm 0.36)\times 10^{-16}~\text{yr}^{-1}~.
	\end{equation}
	The nowadays maximal drift rate of $\delta v$ in the axi-Higgs model is determined by the relation in Eq.~(\ref{eq:vdot}). Numerically, we have  
\begin{eqnarray}
{\delta v}_{\rm max} \simeq 1.7\times 10^{-5}\left(\frac{\delta v_{\rm rec}}{0.01}\right) \left(\frac{1+z}{10}\right)^{3}\left(\frac{m_a}{10^{-30}\,{\rm eV}}\right)^{-2}\; \ , \label{eq:dvdtmax} 
\end{eqnarray}
and hence
\begin{eqnarray}
\frac{d(\delta v)}{d t} \bigg|_{t_0,{\rm max}}  \simeq 1.0\times 10^{-15} \left(\frac{m_a}{10^{-30}\,{\rm eV}}\right)^{-1}  {\rm yr}^{-1}  \; \ . \label{eq:dvmax}
\end{eqnarray}	
	Note that the local drift rate will be zero if we are sitting right on the peak or trough of the axion oscillation. Consequently, it is always possible to have a  $m_a\ll 10^{-29}\,{\rm eV}$ by tuning the local phase of the axion oscillation. To properly take this effect into account, we marginalize the axion-oscillation phase, and present the AC constraints on $m_a$ in the left panel of Fig.~\ref{fig:late}. At 68\% C.L., we exclude the models with $m_a < 1.0 \times 10^{-30}\,{\rm eV}$.

The measurements of the QSs, or more accurately their molecular absorption spectra, can be applied to constrain $\delta v \simeq \delta \mu$ directly.  The richness  of these molecular spectra helps break the degeneracy of the lines shift caused by the Higgs VEV variation $\delta v$ and the redshift $z$. For example, the energy levels of the electronic, vibrational, and rotational modes of the molecules depend on $\mu$ as~\cite{Safronova:2017xyt}:
	\begin{equation}
		E_{\rm el} \propto \mu^1 \ , \quad E_{\rm vib} \propto \mu^{1.5} \ , \quad E_{\rm rot} \propto \mu^{2} \ .
	\end{equation}
Moreover, the spectral lines from the molecular (hyper)fine structure, $\Lambda$-doubling, hindered rotation, and atomic transitions can further break this degeneracy~\cite{Safronova:2017xyt}. Thus we are allowed to measure the axion oscillation amplitude in the distant past directly. 
The typical sensitivity of the QS measurements on $\delta v$ is of order $\sim \mathcal{O}(10^{-5}) - \mathcal O(10^{-6})$~\footnote{The data in~\cite{Levshakov:2020ule} include additionally the contributions from some astrophysical objects other than the QSs such as the QS candidates and dusty star-forming galaxies~\cite{Casey:2014hya}. We will tolerate the inaccuracy of using the terminology of ``quasar'' here, since the limits obtained for $m_a$ in this context do not rely on the identification of these objects directly.}. It is limited by several factors such as Doppler noise and the background emissions~\cite{Levshakov:2020ule}. Although the amplitude of $\delta v$ is expected to be higher at higher $z$, the precisions for the QS measurements become relatively low in this case. Therefore, it is valuable to combine the measurements of the QSs at all redshifts while still keeping in mind the oscillating behavior of the axion in data analysis.

We demonstrate $\delta v_{\rm max}$ as a function of $z$ in the right panel of Fig.~\ref{fig:late}. Here the data points of the QSs are taken from~\cite{Levshakov:2020ule}, with their $z$ ranging from 0.25 to 6.5~\footnote{Different from~\cite{Levshakov:2020ule}, where some data points at different redshifts are averaged as one input (see Fig.~3 in~\cite{Levshakov:2020ule}), we treat these data points individually while drawing the right panel of Fig.~\ref{fig:late}.}. Largely due to the impacts of the data points with $z < 3$, many of which have a central value deviate from $\delta v  =0$ by more than $1\sigma$, the full range of $m_a$ for the axi-Higgs model is excluded at 68\%~C.L, after the axion-oscillation phase is marginalized. This is also true for standard $\Lambda$CDM model. At $95\%$ C.L., however, $m_a$ is allowed to extend to $5.1\times 10^{-31}$~eV from above, a range broader than the AC limit at the same C.L., $i.e.$, $1.6\times 10^{-30}$~eV.

So far, our discussions focus on the single-axion model. For the two-axion model, with the additional axion $a_2$ being the FDM candidate ($m_2 \sim 10^{-22}$ eV), the bounds on $\delta v_{\rm rec}$ is relaxed; in particular, as discussed in Sec.\,\ref{sec:HS8}, $\delta v_{\rm rec} \simeq 0.04$ is most reasonable. So the resulting $a_1$ oscillation amplitude ${\delta v}_{\rm max}$ in Eq.~(\ref{eq:dvdtmax}) is enhanced for $m_1 = m_a$, yielding a stronger signal strength.   
	
The next-generation AC technology will improve its sensitivity on frequency to a level $\sim \mathcal{O}(10^{-18}){\rm yr}^{-1}$. Such developments include new methods for optical lattice clocks~\cite{ushijima2015cryogenic}, optical clocks based on highly charged ions and hyperfine transitions~\cite{Kozlov:2018mbp}, etc. A more challenging approach of using nuclear clocks based on long-lived, low-energy isomer $^{229{\rm m}}$Th may allow us to reach a $\sim \mathcal{O}(10^{-19}){\rm yr}^{-1}$ sensitivity on frequency~\cite{Campbell:2012zzb,Peik:2020cwm}.
To exclude the axi-Higgs model with $m_a=3.3\times 10^{-29}\,{\rm eV}$ at 95\%~C.L., the precision of measuring $|d(\delta v)/d t|$ needs to be $\lesssim 2.2 \times 10^{-18}\,{\rm yr}^{-1}$. Such a precision can be expected for the next one or two decades, using these new technologies. 

We also expect an essential improvement to the precision of measuring the molecular spectra in the near future, from both infrared and radio astronomy. In terms of the infrared observations, the upcoming Thirty Meter Telescope (TMT)~\cite{Skidmore:2015lga} and James Webb Space Telescope (JSWT)~\cite{Behroozi:2020jhj} may push up the precision by more than one order of magnitude. As for the radio astronomy, the upgraded Atacama Large Millimeter/submillimeter Array (ALMA)~\cite{Marsden:2013iyk}, the Five-hundred-meter Aperture Spherical Radio Telescope (FAST)~\cite{Chen:2019ltf}, and the Square Kilometre Array (SKA)~\cite{Carilli:2004nx} may play a complementary role, by measuring new molecular transitions with high precision~\cite{Levshakov:2020ule}. In view of the great potential of the ongoing or the near-future astronomical observations in testing the axi-Higgs model, we make a modest sensitivity projection for the $m_a$ lower limits. To achieve that, we take the uncertainty of each QS data point in Fig.~\ref{fig:late} as the reference precision, and assume all data points to center at $\delta v = 0$ ($i.e.$, assume all data to match with the standard $\Lambda$CDM model perfectly). This immediately yields a ``projected'' lower limit $3.0 \times 10^{-31}$~eV at 68\% C.L. for $m_a$.  Then with an improvement in precision by two orders, which could be anticipated for the said large-scale telescopes due to the advances of the light-collecting technology and the progress on the wavelength-calibration method~\cite{Ubachs:2015fro}, this lower limit will increase to $\sim 3.0  \times 10^{-30}$~eV. We demonstrate these results in Fig.~\ref{fig:overall}.

\subsection*{Remarks}\label{sec:testA}

Driven by the evolution of the axion field, after its condensate, $\delta v(x,t)$ oscillates in the three dimensional space of our universe with a period 
\begin{equation}
\Delta z \simeq 0.83 \left(\frac{1+z}{10}\right)^{2.5}\left(\frac{m_a}{10^{-30}\,{\rm eV}}\right)^{-1}\; \ .
\end{equation}
Potentially this will allow us to correlate the QS data points observed and expected to be observed (and even with the AC measurments), if their redshifts are not very small compared to 10 and the axion mass $m_a$ is $\sim 10^{-30}$~eV. At these redshifts, spatial fluctuations in the phase of $\delta v$ oscillations become insignificant and the noise of these measurements could be largely suppressed. This is somewhat reminiscent of the detection of stochastic gravitational waves using pulsar timing array. In particular, if any evidence on $\delta v (t) \neq 0$ or $ d (\delta v)/dt \neq 0$ is found directly in the near future, such an analysis would be highly valuable for probing the evolution pattern of $\delta v(t)$ and hence its nature.

	\section{Conclusions}\label{sec:con}
	
Motivated theoretically by string theory and experimentally by a series of cosmological and astronomical observations, we propose a model of an axion coupled with the Higgs field, named ``axi-Higgs'', in this paper. In this model, the axion and Higgs fields evolve as a coupled system in the early universe.  The perfect square form of their potential, together with the damping effect of the Higgs decay width, yields the desirable feature of the model: the evolution of the Higgs VEV is driven by the axion evolution, since moments before the BBN. 

The axi-Higgs model is highly predictive. In the single-axion version, it is parametrized by four parameters only: $m_a$, $\delta v_{\rm ini}$, $a_{\rm ini}$ and $f_a$. Amazingly they are all reasonably constrained (see Fig.~\ref{fig:overall}). $\delta v_{\rm ini}=\delta v_{\rm BBN}=\delta v_{\rm rec}$ is imposed to resolve the Li$^7$ puzzle and Hubble tensions. The ICB measurement puts the constraint on $a_{\rm ini}/f_a$. Together with the constraint from addressing the $S_8/\sigma_8$ tension, we obtain the values of $a_{\rm ini}$ and hence of $f_a$. If the $a_{\rm ini}$ value is too small, a fine-tuning is needed to have the favored value for $\delta v_{\rm ini}$ (see Eq.~(\ref{eq:c})). Therefore, the parameters of this model are well-determined. 

A priori, in solving the $^7$Li puzzle, only a $\delta v_\text{BBN} \sim 1 \%$ is enough, while the axion plays no role. In explaining the ICB anomaly, only the axion properties are relevant while the variation of the Higgs VEV plays no role. It is in tackling the Hubble and $S_8/\sigma_8$ tensions that both the axion and $\delta v_{\rm rec}$ come into play (see Eq.\,(\ref{eq:H0}), Eq.\,(\ref{eq:S8}) and Fig.\,{\ref{fig: xv-s8}). Here the axi-Higgs model, in linking them together, provides a simple framework to further explore their connections.  

Comprehensive investigation on the axi-Higgs model would be highly valuable. In its two-axion version, $\delta v_{\rm rec}$ is decoupled from $\delta v_{\rm BBN}$. We are thus allowed to freely vary $\delta v_{\rm rec}$ to a larger value to fit the CMB data, while maintaining $\delta v_{\rm BBN} \sim 1\%$. Together with a larger contribution of the axion to the total matter density today, this may lead to a better resolution to both Hubble and $S_8/\sigma_8$ tensions. Now a more dedicated analysis following this line is available in~\cite{Fung:2021fcj}. But, a full-data analysis is still needed. 

The axi-Higgs model is accessible to the near-future measurements. The axion evolution can be approximately modeled by a damped oscillator. It rolls down to its potential minimum after $H(t)$ drops below $m_a$. Then it starts to oscillate around the minimal point in an underdamped manner. The variation of the Higgs VEV may be detected by the spectral measurements of the QSs, while its oscillating feature could be observed in the AC measurements. With further improvements in the experimental precisions, the axi-Higgs model should be seriously tested.

	\vspace{6mm}

{\bf {\Large Acknowledgements}}
	
%	We thank Luke Hart and Jens Chluba for valuable communications. 
%	This work is supported partly by the Area of Excellence under the Grant No.~AoE/P-404/18-3(6) and partly by the General Research Fund under Grant No.~16305219. Both grants were issued by the Research Grants Council of Hong Kong S.A.R.	
	
We thank Luke Hart and Jens Chluba for valuable communications.
This work is jointly supported by the Collaborative Research Fund under Grant No.~C6017-20G, the Area of Excellence under Grant No.~AoE/P-404/18-3(6) and the General Research Fund under Grant No.~16305219. All grants were issued by the Research Grants Council of Hong Kong S.A.R.

	\appendix

	\appendix

	\section{Redshifts of Recombination and Baryon Drag} \label{rec_appendix}
%	\subsection*{Recombinationnation and baryon drag}
In this work, the redshift of recombination $z_*$ is defined to be the same as that in \textbf{CAMB}~\cite{Lewis:1999bs} and also in~\cite{ Hu:1995en}, at which the optical depth  
	\begin{align}
		\tau (z) = \int^z_0 dz \dfrac{\sigma_T n_e (z)}{(1 + z) H(z)}  
	\end{align}
is equal to one. 
Here the free electron density $n_e(z)$ is given by 
	\begin{align}
		n_e(z) = (1 - Y_P) \dfrac{\rho_{b, 0}}{m_H}  x_e(z) \ , 
	\end{align}
with $\sigma_T$ being the cross section of Thompson scattering and $x_e(z)$ being the fraction of free electrons. We apply the numerical package \textbf{Recfast++}~\cite{Chluba:2010ca, Rubi_o_Mart_n_2010, Chluba_2010_a, Chluba_2010_b} \footnote{\textbf{Recfast++} is a modified version of the original Recfast~\cite{Seager:1999bc}, with a more sophisticated treatment of the recombination effects studied in~\cite{Switzer:2007sn, Grin_2010, Ali_Ha_moud_2010}.} to compute $x_e(z)$, with varying $m_e$. Alternatively, $z_*$ can be determined by maximizing the visibility function
	\begin{align}
		g(z) = H(z) \dfrac{d\tau}{dz} e^{-\tau}  \ .
	\end{align}
We demonstrate the profile of the visibility function in the left panel of Fig.~\ref{vis_drag_func}. It is Gaussian-like, with its width characterizing the thickness of the last scattering surface of the CMB photons. 
	
		\begin{figure*}[!ht]
		\centering
		\includegraphics[scale=0.55]{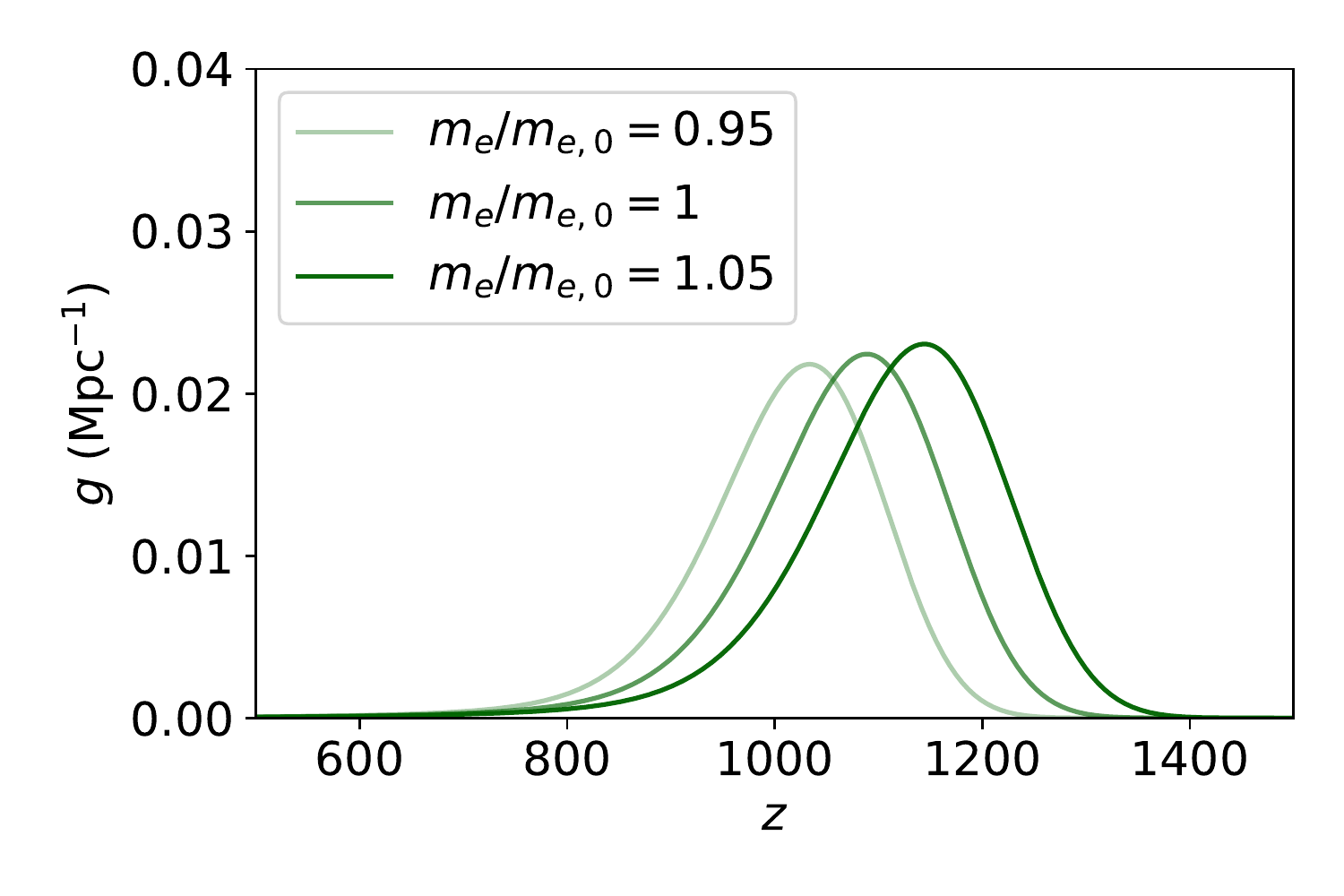}\includegraphics[scale=0.55]{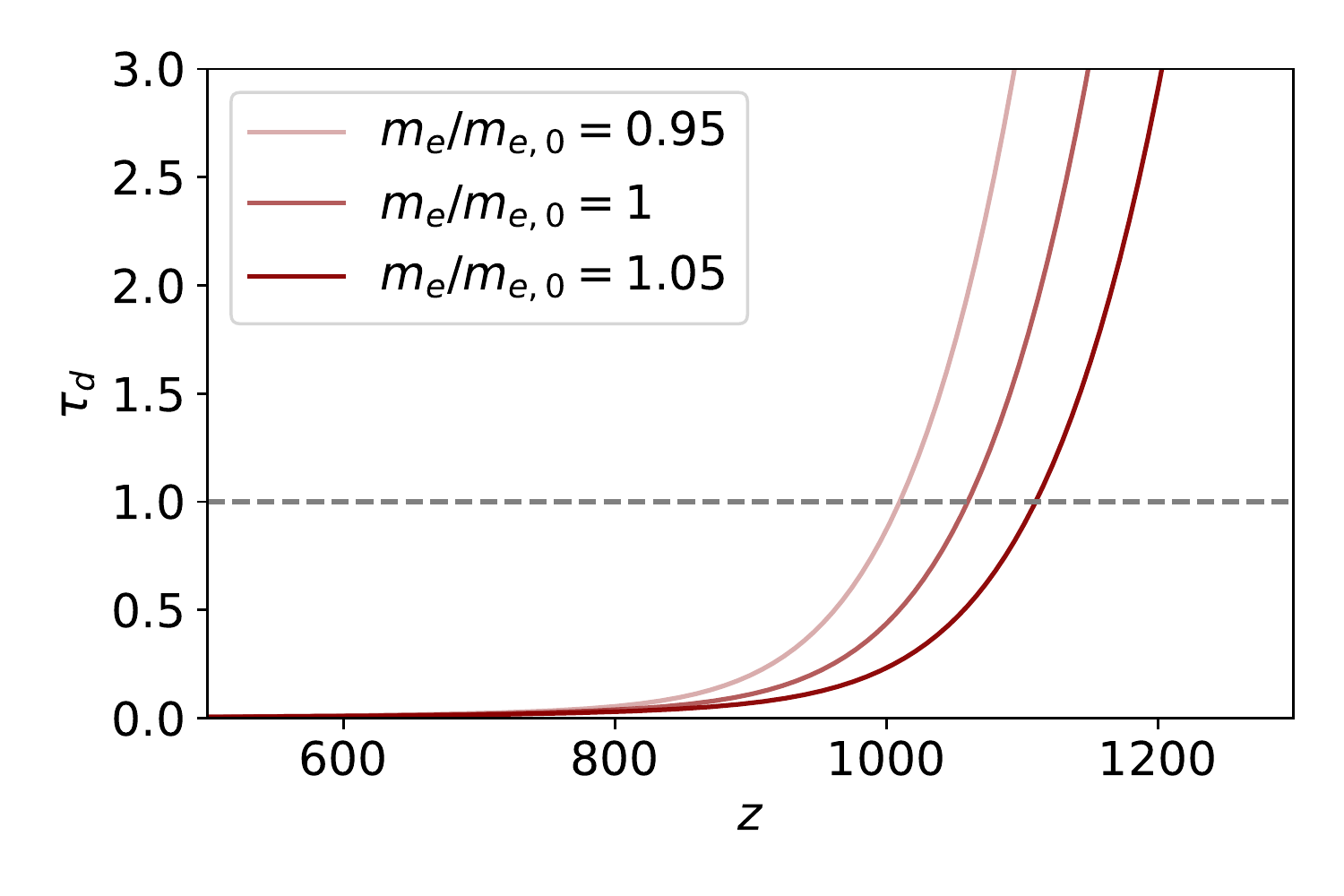}
		\caption{Visibility function and drag depth for different values of $m_e/m_{e,0}$. The variations of $m_e$ are exaggerated for illustration.}
		\label{vis_drag_func}
	\end{figure*}

Similarly the redshift of baryon drag $z_d$ can be defined as the one where the drag depth
	\begin{align}
		\tau_d(z) = \int^z_0 \dfrac{d\tau/dz}{3\rho_b/4\rho_\gamma} 
	\end{align}
is equal to one. 	
The drag depth evolves monotonically as shown in the right panel of Fig.~\ref{vis_drag_func}. At the redshifts below $z_d$, the baryons cease being dragged by the photons in their tight-coupling acoustic oscillations. For $\Lambda$CDM, the value of $z_d$ is often taken to be $r_d \simeq 1.02 \; r_*$. \\

\section{More Details of Calculating $Y_{|X}$}

	As discussed in Sec.~\ref{sec:cmb}, varying the Higgs VEV or $m_e$ upward at recombination can non-trivially impact $z_*$ and $z_d$, yielding bigger values for both, while modifying other cosmological parameters is less relevant, generating a slight change to their values only. In this work we numerically calculate their dependence on these parameters using the centered second-order formulae, namely 
	\begin{align}
		\dfrac{\partial \ln z_{*/d}}{\partial \ln X} = \dfrac{z_{*/d}(\delta X = 0.001) - z_{*/d}(\delta X = - 0.001)}{0.002 \; z_{*/d, \text{P18}}} \ .
	\end{align}
Here $X = v$, $\omega_b$, $\omega_c$ and $h$. The numerical results are summarized in Tab.~\ref{rec_drag_redshift}.
	
	\begin{table}
		\footnotesize
		\centering
		\begin{tabular}{ |c|c|c|c|c| }
			\hline
			$X$ & $v$ & $\omega_b$ & $\omega_c$ &  $h$ \\
			\hline\hline
			$z_{*,\text{P18}}$ & \multicolumn{4}{c|}{$1089.874$}  \\
			\hline
			$z_* (\delta X = 0.001)$ & $1090.984$ & $1089.846$ & $1089.885$ & $1089.874$ \\
			\hline
			$z_* (\delta X = -0.001)$ & $1088.764$ & $1089.903$ & $1089.864$ & $1089.874$ \\
			\hline
			$z_{*||X}$ & $1.0184$ & $-0.026$ & $0.0097$ & $\sim 0$ \\
			\hline\hline
			$z_{d,\text{P18}}$ & \multicolumn{4}{c|}{$1059.947$}  \\
			\hline
			$z_d (\delta X = 0.001)$ & $1060.949$ & $1059.998$ & $1059.956$ & $1059.947$ \\
			\hline
			$z_d (\delta X = -0.001)$ & $1058.946$ & $1059.896$ & $1059.939$ & $1059.947$ \\
			\hline
			$z_{d||X}$ & $0.9450$ & $0.0482$ & $0.0082$ & $\sim 0$ \\
			\hline
		\end{tabular}
		\caption{Numerical values of $z_{*/d ||X}$ with $X = v$, $\omega_b$, $\omega_c$ and $h$, at the reference point.}
		\label{rec_drag_redshift}
	\end{table}

As for the dependence of $r_{*/d}$ and $D_*$ on these cosmological parameters, namely the $Y_{|X}$ values in Tab.~\ref{derivatives},  it is calculated using the formulae below	
	\begin{align}
		&\dfrac{\partial\ln r_{*/d}}{\partial\ln \omega_b} = - \dfrac{\mathcal{D} \omega_b}{2r_{*/d}}  \int_{z_{*/d}}^{\infty} dz \dfrac{c_s(z)}{h(z)} \left[ \dfrac{9}{4} \dfrac{c_s^2(z)}{\omega_\gamma (1 + z)} + \dfrac{(1 + z)^3 - 1}{h^2(z)} \right]; \\
		&\dfrac{\partial\ln r_{*/d}}{\partial\ln \omega_c} = -\dfrac{\mathcal{D} \omega_c}{2r_{*/d}} \int_{z_{*/d}}^{\infty} dz \dfrac{c_s(z)}{h^3(z)} \left[ (1+z)^3 - 1 \right]; \\
		&\dfrac{\partial\ln r_{*/d}}{\partial\ln h} = -\dfrac{\mathcal{D} h^2}{r_{*/d}} \int_{z_{*/d}}^{\infty} dz \dfrac{c_s(z)}{h^3(z)}; \quad \dfrac{\partial\ln r_{*/d}}{\partial\ln z_{*/d}} = - \mathcal{D} \dfrac{z_{*/d}}{r_{*/d}} \dfrac{c_s(z_{*/d})}{h(z_{*/d})} \\
		&\dfrac{\partial\ln D_*}{\partial\ln \omega_b} = - \dfrac{\mathcal{D} \omega_b}{2D_*}  \int^{z_*}_{0} \dfrac{dz}{h^3(z)} \left[ (1+z)^3 - 1 \right]; \\
		&\dfrac{\partial\ln D_*}{\partial\ln \omega_c} = - \dfrac{\mathcal{D} \omega_c}{2D_*}  \int^{z_*}_{0} \dfrac{dz}{h^3(z)} \left[ (1+z)^3 - 1 \right]; \\
		&\dfrac{\partial\ln D_*}{\partial\ln h} = - \dfrac{\mathcal{D} h^2}{D_*}  \int^{z_*}_{0} \dfrac{dz}{h^3(z)}; \quad \dfrac{\partial\ln D_*}{\partial\ln z_*} = \dfrac{\mathcal{D} z_*}{D_* h(z_*)} \ .
	\end{align}
Here $h(z) = \sqrt{\omega_r (1+z)^4 + \omega_m (1+z)^3 + \omega_\Lambda}$ is the dimensionless Hubble parameter. The ``P18" subscript for the reference quantities has been omitted.

	\bibliographystyle{JHEP}
	\bibliography{reference}

\providecommand{\href}[2]{#2}\begingroup\raggedright\begin{thebibliography}{100}

\bibitem{Kneller:2003xf}
J.~P. Kneller and G.~C. McLaughlin, {\it {BBN and Lambda(QCD)}},  {\em Phys.
  Rev. D} {\bf 68} (2003) 103508,
  [\href{http://arxiv.org/abs/nucl-th/0305017}{{\tt nucl-th/0305017}}].

\bibitem{Aghanim:2018eyx}
{\bf Planck} Collaboration, N.~Aghanim et~al., {\it {Planck 2018 results. VI.
  Cosmological parameters}},  {\em Astron. Astrophys.} {\bf 641} (2020) A6,
  [\href{http://arxiv.org/abs/1807.06209}{{\tt arXiv:1807.06209}}].

\bibitem{Verde:2019ivm}
L.~Verde, T.~Treu, and A.~Riess, {\it {Tensions between the Early and the Late
  Universe}},  {\em Nature Astron.} {\bf 3} (7, 2019) 891,
  [\href{http://arxiv.org/abs/1907.10625}{{\tt arXiv:1907.10625}}].

\bibitem{Minami:2020odp}
Y.~Minami and E.~Komatsu, {\it {New Extraction of the Cosmic Birefringence from
  the Planck 2018 Polarization Data}},  {\em Phys. Rev. Lett.} {\bf 125}
  (2020), no.~22 221301, [\href{http://arxiv.org/abs/2011.11254}{{\tt
  arXiv:2011.11254}}].

\bibitem{Troxel:2017xyo}
{\bf DES} Collaboration, M.~A. Troxel et~al., {\it {Dark Energy Survey Year 1
  results: Cosmological constraints from cosmic shear}},  {\em Phys. Rev. D}
  {\bf 98} (2018), no.~4 043528, [\href{http://arxiv.org/abs/1708.01538}{{\tt
  arXiv:1708.01538}}].

\bibitem{Hildebrandt:2016iqg}
H.~Hildebrandt et~al., {\it {KiDS-450: Cosmological parameter constraints from
  tomographic weak gravitational lensing}},  {\em Mon. Not. Roy. Astron. Soc.}
  {\bf 465} (2017) 1454, [\href{http://arxiv.org/abs/1606.05338}{{\tt
  arXiv:1606.05338}}].

\bibitem{Handley:2019wlz}
W.~Handley and P.~Lemos, {\it {Quantifying tensions in cosmological parameters:
  Interpreting the DES evidence ratio}},  {\em Phys. Rev. D} {\bf 100} (2019),
  no.~4 043504, [\href{http://arxiv.org/abs/1902.04029}{{\tt
  arXiv:1902.04029}}].

\bibitem{Li:2005km}
B.~Li and M.-C. Chu, {\it {Big bang nucleosynthesis with an evolving radion in
  the brane world scenario}},  {\em Phys. Rev. D} {\bf 73} (2006) 023509,
  [\href{http://arxiv.org/abs/astro-ph/0511642}{{\tt astro-ph/0511642}}].

\bibitem{Coc:2006sx}
A.~Coc, N.~J. Nunes, K.~A. Olive, J.-P. Uzan, and E.~Vangioni, {\it {Coupled
  Variations of Fundamental Couplings and Primordial Nucleosynthesis}},  {\em
  Phys. Rev. D} {\bf 76} (2007) 023511,
  [\href{http://arxiv.org/abs/astro-ph/0610733}{{\tt astro-ph/0610733}}].

\bibitem{Dent:2007zu}
T.~Dent, S.~Stern, and C.~Wetterich, {\it {Primordial nucleosynthesis as a
  probe of fundamental physics parameters}},  {\em Phys. Rev. D} {\bf 76}
  (2007) 063513, [\href{http://arxiv.org/abs/0705.0696}{{\tt
  arXiv:0705.0696}}].

\bibitem{Browder:2008em}
T.~E. Browder, T.~Gershon, D.~Pirjol, A.~Soni, and J.~Zupan, {\it {New Physics
  at a Super Flavor Factory}},  {\em Rev. Mod. Phys.} {\bf 81} (2009)
  1887--1941, [\href{http://arxiv.org/abs/0802.3201}{{\tt arXiv:0802.3201}}].

\bibitem{Bedaque:2010hr}
P.~F. Bedaque, T.~Luu, and L.~Platter, {\it {Quark mass variation constraints
  from Big Bang nucleosynthesis}},  {\em Phys. Rev. C} {\bf 83} (2011) 045803,
  [\href{http://arxiv.org/abs/1012.3840}{{\tt arXiv:1012.3840}}].

\bibitem{Cheoun:2011yn}
M.-K. Cheoun, T.~Kajino, M.~Kusakabe, and G.~J. Mathews, {\it {Time Dependent
  Quark Masses and Big Bang Nucleosynthesis Revisited}},  {\em Phys. Rev. D}
  {\bf 84} (2011) 043001, [\href{http://arxiv.org/abs/1104.5547}{{\tt
  arXiv:1104.5547}}].

\bibitem{Berengut:2013nh}
J.~Berengut, E.~Epelbaum, V.~Flambaum, C.~Hanhart, U.-G. Meissner, J.~Nebreda,
  and J.~Pelaez, {\it {Varying the light quark mass: impact on the nuclear
  force and Big Bang nucleosynthesis}},  {\em Phys. Rev. D} {\bf 87} (2013),
  no.~8 085018, [\href{http://arxiv.org/abs/1301.1738}{{\tt arXiv:1301.1738}}].

\bibitem{Hall:2014dfa}
L.~J. Hall, D.~Pinner, and J.~T. Ruderman, {\it {The Weak Scale from BBN}},
  {\em JHEP} {\bf 12} (2014) 134, [\href{http://arxiv.org/abs/1409.0551}{{\tt
  arXiv:1409.0551}}].

\bibitem{Heffernan:2017hwa}
M.~Heffernan, P.~Banerjee, and A.~Walker-Loud, {\it {Quantifying the
  sensitivity of Big Bang Nucleosynthesis to isospin breaking with input from
  lattice QCD}},  \href{http://arxiv.org/abs/1706.04991}{{\tt
  arXiv:1706.04991}}.

\bibitem{Mori:2019cfo}
K.~Mori and M.~Kusakabe, {\it {Roles of $^7$Be$(n,p)^7$Li resonances in big
  bang nucleosynthesis with time-dependent quark mass and Li reduction by a
  heavy quark mass}},  {\em Phys. Rev. D} {\bf 99} (2019), no.~8 083013,
  [\href{http://arxiv.org/abs/1901.03943}{{\tt arXiv:1901.03943}}].

\bibitem{Ade:2014zfo}
{\bf Planck} Collaboration, P.~A.~R. Ade et~al., {\it {Planck intermediate
  results - XXIV. Constraints on variations in fundamental constants}},  {\em
  Astron. Astrophys.} {\bf 580} (2015) A22,
  [\href{http://arxiv.org/abs/1406.7482}{{\tt arXiv:1406.7482}}].

\bibitem{Hart:2019dxi}
L.~Hart and J.~Chluba, {\it {Updated fundamental constant constraints from
  Planck 2018 data and possible relations to the Hubble tension}},  {\em Mon.
  Not. Roy. Astron. Soc.} {\bf 493} (2020), no.~3 3255--3263,
  [\href{http://arxiv.org/abs/1912.03986}{{\tt arXiv:1912.03986}}].

\bibitem{Huntemann:2014dya}
N.~Huntemann, B.~Lipphardt, C.~Tamm, V.~Gerginov, S.~Weyers, and E.~Peik, {\it
  {Improved limit on a temporal variation of $m_p/m_e$ from comparisons of
  Yb$^+$ and Cs atomic clocks}},  {\em Phys. Rev. Lett.} {\bf 113} (2014),
  no.~21 210802, [\href{http://arxiv.org/abs/1407.4408}{{\tt
  arXiv:1407.4408}}].

\bibitem{Godun:2014naa}
R.~M. Godun, P.~B.~R. Nisbet-Jones, J.~M. Jones, S.~A. King, L.~A.~M. Johnson,
  H.~S. Margolis, K.~Szymaniec, S.~N. Lea, K.~Bongs, and P.~Gill, {\it
  {Frequency Ratio of Two Optical Clock Transitions in Yb+171 and Constraints
  on the Time Variation of Fundamental Constants}},  {\em Phys. Rev. Lett.}
  {\bf 113} (2014), no.~21 210801, [\href{http://arxiv.org/abs/1407.0164}{{\tt
  arXiv:1407.0164}}].

\bibitem{Lange:2020cul}
R.~Lange, N.~Huntemann, J.~M. Rahm, C.~Sanner, H.~Shao, B.~Lipphardt, C.~Tamm,
  S.~Weyers, and E.~Peik, {\it {Improved limits for violations of local
  position invariance from atomic clock comparisons}},  {\em Phys. Rev. Lett.}
  {\bf 126} (2021), no.~1 011102, [\href{http://arxiv.org/abs/2010.06620}{{\tt
  arXiv:2010.06620}}].

\bibitem{Li:2020rzo}
S.~Y. Li, Y.-C. Qiu, and S.~H.~H. Tye, {\it {Standard Model from A Supergravity
  Model with a Naturally Small Cosmological Constant}},  {\em JHEP} {\bf 05}
  (2021) 181, [\href{http://arxiv.org/abs/2010.10089}{{\tt arXiv:2010.10089}}].

\bibitem{Sumitomo:2013vla}
Y.~Sumitomo, S.~Tye, and S.~S. Wong, {\it {Statistical Distribution of the
  Vacuum Energy Density in Racetrack K{\"a}hler Uplift Models in String
  Theory}},  {\em JHEP} {\bf 07} (2013) 052,
  [\href{http://arxiv.org/abs/1305.0753}{{\tt arXiv:1305.0753}}].

\bibitem{Qiu:2020los}
Y.-C. Qiu and S.~H.~H. Tye, {\it {Linking the Supersymmetric Standard Model to
  the Cosmological Constant}},  {\em JHEP} {\bf 01} (2021) 117,
  [\href{http://arxiv.org/abs/2006.16620}{{\tt arXiv:2006.16620}}].

\bibitem{Coriano:2005own}
C.~Coriano, N.~Irges, and E.~Kiritsis, {\it {On the effective theory of low
  scale orientifold string vacua}},  {\em Nucl. Phys. B} {\bf 746} (2006)
  77--135, [\href{http://arxiv.org/abs/hep-ph/0510332}{{\tt hep-ph/0510332}}].

\bibitem{Preskill:1982cy}
J.~Preskill, M.~B. Wise, and F.~Wilczek, {\it {Cosmology of the Invisible
  Axion}},  {\em Phys. Lett. B} {\bf 120} (1983) 127--132.

\bibitem{Abbott:1982af}
L.~F. Abbott and P.~Sikivie, {\it {A Cosmological Bound on the Invisible
  Axion}},  {\em Phys. Lett. B} {\bf 120} (1983) 133--136.

\bibitem{Dine:1982ah}
M.~Dine and W.~Fischler, {\it {The Not So Harmless Axion}},  {\em Phys. Lett.
  B} {\bf 120} (1983) 137--141.

\bibitem{Hui:2016ltb}
L.~Hui, J.~P. Ostriker, S.~Tremaine, and E.~Witten, {\it {Ultralight scalars as
  cosmological dark matter}},  {\em Phys. Rev. D} {\bf 95} (2017), no.~4
  043541, [\href{http://arxiv.org/abs/1610.08297}{{\tt arXiv:1610.08297}}].

\bibitem{Tye:2016jzi}
S.~H.~H. Tye and S.~S.~C. Wong, {\it {Linking Light Scalar Modes with A Small
  Positive Cosmological Constant in String Theory}},  {\em JHEP} {\bf 06}
  (2017) 094, [\href{http://arxiv.org/abs/1611.05786}{{\tt arXiv:1611.05786}}].

\bibitem{Pitrou:2018cgg}
C.~Pitrou, A.~Coc, J.-P. Uzan, and E.~Vangioni, {\it {Precision big bang
  nucleosynthesis with improved Helium-4 predictions}},  {\em Phys. Rept.} {\bf
  754} (2018) 1--66, [\href{http://arxiv.org/abs/1801.08023}{{\tt
  arXiv:1801.08023}}].

\bibitem{Pogosian:2020ded}
L.~Pogosian, G.-B. Zhao, and K.~Jedamzik, {\it {Recombination-independent
  determination of the sound horizon and the Hubble constant from BAO}},  {\em
  Astrophys. J. Lett.} {\bf 904} (2020), no.~2 L17,
  [\href{http://arxiv.org/abs/2009.08455}{{\tt arXiv:2009.08455}}].

\bibitem{Carroll:1989vb}
S.~M. Carroll, G.~B. Field, and R.~Jackiw, {\it {Limits on a Lorentz and Parity
  Violating Modification of Electrodynamics}},  {\em Phys. Rev. D} {\bf 41}
  (1990) 1231.

\bibitem{Carroll:1991zs}
S.~M. Carroll and G.~B. Field, {\it {The Einstein equivalence principle and the
  polarization of radio galaxies}},  {\em Phys. Rev. D} {\bf 43} (1991) 3789.

\bibitem{Harari:1992ea}
D.~Harari and P.~Sikivie, {\it {Effects of a Nambu-Goldstone boson on the
  polarization of radio galaxies and the cosmic microwave background}},  {\em
  Phys. Lett. B} {\bf 289} (1992) 67--72.

\bibitem{Levshakov:2020ule}
S.~A. Levshakov, M.~G. Kozlov, and I.~I. Agafonova, {\it {Constraints on the
  electron-to-proton mass ratio variation at the epoch of reionization}},  {\em
  Mon. Not. Roy. Astron. Soc.} {\bf 498} (2020), no.~3 3624--3632,
  [\href{http://arxiv.org/abs/2008.11143}{{\tt arXiv:2008.11143}}].

\bibitem{Hu:2000ke}
W.~Hu, R.~Barkana, and A.~Gruzinov, {\it {Cold and fuzzy dark matter}},  {\em
  Phys. Rev. Lett.} {\bf 85} (2000) 1158--1161,
  [\href{http://arxiv.org/abs/astro-ph/0003365}{{\tt astro-ph/0003365}}].

\bibitem{Schive:2014dra}
H.-Y. Schive, T.~Chiueh, and T.~Broadhurst, {\it {Cosmic Structure as the
  Quantum Interference of a Coherent Dark Wave}},  {\em Nature Phys.} {\bf 10}
  (2014) 496--499, [\href{http://arxiv.org/abs/1406.6586}{{\tt
  arXiv:1406.6586}}].

\bibitem{Marsh:2015xka}
D.~J.~E. Marsh, {\it {Axion Cosmology}},  {\em Phys. Rept.} {\bf 643} (2016)
  1--79, [\href{http://arxiv.org/abs/1510.07633}{{\tt arXiv:1510.07633}}].

\bibitem{Aver:2015iza}
E.~Aver, K.~A. Olive, and E.~D. Skillman, {\it {The effects of He I
  \ensuremath{\lambda}10830 on helium abundance determinations}},  {\em JCAP}
  {\bf 07} (2015) 011, [\href{http://arxiv.org/abs/1503.08146}{{\tt
  arXiv:1503.08146}}].

\bibitem{Cooke:2017cwo}
R.~J. Cooke, M.~Pettini, and C.~C. Steidel, {\it {One Percent Determination of
  the Primordial Deuterium Abundance}},  {\em Astrophys. J.} {\bf 855} (2018),
  no.~2 102, [\href{http://arxiv.org/abs/1710.11129}{{\tt arXiv:1710.11129}}].

\bibitem{Sbordone:2010zi}
L.~Sbordone et~al., {\it {The metal-poor end of the Spite plateau. 1: Stellar
  parameters, metallicities and lithium abundances}},  {\em Astron. Astrophys.}
  {\bf 522} (2010) A26, [\href{http://arxiv.org/abs/1003.4510}{{\tt
  arXiv:1003.4510}}].

\bibitem{Fields:2011zzb}
B.~D. Fields, {\it {The primordial lithium problem}},  {\em Ann. Rev. Nucl.
  Part. Sci.} {\bf 61} (2011) 47--68,
  [\href{http://arxiv.org/abs/1203.3551}{{\tt arXiv:1203.3551}}].

\bibitem{Hayakawa:2020bjr}
S.~Hayakawa et~al., {\it {Experimental Study on the 7Be\((n,p)\)7Li and the
  7Be\((n,\alpha )\)4He Reactions for Cosmological Lithium Problem}},  {\em JPS
  Conf. Proc.} {\bf 31} (2020) 011036.

\bibitem{Ishikawa:2020fbm}
S.~Ishikawa et~al., {\it {Experimental Study of the 7Be\((n,p_{1})\)7Li*
  Reaction for the Cosmological Lithium Problem}},  {\em JPS Conf. Proc.} {\bf
  31} (2020) 011037.

\bibitem{Clara:2020efx}
M.~Clara and C.~Martins, {\it {Primordial nucleosynthesis with varying
  fundamental constants: Improved constraints and a possible solution to the
  Lithium problem}},  {\em Astron. Astrophys.} {\bf 633} (2020) L11,
  [\href{http://arxiv.org/abs/2001.01787}{{\tt arXiv:2001.01787}}].

\bibitem{Iliadis:2020jtc}
C.~Iliadis and A.~Coc, {\it {Thermonuclear reaction rates and primordial
  nucleosynthesis}},  {\em Astrophys. J.} {\bf 901} (2020), no.~2 127,
  [\href{http://arxiv.org/abs/2008.12200}{{\tt arXiv:2008.12200}}].

\bibitem{Gupta:2020wgz}
R.~P. Gupta, {\it {Do varying physical constants provide solution to the
  lithium problem?}},  \href{http://arxiv.org/abs/2010.13628}{{\tt
  arXiv:2010.13628}}.

\bibitem{Zyla:2020zbs}
{\bf Particle Data Group} Collaboration, P.~Zyla et~al., {\it {Review of
  Particle Physics}},  {\em PTEP} {\bf 2020} (2020), no.~8 083C01.

\bibitem{Fields:2019pfx}
B.~D. Fields, K.~A. Olive, T.-H. Yeh, and C.~Young, {\it {Big-Bang
  Nucleosynthesis After Planck}},  {\em JCAP} {\bf 03} (2020) 010,
  [\href{http://arxiv.org/abs/1912.01132}{{\tt arXiv:1912.01132}}].

\bibitem{Walker-Loud:2014iea}
A.~Walker-Loud, {\it {Nuclear Physics Review}},  {\em PoS} {\bf LATTICE2013}
  (2014) 013, [\href{http://arxiv.org/abs/1401.8259}{{\tt arXiv:1401.8259}}].

\bibitem{Flambaum:2007mj}
V.~Flambaum and R.~B. Wiringa, {\it {Dependence of nuclear binding on hadronic
  mass variation}},  {\em Phys. Rev. C} {\bf 76} (2007) 054002,
  [\href{http://arxiv.org/abs/0709.0077}{{\tt arXiv:0709.0077}}].

\bibitem{Berengut:2009js}
J.~Berengut, V.~Flambaum, and V.~Dmitriev, {\it {Effect of quark-mass variation
  on big bang nucleosynthesis}},  {\em Phys. Lett. B} {\bf 683} (2010)
  114--118, [\href{http://arxiv.org/abs/0907.2288}{{\tt arXiv:0907.2288}}].

\bibitem{Lesgourgues:2012uu}
J.~Lesgourgues and S.~Pastor, {\it {Neutrino mass from Cosmology}},  {\em Adv.
  High Energy Phys.} {\bf 2012} (2012) 608515,
  [\href{http://arxiv.org/abs/1212.6154}{{\tt arXiv:1212.6154}}].

\bibitem{Hart:2017ndk}
L.~Hart and J.~Chluba, {\it {New constraints on time-dependent variations of
  fundamental constants using Planck data}},  {\em Mon. Not. Roy. Astron. Soc.}
  {\bf 474} (2018), no.~2 1850--1861,
  [\href{http://arxiv.org/abs/1705.03925}{{\tt arXiv:1705.03925}}].

\bibitem{Alam:2020sor}
{\bf eBOSS} Collaboration, S.~Alam et~al., {\it {The Completed SDSS-IV extended
  Baryon Oscillation Spectroscopic Survey: Cosmological Implications from two
  Decades of Spectroscopic Surveys at the Apache Point observatory}},
  \href{http://arxiv.org/abs/2007.08991}{{\tt arXiv:2007.08991}}.

\bibitem{Zhao:2020tis}
G.-B. Zhao et~al., {\it {The Completed SDSS-IV extended Baryon Oscillation
  Spectroscopic Survey: a multi-tracer analysis in Fourier space for measuring
  the cosmic structure growth and expansion rate}},
  \href{http://arxiv.org/abs/2007.09011}{{\tt arXiv:2007.09011}}.

\bibitem{Wang:2020tje}
Y.~Wang et~al., {\it {The clustering of the SDSS-IV extended Baryon Oscillation
  Spectroscopic Survey DR16 luminous red galaxy and emission line galaxy
  samples: cosmic distance and structure growth measurements using multiple
  tracers in configuration space}},  {\em Mon. Not. Roy. Astron. Soc.} {\bf
  498} (2020), no.~3 3470--3483, [\href{http://arxiv.org/abs/2007.09010}{{\tt
  arXiv:2007.09010}}].

\bibitem{Hou:2020rse}
J.~Hou et~al., {\it {The Completed SDSS-IV extended Baryon Oscillation
  Spectroscopic Survey: BAO and RSD measurements from anisotropic clustering
  analysis of the Quasar Sample in configuration space between redshift 0.8 and
  2.2}},  {\em Mon. Not. Roy. Astron. Soc.} {\bf 500} (2020), no.~1 1201--1221,
  [\href{http://arxiv.org/abs/2007.08998}{{\tt arXiv:2007.08998}}].

\bibitem{duMasdesBourboux:2020pck}
H.~du~Mas~des Bourboux et~al., {\it {The Completed SDSS-IV Extended Baryon
  Oscillation Spectroscopic Survey: Baryon Acoustic Oscillations with
  Ly\ensuremath{\alpha} Forests}},  {\em Astrophys. J.} {\bf 901} (2020), no.~2
  153, [\href{http://arxiv.org/abs/2007.08995}{{\tt arXiv:2007.08995}}].

\bibitem{Beutler_2011}
F.~Beutler, C.~Blake, M.~Colless, D.~H. Jones, L.~Staveley-Smith, L.~Campbell,
  Q.~Parker, W.~Saunders, and F.~Watson, {\it The 6df galaxy survey: baryon
  acoustic oscillations and the local hubble constant},  {\em Monthly Notices
  of the Royal Astronomical Society} {\bf 416} (Jul, 2011) 3017–3032.

\bibitem{Ross:2014qpa}
A.~J. Ross, L.~Samushia, C.~Howlett, W.~J. Percival, A.~Burden, and M.~Manera,
  {\it {The clustering of the SDSS DR7 main Galaxy sample \textendash{} I. A 4
  per cent distance measure at $z = 0.15$}},  {\em Mon. Not. Roy. Astron. Soc.}
  {\bf 449} (2015), no.~1 835--847, [\href{http://arxiv.org/abs/1409.3242}{{\tt
  arXiv:1409.3242}}].

\bibitem{Lewis:2019xzd}
A.~Lewis, {\it {GetDist: a Python package for analysing Monte Carlo samples}},
  \href{http://arxiv.org/abs/1910.13970}{{\tt arXiv:1910.13970}}.

\bibitem{Riess:2019cxk}
A.~G. Riess, S.~Casertano, W.~Yuan, L.~M. Macri, and D.~Scolnic, {\it {Large
  Magellanic Cloud Cepheid Standards Provide a 1\% Foundation for the
  Determination of the Hubble Constant and Stronger Evidence for Physics beyond
  $\Lambda$CDM}},  {\em Astrophys. J.} {\bf 876} (2019), no.~1 85,
  [\href{http://arxiv.org/abs/1903.07603}{{\tt arXiv:1903.07603}}].

\bibitem{Wong:2019kwg}
K.~C. Wong et~al., {\it {H0LiCOW \textendash{} XIII. A 2.4 per cent measurement
  of H0 from lensed quasars: 5.3\ensuremath{\sigma} tension between early- and
  late-Universe probes}},  {\em Mon. Not. Roy. Astron. Soc.} {\bf 498} (2020),
  no.~1 1420--1439, [\href{http://arxiv.org/abs/1907.04869}{{\tt
  arXiv:1907.04869}}].

\bibitem{Reid:2008nm}
M.~Reid, J.~Braatz, J.~Condon, L.~Greenhill, C.~Henkel, and K.~Lo, {\it {The
  Megamaser Cosmology Project: I. VLBI observations of UGC 3789}},  {\em
  Astrophys. J.} {\bf 695} (2009) 287--291,
  [\href{http://arxiv.org/abs/0811.4345}{{\tt arXiv:0811.4345}}].

\bibitem{Freedman:2019jwv}
W.~L. Freedman et~al., {\it {The Carnegie-Chicago Hubble Program. VIII. An
  Independent Determination of the Hubble Constant Based on the Tip of the Red
  Giant Branch}},  \href{http://arxiv.org/abs/1907.05922}{{\tt
  arXiv:1907.05922}}.

\bibitem{Potter:2018}
C.~Potter, J.~B. Jensen, J.~Blakeslee, et~al., {\it Calibrating the type ia
  supernova distance scale using surface brightness fluctuations},  {\em
  American Astronomical Society Meeting Abstracts \#} {\bf 232} (2018) 232.

\bibitem{Huang:2018dbn}
C.~D. Huang et~al., {\it {A Near-infrared Period\textendash{}Luminosity
  Relation for Miras in NGC 4258, an Anchor for a New Distance Ladder}},  {\em
  Astrophys. J.} {\bf 857} (2018), no.~1 67,
  [\href{http://arxiv.org/abs/1801.02711}{{\tt arXiv:1801.02711}}].

\bibitem{Gil-Marin:2015nqa}
H.~Gil-Mar\'\i{}n et~al., {\it {The clustering of galaxies in the SDSS-III
  Baryon Oscillation Spectroscopic Survey: BAO measurement from the
  LOS-dependent power spectrum of DR12 BOSS galaxies}},  {\em Mon. Not. Roy.
  Astron. Soc.} {\bf 460} (2016), no.~4 4210--4219,
  [\href{http://arxiv.org/abs/1509.06373}{{\tt arXiv:1509.06373}}].

\bibitem{Hu:1995kot}
W.~Hu, N.~Sugiyama, and J.~Silk, {\it {The Physics of microwave background
  anisotropies}},  {\em Nature} {\bf 386} (1997) 37--43,
  [\href{http://arxiv.org/abs/astro-ph/9604166}{{\tt astro-ph/9604166}}].

\bibitem{Hu:2000ti}
W.~Hu, M.~Fukugita, M.~Zaldarriaga, and M.~Tegmark, {\it {CMB observables and
  their cosmological implications}},  {\em Astrophys. J.} {\bf 549} (2001) 669,
  [\href{http://arxiv.org/abs/astro-ph/0006436}{{\tt astro-ph/0006436}}].

\bibitem{Hu:2001bc}
W.~Hu and S.~Dodelson, {\it {Cosmic Microwave Background Anisotropies}},  {\em
  Ann. Rev. Astron. Astrophys.} {\bf 40} (2002) 171--216,
  [\href{http://arxiv.org/abs/astro-ph/0110414}{{\tt astro-ph/0110414}}].

\bibitem{Kachru:2003aw}
S.~Kachru, R.~Kallosh, A.~D. Linde, and S.~P. Trivedi, {\it {De Sitter vacua in
  string theory}},  {\em Phys. Rev. D} {\bf 68} (2003) 046005,
  [\href{http://arxiv.org/abs/hep-th/0301240}{{\tt hep-th/0301240}}].

\bibitem{Cribiori:2019hod}
N.~Cribiori, C.~Roupec, T.~Wrase, and Y.~Yamada, {\it {Supersymmetric
  anti-D3-brane action in the Kachru-Kallosh-Linde-Trivedi setup}},  {\em Phys.
  Rev. D} {\bf 100} (2019), no.~6 066001,
  [\href{http://arxiv.org/abs/1906.07727}{{\tt arXiv:1906.07727}}].

\bibitem{Parameswaran:2020ukp}
S.~Parameswaran and F.~Tonioni, {\it {Non-supersymmetric String Models from
  Anti-D3-/D7-branes in Strongly Warped Throats}},
  \href{http://arxiv.org/abs/2007.11333}{{\tt arXiv:2007.11333}}.

\bibitem{Vercnocke:2016fbt}
B.~Vercnocke and T.~Wrase, {\it {Constrained superfields from an anti-D3-brane
  in KKLT}},  {\em JHEP} {\bf 08} (2016) 132,
  [\href{http://arxiv.org/abs/1605.03961}{{\tt arXiv:1605.03961}}].

\bibitem{Antoniadis:2010hs}
I.~Antoniadis, E.~Dudas, D.~Ghilencea, and P.~Tziveloglou, {\it {Non-linear
  MSSM}},  {\em Nucl. Phys. B} {\bf 841} (2010) 157--177,
  [\href{http://arxiv.org/abs/1006.1662}{{\tt arXiv:1006.1662}}].

\bibitem{Kallosh:2015nia}
R.~Kallosh, F.~Quevedo, and A.~M. Uranga, {\it {String Theory Realizations of
  the Nilpotent Goldstino}},  {\em JHEP} {\bf 12} (2015) 039,
  [\href{http://arxiv.org/abs/1507.07556}{{\tt arXiv:1507.07556}}].

\bibitem{GarciadelMoral:2017vnz}
M.~P. Garcia~del Moral, S.~Parameswaran, N.~Quiroz, and I.~Zavala, {\it
  {Anti-D3 branes and moduli in non-linear supergravity}},  {\em JHEP} {\bf 10}
  (2017) 185, [\href{http://arxiv.org/abs/1707.07059}{{\tt arXiv:1707.07059}}].

\bibitem{Komargodski:2009rz}
Z.~Komargodski and N.~Seiberg, {\it {From Linear SUSY to Constrained
  Superfields}},  {\em JHEP} {\bf 09} (2009) 066,
  [\href{http://arxiv.org/abs/0907.2441}{{\tt arXiv:0907.2441}}].

\bibitem{Dudas:2019pls}
E.~Dudas and S.~L\"ust, {\it {An update on moduli stabilization with antibrane
  uplift}},  \href{http://arxiv.org/abs/1912.09948}{{\tt arXiv:1912.09948}}.

\bibitem{Uzan:2010pm}
J.-P. Uzan, {\it {Varying Constants, Gravitation and Cosmology}},  {\em Living
  Rev. Rel.} {\bf 14} (2011) 2, [\href{http://arxiv.org/abs/1009.5514}{{\tt
  arXiv:1009.5514}}].

\bibitem{Safronova:2017xyt}
M.~S. Safronova, D.~Budker, D.~DeMille, D.~F.~J. Kimball, A.~Derevianko, and
  C.~W. Clark, {\it {Search for New Physics with Atoms and Molecules}},  {\em
  Rev. Mod. Phys.} {\bf 90} (2018), no.~2 025008,
  [\href{http://arxiv.org/abs/1710.01833}{{\tt arXiv:1710.01833}}].

\bibitem{Fujita:2020ecn}
T.~Fujita, K.~Murai, H.~Nakatsuka, and S.~Tsujikawa, {\it {Detection of
  isotropic cosmic birefringence and its implications for axion-like particles
  including dark energy}},  \href{http://arxiv.org/abs/2011.11894}{{\tt
  arXiv:2011.11894}}.

\bibitem{Andriolo:2018dee}
S.~Andriolo, S.~Y. Li, and S.~H.~H. Tye, {\it {The Cosmological Constant and
  the Electroweak Scale}},  {\em JHEP} {\bf 10} (2019) 212,
  [\href{http://arxiv.org/abs/1812.04873}{{\tt arXiv:1812.04873}}].

\bibitem{DiValentino:2018gcu}
E.~Di~Valentino and S.~Bridle, {\it {Exploring the Tension between Current
  Cosmic Microwave Background and Cosmic Shear Data}},  {\em Symmetry} {\bf 10}
  (2018), no.~11 585.

\bibitem{Heymans_2021}
C.~Heymans, T.~Tröster, M.~Asgari, C.~Blake, H.~Hildebrandt, B.~Joachimi,
  K.~Kuijken, C.-A. Lin, A.~G. Sánchez, J.~L. van~den Busch, and et~al., {\it
  Kids-1000 cosmology: Multi-probe weak gravitational lensing and spectroscopic
  galaxy clustering constraints},  {\em Astron. Astrophys.} {\bf 646} (Feb,
  2021) A140.

\bibitem{Rykoff:2013ovv}
{\bf SDSS} Collaboration, E.~S. Rykoff et~al., {\it {redMaPPer I: Algorithm and
  SDSS DR8 Catalog}},  {\em Astrophys. J.} {\bf 785} (2014) 104,
  [\href{http://arxiv.org/abs/1303.3562}{{\tt arXiv:1303.3562}}].

\bibitem{Hlozek:2014lca}
R.~Hlozek, D.~Grin, D.~J.~E. Marsh, and P.~G. Ferreira, {\it {A search for
  ultralight axions using precision cosmological data}},  {\em Phys. Rev. D}
  {\bf 91} (2015), no.~10 103512, [\href{http://arxiv.org/abs/1410.2896}{{\tt
  arXiv:1410.2896}}].

\bibitem{Hlozek:2017zzf}
R.~Hlozek, D.~J.~E. Marsh, and D.~Grin, {\it {Using the Full Power of the
  Cosmic Microwave Background to Probe Axion Dark Matter}},  {\em Mon. Not.
  Roy. Astron. Soc.} {\bf 476} (2018), no.~3 3063--3085,
  [\href{http://arxiv.org/abs/1708.05681}{{\tt arXiv:1708.05681}}].

\bibitem{Kobayashi:2017jcf}
T.~Kobayashi, R.~Murgia, A.~De~Simone, V.~Ir\v{s}i\v{c}, and M.~Viel, {\it
  {Lyman-$\alpha$ constraints on ultralight scalar dark matter: Implications
  for the early and late universe}},  {\em Phys. Rev. D} {\bf 96} (2017),
  no.~12 123514, [\href{http://arxiv.org/abs/1708.00015}{{\tt
  arXiv:1708.00015}}].

\bibitem{scalar-field-structure-formation}
M.~I. {Khlopov}, B.~A. {Malomed}, and I.~B. {Zeldovich}, {\it {Gravitational
  instability of scalar fields and formation of primordial black holes}},  {\em
  mnras} {\bf 215} (Aug., 1985) 575--589.

\bibitem{Marsh:2010wq}
D.~J.~E. Marsh and P.~G. Ferreira, {\it {Ultra-Light Scalar Fields and the
  Growth of Structure in the Universe}},  {\em Phys. Rev. D} {\bf 82} (2010)
  103528, [\href{http://arxiv.org/abs/1009.3501}{{\tt arXiv:1009.3501}}].

\bibitem{Lepora:1998ix}
N.~F. Lepora, {\it {Cosmological birefringence and the microwave background}},
  \href{http://arxiv.org/abs/gr-qc/9812077}{{\tt gr-qc/9812077}}.

\bibitem{Lue:1998mq}
A.~Lue, L.-M. Wang, and M.~Kamionkowski, {\it {Cosmological signature of new
  parity violating interactions}},  {\em Phys. Rev. Lett.} {\bf 83} (1999)
  1506--1509, [\href{http://arxiv.org/abs/astro-ph/9812088}{{\tt
  astro-ph/9812088}}].

\bibitem{Antonucci:1993sg}
R.~Antonucci, {\it {Unified models for active galactic nuclei and quasars}},
  {\em Ann. Rev. Astron. Astrophys.} {\bf 31} (1993) 473--521.

\bibitem{Liu:2019brz}
T.~Liu, G.~Smoot, and Y.~Zhao, {\it {Detecting axionlike dark matter with
  linearly polarized pulsar light}},  {\em Phys. Rev. D} {\bf 101} (2020),
  no.~6 063012, [\href{http://arxiv.org/abs/1901.10981}{{\tt
  arXiv:1901.10981}}].

\bibitem{Caputo:2019tms}
A.~Caputo, L.~Sberna, M.~Frias, D.~Blas, P.~Pani, L.~Shao, and W.~Yan, {\it
  {Constraints on millicharged dark matter and axionlike particles from timing
  of radio waves}},  {\em Phys. Rev. D} {\bf 100} (2019), no.~6 063515,
  [\href{http://arxiv.org/abs/1902.02695}{{\tt arXiv:1902.02695}}].

\bibitem{Fujita:2018zaj}
T.~Fujita, R.~Tazaki, and K.~Toma, {\it {Hunting Axion Dark Matter with
  Protoplanetary Disk Polarimetry}},  {\em Phys. Rev. Lett.} {\bf 122} (2019),
  no.~19 191101, [\href{http://arxiv.org/abs/1811.03525}{{\tt
  arXiv:1811.03525}}].

\bibitem{Chen:2019fsq}
Y.~Chen, J.~Shu, X.~Xue, Q.~Yuan, and Y.~Zhao, {\it {Probing Axions with Event
  Horizon Telescope Polarimetric Measurements}},  {\em Phys. Rev. Lett.} {\bf
  124} (2020), no.~6 061102, [\href{http://arxiv.org/abs/1905.02213}{{\tt
  arXiv:1905.02213}}].

\bibitem{Feng:2004mq}
B.~Feng, H.~Li, M.~Li, and X.~Zhang, {\it {Gravitational leptogenesis and its
  signatures in CMB}},  {\em Phys. Lett.} {\bf B620} (2005) 27--32,
  [\href{http://arxiv.org/abs/hep-ph/0406269}{{\tt hep-ph/0406269}}].

\bibitem{Liu:2006uh}
G.-C. Liu, S.~Lee, and K.-W. Ng, {\it {Effect on cosmic microwave background
  polarization of coupling of quintessence to pseudoscalar formed from the
  electromagnetic field and its dual}},  {\em Phys. Rev. Lett.} {\bf 97} (2006)
  161303, [\href{http://arxiv.org/abs/astro-ph/0606248}{{\tt
  astro-ph/0606248}}].

\bibitem{Guena:2012zz}
J.~Guena, M.~Abgrall, D.~Rovera, P.~Rosenbusch, M.~E. Tobar, P.~Laurent,
  A.~Clairon, and S.~Bize, {\it {Improved Tests of Local Position Invariance
  Using Rb-87 and Cs-133 Fountains}},  {\em Phys. Rev. Lett.} {\bf 109} (2012)
  080801, [\href{http://arxiv.org/abs/1205.4235}{{\tt arXiv:1205.4235}}].

\bibitem{Casey:2014hya}
C.~M. Casey, D.~Narayanan, and A.~Cooray, {\it {Dusty Star-Forming Galaxies at
  High Redshift}},  {\em Phys. Rept.} {\bf 541} (2014) 45--161,
  [\href{http://arxiv.org/abs/1402.1456}{{\tt arXiv:1402.1456}}].

\bibitem{ushijima2015cryogenic}
I.~Ushijima, M.~Takamoto, M.~Das, T.~Ohkubo, and H.~Katori, {\it Cryogenic
  optical lattice clocks},  {\em Nature Photonics} {\bf 9} (2015), no.~3
  185--189.

\bibitem{Kozlov:2018mbp}
M.~G. Kozlov, M.~S. Safronova, J.~R. Crespo L\'opez-Urrutia, and P.~O. Schmidt,
  {\it {Highly charged ions: Optical clocks and applications in fundamental
  physics}},  {\em Rev. Mod. Phys.} {\bf 90} (2018), no.~4 045005,
  [\href{http://arxiv.org/abs/1803.06532}{{\tt arXiv:1803.06532}}].

\bibitem{Campbell:2012zzb}
C.~J. Campbell, A.~G. Radnaev, A.~Kuzmich, V.~A. Dzuba, V.~V. Flambaum, and
  A.~Derevianko, {\it {A Single-Ion Nuclear Clock for Metrology at the 19th
  Decimal Place}},  {\em Phys. Rev. Lett.} {\bf 108} (2012) 120802,
  [\href{http://arxiv.org/abs/1110.2490}{{\tt arXiv:1110.2490}}].

\bibitem{Peik:2020cwm}
E.~Peik, T.~Schumm, M.~S. Safronova, A.~P\'alffy, J.~Weitenberg, and P.~G.
  Thirolf, {\it {Nuclear clocks for testing fundamental physics}},
  \href{http://arxiv.org/abs/2012.09304}{{\tt arXiv:2012.09304}}.

\bibitem{Skidmore:2015lga}
{\bf TMT International Science Development Teams \& TMT Science Advisory
  Committee} Collaboration, W.~Skidmore et~al., {\it {Thirty Meter Telescope
  Detailed Science Case: 2015}},  {\em Res. Astron. Astrophys.} {\bf 15}
  (2015), no.~12 1945--2140, [\href{http://arxiv.org/abs/1505.01195}{{\tt
  arXiv:1505.01195}}].

\bibitem{Behroozi:2020jhj}
P.~Behroozi et~al., {\it {The Universe at z \ensuremath{>} 10: predictions for
  $JWST$ from the universemachine DR1}},  {\em Mon. Not. Roy. Astron. Soc.}
  {\bf 499} (2020), no.~4 5702--5718,
  [\href{http://arxiv.org/abs/2007.04988}{{\tt arXiv:2007.04988}}].

\bibitem{Marsden:2013iyk}
D.~W. Marsden et~al., {\it {The Atacama Cosmology Telescope: Dusty Star-Forming
  Galaxies and Active Galactic Nuclei in the Southern Survey}},  {\em Mon. Not.
  Roy. Astron. Soc.} {\bf 439} (2014), no.~2 1556--1574,
  [\href{http://arxiv.org/abs/1306.2288}{{\tt arXiv:1306.2288}}].

\bibitem{Chen:2019ltf}
X.~Chen, S.~P. Ellingsen, and Y.~Mei, {\it {Astrophysical constraints on the
  proton-to-electron mass ratio with FAST}},  {\em Res. Astron. Astrophys.}
  {\bf 19} (2019) 18, [\href{http://arxiv.org/abs/1904.03871}{{\tt
  arXiv:1904.03871}}].

\bibitem{Carilli:2004nx}
C.~L. Carilli and S.~Rawlings, {\it {Science with the Square Kilometer Array:
  Motivation, key science projects, standards and assumptions}},  {\em New
  Astron. Rev.} {\bf 48} (2004) 979,
  [\href{http://arxiv.org/abs/astro-ph/0409274}{{\tt astro-ph/0409274}}].

\bibitem{Ubachs:2015fro}
W.~Ubachs, J.~Bagdonaite, E.~J. Salumbides, M.~T. Murphy, and L.~Kaper, {\it
  {Search for a drifting proton--electron mass ratio from H$_2$}},  {\em Rev.
  Mod. Phys.} {\bf 88} (2016) 021003,
  [\href{http://arxiv.org/abs/1511.04476}{{\tt arXiv:1511.04476}}].

\bibitem{Fung:2021fcj}
L.~W. Fung, L.~Li, T.~Liu, H.~N. Luu, Y.-C. Qiu, and S.~H.~H. Tye, {\it {The
  Hubble Constant in the Axi-Higgs Universe}},
  \href{http://arxiv.org/abs/2105.01631}{{\tt arXiv:2105.01631}}.

\bibitem{Lewis:1999bs}
A.~Lewis, A.~Challinor, and A.~Lasenby, {\it {Efficient computation of CMB
  anisotropies in closed FRW models}},  {\em Astrophys. J.} {\bf 538} (2000)
  473--476, [\href{http://arxiv.org/abs/astro-ph/9911177}{{\tt
  astro-ph/9911177}}].

\bibitem{Hu:1995en}
W.~Hu and N.~Sugiyama, {\it {Small scale cosmological perturbations: An
  Analytic approach}},  {\em Astrophys. J.} {\bf 471} (1996) 542--570,
  [\href{http://arxiv.org/abs/astro-ph/9510117}{{\tt astro-ph/9510117}}].

\bibitem{Chluba:2010ca}
J.~Chluba and R.~Thomas, {\it {Towards a complete treatment of the cosmological
  recombination problem}},  {\em Mon. Not. Roy. Astron. Soc.} {\bf 412} (2011)
  748, [\href{http://arxiv.org/abs/1010.3631}{{\tt arXiv:1010.3631}}].

\bibitem{Rubi_o_Mart_n_2010}
J.~A. Rubiño-Martín, J.~Chluba, W.~A. Fendt, and B.~D. Wandelt, {\it
  Estimating the impact of recombination uncertainties on the cosmological
  parameter constraints from cosmic microwave background experiments},  {\em
  Monthly Notices of the Royal Astronomical Society} {\bf 403} (Mar, 2010)
  439–452.

\bibitem{Chluba_2010_a}
J.~Chluba, {\it Could the cosmological recombination spectrum help us
  understand annihilating dark matter?},  {\em Monthly Notices of the Royal
  Astronomical Society} {\bf 402} (Feb, 2010) 1195–1207.

\bibitem{Chluba_2010_b}
J.~Chluba, G.~M. Vasil, and L.~J. Dursi, {\it Recombinations to the rydberg
  states of hydrogen and their effect during the cosmological recombination
  epoch},  {\em Monthly Notices of the Royal Astronomical Society} {\bf 407}
  (Jul, 2010) 599–612.

\bibitem{Seager:1999bc}
S.~Seager, D.~D. Sasselov, and D.~Scott, {\it {A new calculation of the
  recombination epoch}},  {\em Astrophys. J. Lett.} {\bf 523} (1999) L1--L5,
  [\href{http://arxiv.org/abs/astro-ph/9909275}{{\tt astro-ph/9909275}}].

\bibitem{Switzer:2007sn}
E.~R. Switzer and C.~M. Hirata, {\it {Primordial helium recombination. 1.
  Feedback, line transfer, and continuum opacity}},  {\em Phys. Rev. D} {\bf
  77} (2008) 083006, [\href{http://arxiv.org/abs/astro-ph/0702143}{{\tt
  astro-ph/0702143}}].

\bibitem{Grin_2010}
D.~Grin and C.~M. Hirata, {\it Cosmological hydrogen recombination: The effect
  of extremely high-nstates},  {\em Physical Review D} {\bf 81} (Apr, 2010).

\bibitem{Ali_Ha_moud_2010}
Y.~Ali-Haïmoud and C.~M. Hirata, {\it Ultrafast effective multilevel atom
  method for primordial hydrogen recombination},  {\em Physical Review D} {\bf
  82} (Sep, 2010).

\end{thebibliography}\endgroup

\end{document}